\documentclass[submission, Phys]{SciPost}

\hypersetup{
    colorlinks,
    linkcolor={red!50!black},
    citecolor={blue!50!black},
    urlcolor={blue!80!black}
}

\usepackage[bitstream-charter]{mathdesign}
\DeclareSymbolFont{usualmathcal}{OMS}{cmsy}{m}{n}
\DeclareSymbolFontAlphabet{\mathcal}{usualmathcal}

\begin{document}

\begin{center}{\Large \textbf{
$q$th-root non-Hermitian Floquet topological insulators\\
}}\end{center}

\begin{center}
Longwen Zhou\textsuperscript{1$\star$},
Raditya Weda Bomantara\textsuperscript{2$\dagger$} and
Shenlin Wu\textsuperscript{1}
\end{center}

\begin{center}
{\bf 1} College of Physics and Optoelectronic Engineering, Ocean University of China, Qingdao 266100, China
\\
{\bf 2} Centre for Engineered Quantum Systems, School of Physics, University of Sydney, Sydney, New South Wales 2006, Australia
\\
${}^\star$ {\small \sf zhoulw13@u.nus.edu}
${}^\dagger$ {\small \sf Raditya.Bomantara@sydney.edu.au}
\end{center}

\begin{center}
\today
\end{center}


\section*{Abstract}
{\bf
Floquet phases of matter have attracted great attention due to their
dynamical and topological nature that are unique to nonequilibrium
settings. In this work, we introduce a generic way of taking any integer $q$th-root of the evolution operator $U$ that describes
Floquet topological matter. We further apply our $q$th-rooting procedure to obtain $2^n$th- and $3^n$th-root first- and second-order
non-Hermitian Floquet topological insulators~(FTIs). There, we explicitly demonstrate the presence of multiple edge and
corner modes at fractional quasienergies $\pm(0,1,...2^{n})\pi/2^{n}$ and $\pm(0,1,...,3^{n})\pi/3^{n}$, 
whose numbers are highly controllable and capturable by the topological invariants of their parent systems. Notably, we
observe non-Hermiticity induced fractional-quasienergy corner modes
and the coexistence of non-Hermitian skin effect with fractional-quasienergy edge states. Our findings thus establish a framework of constructing
an intriguing class of topological matter in Floquet open systems.
}

\vspace{10pt}
\noindent\rule{\textwidth}{1pt}
\tableofcontents\thispagestyle{fancy}
\noindent\rule{\textwidth}{1pt}
\vspace{10pt}

\section{Introduction}
\label{sec:Int}
Periodically driven (Floquet) systems have attracted perennial interest owing
to their fascinating dynamical, topological and transport properties
(see Refs.~\cite{FTPRev1,FTPRev2,FTPRev3,FTPRev4,FTPRev5,FTPRev6} for reviews).
Theoretical classifications of Floquet matter 
have been achieved for both free \cite{FTPClass1,FTPClass2,FTPClass3}
and interacting \cite{FTPClass4,FTPClass5,FTPClass6} systems. Experimental
observations of Floquet phases have also been made in cold atoms \cite{FCAExp1,FCAExp2,FCAExp3},
photonics \cite{FPHExp1,FPHExp2,FPHExp3} and solid state materials
\cite{FSSExp1,FSSExp2,FSSExp3}, boosting the developments of new
ideas in ultrafast electronics \cite{FTPRev4} and topological quantum
computing \cite{FQC1,FQC2,FQC3}.

Recently, square-root topological phase is discovered 
\cite{SRTP1}, whose topological properties are inherited
from its squared parent model through a process analogous to the transition
from Klein-Gordon \cite{Gordon,Klein} to Dirac equations \cite{Dirac}
in relativistic quantum mechanics. In-gap 
edge modes are
found in tight-binding models of square-root topological insulators,
superconductors and semimetals \cite{SRTP2,SRTP3,SRTP4,SRTP5,SRTP6,SRTP7,SRTP8,SRTP9,SRTP10,SRTP11,SRTP12,SRTP13,SRTP14}.
Moreover, general rules of constructing $2^{n}$th-root topological
phases \cite{SRTP7,SRTP13,SRTP14} and their symmetry classifications \cite{SRTP11} are proposed.
Experimental evidence of square-root topological phases are reported
in photonic \cite{SRTP3}, electric \cite{SRTP4} and acoustic \cite{SRTP5}
systems.

In a periodically driven system, the central object for the description
of topological properties is the Floquet operator, which is the evolution
operator of the system over a complete driving period $T$. Taking
the square-root of such a propagator for the purpose of generating
its topological descendant is, however, a highly nontrivial task.
This can be seen by writing the Floquet operator as $U={\cal T}e^{-\frac{i}{\hbar}\int_{0}^{T}H(t)dt}=e^{-i\frac{T}{\hbar}H_{{\rm eff}}}$,
where ${\cal T}$ is the time-ordering operator, $H(t)=H(t+T)$ is
the time-periodic Hamiltonian of the system, and $H_{{\rm eff}}$
is the Floquet effective Hamiltonian obtained by formally working
out the time-ordered product in ${\cal T}e^{-\frac{i}{\hbar}\int_{0}^{T}H(t)dt}$.
We may now take the square-root of $U$ 
naively as $\sqrt{U}=e^{-i\frac{T}{\hbar}\frac{H_{{\rm eff}}}{2}}$.
However, such a trial of generating square-root Floquet topological
phases tends out to be problematic and useless. First, the exact form
of $H_{{\rm eff}}$ can be rather complicated (usually including driving-induced
long-range coupling terms), not physically obtainable, or even insufficient
to describe Floquet phases with no static counterparts such as those possessing anomalous Floquet edge modes \cite{AFTI1,AFTI2,AFTI3}. 
Second, there are no transparent ways to find $H_{{\rm eff}}$ from
$H(t)$, i.e., the parameters in $H_{{\rm eff}}$ are usually nonlinear
combinations of physical parameters in $H(t)$, such that simply reducing
the parameters of $H(t)$ by half could not yield $H_{{\rm eff}}/2$.
Even obtained, the $H_{{\rm eff}}$ and $H_{{\rm eff}}/2$ describe
essentially the same physical system up to a global constant, and
no new physics are expected to emerge following such a halving process. Therefore,
the straightforward operation, $\sqrt{U}=e^{-i\frac{T}{\hbar}\frac{H_{{\rm eff}}}{2}}$,
does not generate a desired square-root of the parent system 
$U$. 

To resolve this puzzle, a nontrivial route of taking the
square-root for $U$ is introduced \cite{FSRTP1}, which closely
follows the original idea of Dirac by adding internal degrees of freedom
for electrons before taking the square-root of their relativistic
wave equation. However, the general applicability of this idea to
the construction of Floquet models beyond taking $2^n$th-root has not been revealed. Motivated by this gap of knowledge, we propose a generic procedure to yield a variety of $q$th-root Floquet phases, where $q$ is \emph{any} arbitrary integer, not necessarily in the form of $2^n$. This is achieved by utilizing a $Z_q$ generalization of Pauli matrices as ancillary degrees of freedom. While our construction is applicable to any periodically driven systems, we focus on two timely examples of non-Hermitian Floquet matter as case studies.

The concept of topological matter has been generalized to non-Hermitian systems in recent years (see Refs.~\cite{NHRev1,NHRev2,NHRev3,NHRev4,NHRev5} for reviews).
In the presence of gain and loss or nonreciprocal effects, unique topological phenomena without
any counterparts in closed systems could emerge, such as the non-Hermitian skin effect (NHSE) \cite{NHSE1,NHSE2,NHSE3,NHSE4,NHSE5,NHSE6,NHSE7} and exceptional topological phases \cite{NHRev5}. 
The interplay between time-periodic drivings and non-Hermitian effects could further induce
intriguing phases in out-of-equilibrium situations, like the non-Hermitian Floquet topological
insulators \cite{NHFTI1,NHFTI2,NHFTI3,NHFTI4,NHFTI5,NHFTI6,NHFTI7,NHFTI8,NHFTI9,NHFTI10,NHFTI11,NHFTI12}, superconductors \cite{NHFTSC1,NHFTSC2}, semimetals \cite{NHFSM1,NHFSM2,NHFSM3,NHFSM4} and quasicrystals \cite{NHFQC1,NHFQC2,NHFQC3}. As reported in this paper, applying our $q$th-rooting procedure to such non-Hermitian Floquet phases yields even more exotic features absent in their original counterparts, such as fractional-quasienergy topological edge and corner modes.

This paper is structured as follows. In Sec.~\ref{sec:The}, we recap the strategy of Ref.~\cite{FSRTP1}, generalize it to the construction of \emph{any} $q$th-root Floquet system, and elaborate the application of this general construction for the case of $q=3$. In Sec.~\ref{sec:Mod}, we introduce two typical models of first- and second-order non-Hermitian Floquet topological insulators, whose square- and cubic-root descendants
are studied in detail in Sec.~\ref{sec:Res} as an application of our method. In Sec.~\ref{sec:Sum}, we
sum up our results and discuss potential future directions.

\section{Theory}
\label{sec:The}

We first review the approach to take the nontrivial square-root of a Floquet
system.
We set the Planck constant $\hbar=1$ and driving period $T=1$ throughout. Following Ref.~\cite{FSRTP1}, we write the one-period evolution (Floquet) operator of \emph{any} time-periodic system as
\begin{equation}
	U=U_{1}U_{2}=\left(\mathcal{T}e^{-\mathrm{i} \int_{0}^{1/2} H(t+1/2) dt} \right) \left(\mathcal{T}e^{-\mathrm{i} \int_{0}^{1/2} H(t) dt} \right), \label{eq:Ut1}
\end{equation}
where $H(t)$ is the system's Hamiltonian. The procedure of Ref.~\cite{FSRTP1} is to first enlarge Hilbert space
of $H(t)$ by introducing a pseudospin-$1/2$ degree of freedom
with the corresponding Pauli matrices $\tau_{x,y,z}$. A two-step
Hamiltonian is next defined in the enlarged Hilbert space as
\begin{equation}
	H_{1/2}(t)=\begin{cases}
		\pi\tau_{y}\otimes\mathbb{I}_{0} & t\in[\ell,\ell+\frac{1}{2})\\
		\begin{array}{c}
		\frac{\tau_{0}+\tau_{z}}{2}\otimes H(t)
		+\frac{\tau_{0}-\tau_{z}}{2}\otimes H(t+1/2)
		\end{array} & t\in[\ell+\frac{1}{2},\ell+1)
	\end{cases},\label{eq:H1ov2}
\end{equation}
where $\ell\in\mathbb{Z}$. $\tau_{0}$ is the identity in the pseudospin-$1/2$
subspace. $\mathbb{I}_{0}$ is the identity in the Hilbert space
of $H(t)$. The Floquet operator of the evolution in the enlarged
Hilbert space reads
\begin{equation}
	U_{1/2}=\begin{pmatrix}\mathcal{T}e^{-\mathrm{i} \int_{1/2}^{1} H(t) dt} & 0\\
	0 & \mathcal{T}e^{-\mathrm{i} \int_{1/2}^{1} H(t+1/2) dt}
	\end{pmatrix} e^{-i\frac{\pi}{2}\tau_{y}\otimes\mathbb{I}_{0}}.\label{eq:U1ov2}
\end{equation}
Note that $\mathcal{T}e^{-\mathrm{i} \int_{1/2}^{1} H(t) dt}=\mathcal{T}e^{-\mathrm{i} \int_{0}^{1/2} H(t+1/2) dt}=U_1$ and $\mathcal{T}e^{-\mathrm{i} \int_{1/2}^{1} H(t+1/2) dt}=\mathcal{T}e^{-\mathrm{i} \int_{0}^{1/2} H(t) dt}=U_2$. Performing the Taylor expansion and introducing $\tau_{\pm}=(\tau_{x}\pm i\tau_{y})/2$,
we find 
\begin{equation}
    U_{1/2}=\tau_{-}\otimes U_{2}-\tau_{+}\otimes U_{1}\label{eq:Uhalf}
\end{equation}
and
\begin{equation}
	U_{1/2}^{2}=e^{i\pi}\begin{pmatrix}U_{1}U_{2} & 0\\
		0 & U_{2}U_{1}
	\end{pmatrix}.\label{eq:U1ov2SQ}
\end{equation}
Since $U_{1}U_{2}=U$ and $U_{2}U_{1}=U_{2}UU_{2}^{-1}$ are related
by a similarity transformation, they describe the same parent Floquet
system up to a half-period shift of the initial evolution time. The
system described by $U_{1/2}^{2}$ can thus be viewed as two equivalent
copies of $U$ up to a global phase shift $\pi$. Therefore, $U_{1/2}^{2}$
and $U$ are expected to share the same topological features concerning
the stroboscopic dynamics. We could view $U_{1/2}$ as a nontrivial
square-root of $U$ in the spirit of Dirac's taking square-root to
reach his equation for electrons \cite{Dirac}. The dynamical and topological properties
of $U$ can further be carried over to $U_{1/2}$, which are confirmed
by explicit studies of Floquet topological superconductors and time
crystals \cite{FSRTP1}.

Iterating the same procedure, we can construct
the $2^{n}$th-root of $U$, i.e., $U_{1/2^{n}}$ for any $n\in\mathbb{Z}^{+}$.
For example, we could generate $U_{1/4}$ by letting $H'_{1}(t)=\frac{\tau_{0}+\tau_{z}}{2}\otimes H(t)+\frac{\tau_{0}-\tau_{z}}{2}\otimes H(t+\frac{1}{2})$,
$H'_{2}=\pi\tau_{y}\otimes\mathbb{I}_{0}$, and defining in a further
enlarged Hilbert space
\begin{equation}
	H_{1/4}(t)=\begin{cases}
		\pi\tau'_{y}\otimes\mathbb{I}'_{0} & t\in[\ell,\ell+\frac{1}{2})\\
		\frac{\tau'_{0}+\tau'_{z}}{2}\otimes H'_{1}(t)+\frac{\tau'_{0}-\tau'_{z}}{2}\otimes H'_{2} & t\in[\ell+\frac{1}{2},\ell+1)
	\end{cases},\label{eq:H1ov4}
\end{equation}
where $\tau'_{y,z}$ and $\tau'_{0}$ are Pauli matrices and identity matrix
acting in the subspace of an extra pseudospin-$1/2$. $\mathbb{I}'_{0}$
denotes the identity in the Hilbert space of $H'_{1,2}$. The resulting
Floquet operator,
\begin{equation}
	U_{1/4}=\left(\mathcal{T}e^{-\mathrm{i} \int_{1/2}^{1} \left(\frac{\tau'_{0}+\tau'_{z}}{2}\otimes H'_{1}(t)+\frac{\tau'_{0}-\tau'_{z}}{2}\otimes H'_{2}\right) dt} \right)
	e^{-i\frac{\pi}{2}\tau'_{y}\otimes\mathbb{I}'_{0}},\label{eq:U1ov4}
\end{equation}
then defines the nontrivial $4$th-root of $U$. It is straightforward
to verify that
\begin{equation}
	U_{1/4}^{4}=e^{i\pi}\begin{pmatrix}U_{1}U_{2} & 0 & 0 & 0\\
		0 & U_{2}U_{1} & 0 & 0\\
		0 & 0 & U_{2}U_{1} & 0\\
		0 & 0 & 0 & U_{1}U_{2}
	\end{pmatrix},\label{eq:U4}
\end{equation}
whose diagonal blocks contain four equivalent copies of $U$ up to a unitary transformation and a
global phase $\pi$.

The extension of the above approach to find any $q$th-root
of $U$ 
can be achieved by introducing higher-dimensional pseudospin
degrees of freedom, i.e., the generalized $q\times q$ Pauli matrices
\begin{equation}
	\left[\eta_{x} \right]_{i,j}=\delta_{i,j-1} +\delta_{i,q} \delta_{j,1} ,\qquad \left[ \eta_{z}\right]_{i,j}=\omega^{j-1} \delta_{i,j},\label{eq:Pauliq}
\end{equation}
where $\omega=e^{\mathrm{i} 2\pi/q}$. These operators satisfy 
\begin{equation}
	\eta_x\eta_z = \omega \eta_z \eta_x,\quad\eta_{x}\eta_{x}^{\dagger}=\eta_{z}\eta_{z}^{\dagger}=\eta_{0},\quad\eta_{x}^{q}=\eta_{z}^{q}=\eta_{0},\label{eq:LamId}
\end{equation}
where $\eta_{0}$ is the identity matrix acting in the pseudospin subspace. Our $q$th-rooting procedure can then be executed in two steps. First, given any time-periodic Hamiltonian $H(t)$, we divide the Floquet operator into $q$ time-steps, i.e.,
\begin{equation}
    U=\prod_{\ell=1}^q U_\ell=U_1\cdots U_q \;, \;\;\;\; U_\ell = \mathcal{T} e^{-\mathrm{i} \int_0^{1/q} H\left(t+\frac{q-\ell}{q}\right) dt }\;.
\end{equation}
Next, we define a two-step Hamiltonian of the form
\begin{equation}
	H_{1/q}(t)=\begin{cases}
		\sum_{j=0}^{q-1} M_j^{(q)} \eta_{x}^j \otimes\mathbb{I}_{0} & t\in[\ell,\ell+\frac{1}{2})\\
		\sum_{j=1}^{q} \mathcal{P}_{j}^{(q)}\otimes \tilde{H}_j(t) & t\in[\ell+\frac{1}{2},\ell+1)
	\end{cases},\label{eq:H1ovq}
\end{equation}
where $M_0^{(q)}=M_0^{(q)\dagger}$ and $M_{j\neq 0}^{(q)}=M_{q-j}^{(q)\dagger}$ for hermiticity. $M_j^{(q)}$ are further chosen such that $e^{-\mathrm{i} \sum_{j=0}^{q-1} M_j^{(q)} \eta_{x}^j}= \eta_x $. $\mathbb{I}_{0}$ is the identity in the Hilbert space of $H(t)$,  $\mathcal{P}_{j}^{(q)}=\frac{\sum_{k=0}^{q-1}\omega^{k (j-1)} \eta_z^k}{q}$, and $\tilde{H}_j(t)= \frac{2H\left(2(t+\frac{q-j}{q})/q\right)}{q}$. It then follows that the associated Floquet operator is
\begin{equation}
    [U_{1/q}]_{i,j} = U_{i} \delta_{i,j-1} + U_q \delta_{i,q} \delta_{j,1} \;,
\end{equation}
such that $U_{1/q}^q$ is block diagonal and consists of all $q$ permutations of $\prod_{\ell=1}^q U_\ell$. This shows that the resulting system indeed represents the $q$th-rooted version of $U$. 

Intuitively, the above construction can be understood as follows. First, note that Eq.~(\ref{eq:H1ovq}) is defined on a system consisting of $n$ subsystems. Consider a particle initially living in the $j$th subsystem. During the first half of the period, evolution under Eq.~(\ref{eq:H1ovq}) amounts to transporting the particle towards subsystem $j-1\;\;{\rm mod}\;\;q$. By noting that $\mathcal{P}_{j}^{(q)}$ represents a projection onto the $j$th subsystem, it then follows that Eq.~(\ref{eq:H1ovq}) further evolves the particle under $H_{j-1}(t)$ during the second half of the period. In the next Floquet cycle, the particle continues moving to subsystem $j-2\;\;{\rm mod}\;\;q$, followed by the half-period evolution under $H_{j-2}(t)$. As the process continues, at the end of $q$ periods, the particle returns to the subsystem $j\;\;{\rm mod}\;\;q$, while accumulating $U_{j} \cdots U_q U_1\cdots U_{j-2}U_{j-1}$, which is unitarily equivalent to $U$.


\begin{figure}
	\begin{centering}
		\includegraphics[scale=0.6]{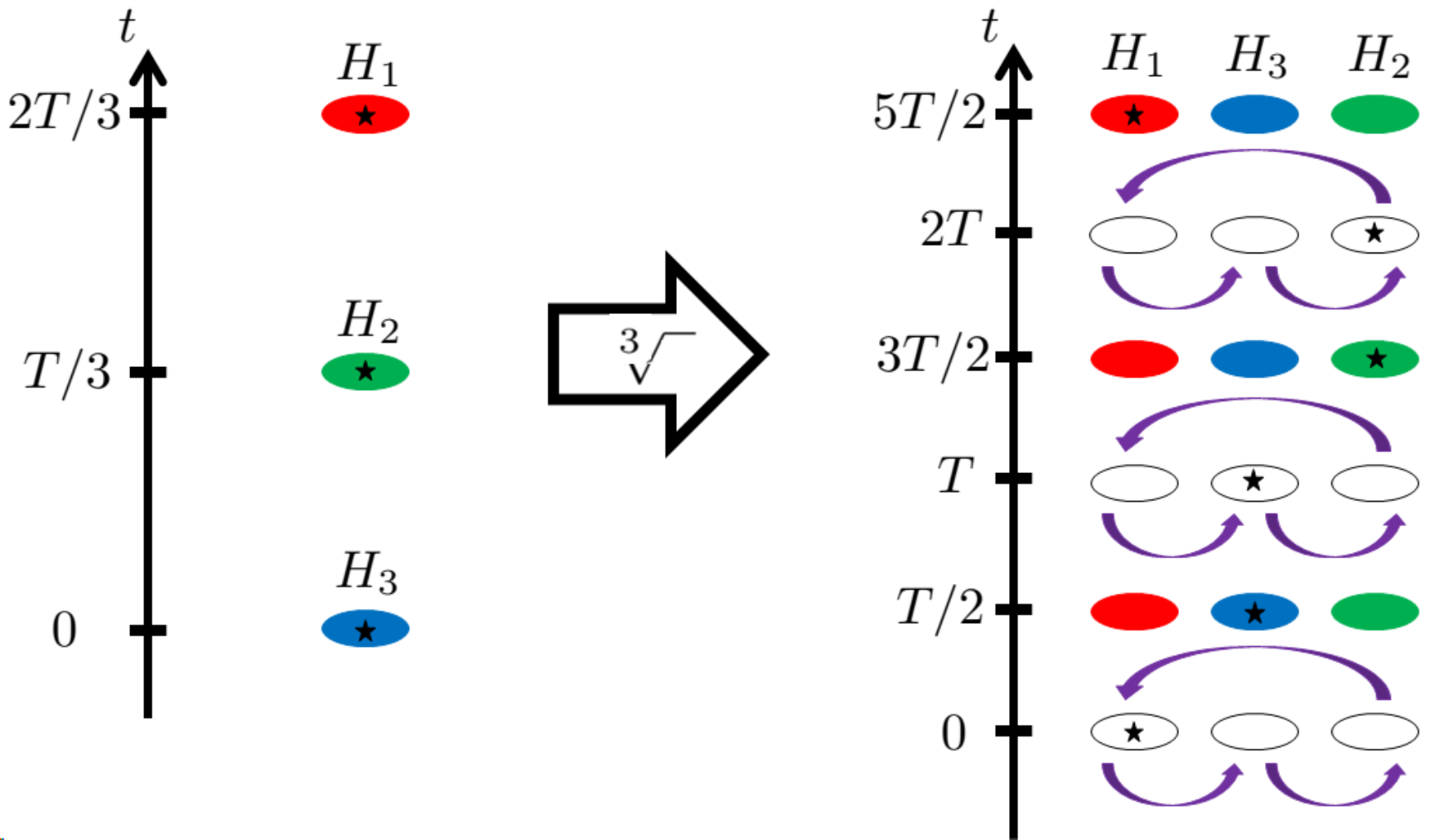}
		\par\end{centering}
	\caption{Schematic of a cubic-root system obtained from the procedure of Eq.~(\ref{eq:H1ovq}). In the parent system, a particle (marked by black star) evolves under $H_3$, $H_2$, and $H_1$ over the course of a period. In the corresponding cubic-root system, a particle living in a given subsystem effectively evolves under the same three Hamiltonians only when viewed over the course of three periods.\label{fig:cbcrootscheme}}
\end{figure}

Having demonstrated the generality of our construction, we will focus on square-root and cubic-root systems for brevity in the remainder of this paper. To this end, we will now present an explicit application of the above construction to obtain a nontrivial cubic-root of a system relevant to the case studies below. Specifically, such a parent system follows a three-step periodically quenched drive, whose time-dependent Hamiltonian takes the form
\begin{equation}
H(t)=\begin{cases}
H_{1} & t\in[\ell+2/3,\ell+1)\\
H_{2} & t\in[\ell+1/3,\ell+2/3)\\
H_{3} & t\in[\ell,\ell+1/3)
\end{cases}\qquad\ell\in\mathbb{Z},\label{eq:Ht2}
\end{equation}
with the corresponding Floquet operator of
\begin{equation}
U=e^{-i\frac{H_{1}}{3}}e^{-i\frac{H_{2}}{3}}e^{-i\frac{H_{3}}{3}}.\label{eq:Ut2}
\end{equation}
Note that the Floquet operator associated with a system following a two-step periodically quenched drive, whose Hamiltonian switches between $h_1$ and $h_2$ after every half period, can also be cast into the form of Eq.~(\ref{eq:Ut2}) by shifting the initial evolution time from $t=0$ to $3/4$ and identifying $H_1,H_3=3 h_1/4$ and $H_2=3h_2/2$. In both cases, the cubic-root of $U$ in Eq.~(\ref{eq:Ut2}) can be obtained according to Eq.~(\ref{eq:H1ovq}) with $M_0^{(3)}=0$ and $M_2^{(3)}=-M_1^{(3)}=\frac{4\pi i}{3\sqrt{3}}$, which leads to $e^{-\mathrm{i} \left(M_1^{(3)}\eta_x + M_2^{(3)}\eta_x^2\right)}=\eta_x$. We further identify $\tilde{H}_j=\frac{2H_{3-j}}{3}$ with $j=0,1,2$, as well as $\eta_{x,z}$ in the explicit matrix forms
\begin{equation}
\eta_{x}=\begin{pmatrix}0 & 1 & 0\\
0 & 0 & 1\\
1 & 0 & 0
\end{pmatrix},\qquad\eta_{z}=\begin{pmatrix}1 & 0 & 0\\
0 & \omega & 0\\
0 & 0 & \omega^{2}
\end{pmatrix}.\label{eq:Pauli3}
\end{equation}
Indeed, it can be directly verified that Eq.~(\ref{eq:Pauli3}) satisfies the algebra of Eq.~(\ref{eq:LamId}). The corresponding Floquet operator of the cubic-root model then reads
\begin{equation}
U_{1/3}=\begin{pmatrix}0 & e^{-i\frac{{\tilde H}_{1}}{3}} & 0\\
0 & 0 & e^{-i\frac{{\tilde H}_{2}}{3}}\\
e^{-i\frac{{\tilde H}_{3}}{3}} & 0 & 0
\end{pmatrix},\label{eq:U1ov3}
\end{equation}
with
\begin{equation}
U_{1/3}^{3}=\begin{pmatrix}e^{-i\frac{{\tilde H}_{1}}{3}}e^{-i\frac{{\tilde H}_{2}}{3}}e^{-i\frac{{\tilde H}_{3}}{3}} & 0 & 0\\
0 & e^{-i\frac{{\tilde H}_{2}}{3}}e^{-i\frac{{\tilde H}_{3}}{3}}e^{-i\frac{{\tilde H}_{1}}{3}} & 0\\
0 & 0 & e^{-i\frac{{\tilde H}_{3}}{3}}e^{-i\frac{{\tilde H}_{1}}{3}}e^{-i\frac{{\tilde H}_{2}}{3}}
\end{pmatrix}.\label{eq:U1ov3CB}
\end{equation}
That is, the three diagonal blocks of $U_{1/3}^{3}$ only differ from one another
by the starting time of evolution and describe equivalent Floquet
systems in stroboscopic dynamics concerning the spectral and topological properties.
This implies that $U_{1/3}$ is indeed a nontrivial cubic-root of the
parent system $U$ in Eq.~(\ref{eq:Ut2}). The presented cubic root procedure is schematically depicted in Fig.~\ref{fig:cbcrootscheme}.

We now discuss how the rooted Floquet system could inherit the symmetry
protected edge states of the parent model while altering their quasienergies 
to rational fractions of $2\pi$. A key symmetry that is relevant to
the topological characterization of the parent systems considered in this work
is the chiral symmetry (CS). If a general Floquet operator $U$ possesses
the CS, there is a unitary operator $\Gamma$ such that $\Gamma U\Gamma^{\dagger}=U^{-1}$.
If $U$ has an eigenstate $|\psi\rangle$ with quasienergy $E$, i.e.,
$U|\psi\rangle=e^{-iE}|\psi\rangle$, its CS implies that $\Gamma U\Gamma^{\dagger}(\Gamma|\psi\rangle)=e^{-iE}(\Gamma|\psi\rangle)$
or $U(\Gamma|\psi\rangle)=e^{-i(-E)}(\Gamma|\psi\rangle)$. Therefore,
$\Gamma|\psi\rangle$ is an eigenstate of $U$ with quasienergy $-E$.
Now if there is an eigenstate $|\psi\rangle$ with $E=0$ or $\pi$,
the CS enforces the presence of another eigenstate $\Gamma|\psi\rangle$
also at $E=0$ or $\pi$ ($E=\pm\pi$ are identified as the same quasienergy since $E$ is defined$\mod2\pi$), yielding eigenstate degeneracy
at the center or boundary of the quasienergy Brillouin zone $E\in[-\pi,\pi]$.
If such eigenmodes appear at the edge or corner of the system,
we obtain CS-protected degenerate edge or corner modes of $U$. 

For the square-root system $U_{1/2}$, we already see that $U_{1/2}^{2}$
is block diagonal and its two diagonal blocks share the same spectral
and topological properties with the parent model $U$. If $|\psi'\rangle$
is an eigenstate of $U_{1/2}$ with quasienergy $E'$, i.e., $U_{1/2}|\psi'\rangle=e^{-iE'}|\psi'\rangle$,
it is straightforward to see that $U_{1/2}^{2}|\psi'\rangle=e^{-iE'}U_{1/2}|\psi'\rangle=e^{-i2E'}|\psi'\rangle$.
$U_{1/2}$ and $U_{1/2}^{2}$ thus share the same eigenbasis. When
the parent model $U$ possesses the CS $\Gamma$, the diagonal blocks
of $U_{1/2}^{2}$ possess the CS, such that $U_{1/2}^{2}$ is chiral
symmetric with respect to $\Gamma'=\tau_{z}\otimes\Gamma$. Degenerate
topological edge/corner modes of $U_{1/2}^{2}$ can thus only appear at $E=0,\pm\pi=2E'\mod2\pi$.
This implies that in the square-root system $U_{1/2}$, we could only
find topological edge/corner modes at the quasienergies $E'=0,\pm\pi/2,\pm\pi$,
which are indeed protected by the CS $\Gamma$ of the parent model. Interestingly,
the degenerate eigenmodes at $E=\pm\pi/2$ are present only in the $U_{1/2}$ and are thus 
unique to the square-root Floquet system.

While the topological protection of the edge/corner modes in $U_{1/2}$  can be understood from the presence of chiral symmetry in its corresponding parent system $U$, it would also be insightful to discuss the protecting symmetries that arise at the level of $U_{1/2}$ directly. To this end, we first note that $\Gamma'=\tau_z \otimes \Gamma$ is also a chiral symmetry with respect to $\tilde{U}_{1/2}=e^{-\mathrm{i} \frac{\pi}{4}} U_{1/2} e^{\mathrm{i} \frac{\pi}{4}}$, i.e., $U_{1/2}$ under the shift in the initial time from $t=0$ to $t=1/4$. Similar to its parent counterpart, such a chiral symmetry is responsible for protecting $E'=0,\pm \pi$ quasienergy edge states in the square-root system. Next, we identify an additional symmetry $\Gamma_{1/2}'=\tau_z \otimes \mathcal{I}$ ($\mathcal{I}$ being the identity operator), which acts only within the enlarged degree of freedom and is thus referred to as the ``subchiral" symmetry \cite{subchiral}. Such a symmetry operates as $\Gamma_{1/2}' U_{1/2} \Gamma_{1/2}'^\dagger = -U_{1/2}$. Consequently, if $|\psi'\rangle$ is a quasienergy $E$ eigenstate of $U_{1/2}$, then $\Gamma_{1/2}' |\psi' \rangle$ is a quasienergy $E\pm \pi$ eigenstate of $U_{1/2}$. Indeed,
\begin{equation}
    U_{1/2} \Gamma_{1/2}' |\psi' \rangle = -\Gamma_{1/2}' U_{1/2} \Gamma_{1/2}'^\dagger \Gamma_{1/2}' |\psi' \rangle = e^{-\mathrm{i} (E\pm \pi)} \Gamma_{1/2}' |\psi' \rangle \;. \nonumber
\end{equation}
In this case, a quasienergy which satisfies $E\pm \pi = -E$, i.e., $E=\pm \pi/2$, is necessarily twofold degenerate due to the product $\Gamma' \Gamma_{1/2}'$. The associated quasienergy eigenstates could further be chosen to be simultaneous $\pm 1$ eigenstates of $\Gamma' \Gamma_{1/2}'$. This is automatically the case for the quasienergy $\pm \pi/2$ edge/corner states. In particular, since $\pm 1$ eigenstates of $\Gamma' \Gamma_{1/2}'$ correspond to states localized at two opposite edges/corners, the discreteness of $\Gamma' \Gamma_{1/2}'$ eigenstates pins such edge/corner states at quasienergy $\pm \pi/2$ in the presence of symmetry-preserving perturbations. This completes the symmetry protection analysis of quasienergy $\pm \pi/2$ edge/corner states in the square-root system.

The above argument can be easily extended to conclude that, for any $q$th-root version of the system, $U_{1/q}^{q}$ also possesses the CS with respect to $\Gamma''=\eta_{z}\otimes\Gamma$. A generalized ``subchiral" symmetry can further be identified as $\Gamma_{1/q}''=\eta_z\otimes \mathcal{I}$, which operates as $\Gamma_{1/q}''U_{1/q}\Gamma_{1/q}''^\dagger =\omega^\dagger U_{1/q}$ and thus forces the quasienergies of $U_{1/q}$ to form a cluster of $E-2\pi j/q$ with $j=0,1,\cdots,q-1$. In the case of cubic-root Floquet systems, which are explicitly studied below, both symmetries lead to the protection of degenerate edge/corner modes at $E''=0,\pm\pi/3,\pm2\pi/3,\pm\pi$. Following
the same routine, we can deduce that if the parent model $U$ possesses
the CS $\Gamma$, the existence of its edge/corner modes at the quasienergies
$E=0,\pi$ guarantees the presence of degenerate edge/corner states
at the quasienergies $(0,1,...2^{n})\pi/2^{n}$ and $(0,1,...,3^{n})\pi/3^{n}$
of the systems described by $U_{1/2^{n}}$ and $U_{1/3^{n}}$, respectively.
Notably, the boundary modes appearing at the fractional quasienergies $p\pi/q$ with $p<q$
and $(p,q)$ being co-prime integers are, to the best of our knowledge, not identified by previous
studies on the symmetry classification and bulk-boundary correspondence of Floquet systems.
They are thus a unique product of our $q$th-root procedure operated on Floquet operators. 
In Sec.~\ref{sec:Res}, we will apply our theory to explicitly construct
square/cubic-root first- and second-order non-Hermitian FTIs based
on the parent models defined in the following section.

\section{Parent models}
\label{sec:Mod}
In this section, we introduce two non-Hermitian Floquet topological insulator (FTI)
models that will be taking square- and cubic-roots.
Detailed investigations of these parent models can be found in Refs.~\cite{NHFTI5} and \cite{NHFTI8}.
All system parameters below are assumed to be properly scaled and set in dimensionless units.

The first model of our consideration describes a non-Hermitian FTI with rich topological phase
diagrams and arbitrarily many degenerate edge modes
in the presence of Floquet NHSE \cite{NHFTI8}.
Its time-dependent Hamiltonian is $H(t)=H_{1}$ for $t\in[\ell+1/2,\ell+1)$
and $H(t)=H_{2}$ for $t\in[\ell,\ell+1/2)$, where $t$ denotes time,
$\ell\in\mathbb{Z}$, and
\begin{alignat}{1}
	H_{1}= & \sum_{n}J_{2}(i|n+1\rangle\langle n|+{\rm H.c.})\otimes\sigma_{y}\nonumber\\
	+& \sum_{n}i\lambda(|n\rangle\langle n+1|+{\rm H.c.})\otimes\sigma_{y},\label{eq:H1}
\end{alignat}
\begin{alignat}{1}
	H_{2}= & \sum_{n}[2\mu|n\rangle\langle n|+J_{1}(|n\rangle\langle n+1|+{\rm H.c.})]\otimes\sigma_{x}\nonumber\\
	+& \sum_{n}i\lambda(i|n+1\rangle\langle n|+{\rm H.c.})\otimes\sigma_{x}.\label{eq:H2}
\end{alignat}
Here $n\in\mathbb{Z}$ is the unit cell index. $\sigma_{x,y,z}$ are
Pauli matrices acting on the two sublattices in each unit cell. $J_{1,2}$
and $i\lambda$ describe symmetric and asymmetric parts of intercell
hopping amplitudes. $\mu$ is the intracell coupling strength. The
Floquet operator $U=e^{-i\frac{1}{2}H_{1}}e^{-i\frac{1}{2}H_{2}}$
that governs the evolution of the system over a driving period (e.g.,
from $t=0$ to $1$) is nonunitary once $\lambda\neq0$. This yields
a model that could possess non-Hermitian FTI phases,
which are characterized by integer or half-integer quantized topological invariants
under the periodic boundary conditions (PBC)~\cite{NHFTI8}. 
Under the open boundary conditions (OBC), the CS of the model $\Gamma={\mathbb I}_N\otimes\sigma_{z}$
($N$ is the number of unit cells and ${\mathbb I}_N$ is an $N\times N$ identity)
allows multiple edge modes to appear in pairs at the quasienergies
zero and $\pi$, whose numbers can be determined by the OBC bulk winding
numbers $\nu_{0}$ and $\nu_{\pi}$ (see Sec.~\ref{sec:app1} for their definitions).
These edge modes are further found to coexist with sufficient amounts
of bulk states localized around both edges of the system due to the NHSE~\cite{NHFTI8}. 

The second model that we will employ describes a non-Hermitian Floquet
second-order topological insulator (FSOTI), which could possess multiple
quartets of corner-localized states at real quasienergies zero and
$\pi$~\cite{NHFTI5}. The Hamiltonian of the model takes the form
of $H(t)={\cal H}_{1}$ for $t\in[\ell+1/2,\ell+1)$ and $H(t)={\cal H}_{2}$
for $t\in[\ell,\ell+1/2)$ with $\ell\in\mathbb{Z}$. Here 
\begin{equation}
	{\cal H}_{1(2)}={\cal H}_{x}\otimes\mathbb{I}_{y}+\mathbb{I}_{x}\otimes{\cal H}_{y1(y2)},\label{eq:H12}
\end{equation}
\begin{equation}
	{\cal H}_{x}=\Delta\sum_{m,n}(|m,n\rangle\langle m+1,n|\otimes\sigma_{-}+{\rm H.c.}),\label{eq:Hx}
\end{equation}
\begin{equation}
	{\cal H}_{y1}=\sum_{m,n}(iJ_{2}|m,n+1\rangle\langle m,n|+{\rm H.c.}+2\mu|m,n\rangle\langle m,n|)\otimes\sigma_{z},\label{eq:Hy1}
\end{equation}
\begin{equation}
	{\cal H}_{y2}=J_{1}\sum_{m,n}(|m,n\rangle\langle m,n+1|+{\rm H.c.})\otimes\sigma_{x}.\label{eq:Hy2}
\end{equation}
The $\mathbb{I}_{x}$ and $\mathbb{I}_{y}$ are identity matrices
for the basis along $x$ and $y$ directions of the lattice. $\sigma_{x,y,z}$
are Pauli matrices and $\sigma_{-}=(\sigma_{x}-i\sigma_{y})/2$. $m,n\in\mathbb{Z}$
are unit cell indices along the two spatial dimensions. $\Delta$
and $J_{1,2}$ describe hopping amplitudes between nearest neighbor
cells along the $x$ and $y$ directions. $\mu$ characterizes the strength
of an onsite potential bias. Gain and loss are introduced to make
the system non-Hermitian by setting $\mu=u+iv$, with $u,v\in\mathbb{R}$
and $v\neq0$. The Floquet operator of the system takes the form ${\cal U}=e^{-i\frac{1}{2}{\cal H}_{1}}e^{-i\frac{1}{2}{\cal H}_{2}}$,
whose spectrum under the OBC features fourfold degenerate topological
corner modes at zero and $\pi$ quasienergies. The numbers of these
corner modes $n_{0}$ and $n_{\pi}$ are related to a pair of bulk
topological winding numbers $\nu_{0}$ and $\nu_{\pi}$ of ${\cal U}$ (see Sec.~\ref{sec:app2} for their definitions)
through a bulk-corner correspondence relation $(n_{0},n_{\pi})=4(|\nu_{0}|,|\nu_{\pi}|)$ \cite{NHFTI5}. 
The fourfold degeneracy of Floquet
corner modes at the quasienergies $E=0,\pi$ is protected by the CS
$\Gamma=\sigma_{z}\otimes\sigma_{y}$ of the two-dimensional system
described by ${\cal U}$ under the PBC \cite{NHFTI5}. 

Applying the procedure of Sec.~\ref{sec:The}, we will obtain the
square and cubic roots of the two Floquet models introduced here,
and unveil the intriguing topological features of the resulting systems
in the following section. As will be demonstrated, our $q$th-root procedure
endows the non-Hermitian Floquet phases in the above two parent models with
even richer topological properties.

\section{Results}
\label{sec:Res}

In Sec.~\ref{subsec:TI}, we present square- and cubic-root non-Hermitian
FTIs generated by the first model in Sec.~\ref{sec:Mod}, which will
be shown to possess multiple and tunable numbers of degenerate edge
modes with the quasienergies $\pi/2$, $\pi/3$ and $2\pi/3$ that
could survive under the NHSE. In Sec.~\ref{subsec:SOTI}, we discuss
square- and cubic-root non-Hermitian FSOTIs yielded by the
second model in Sec.~\ref{sec:Mod}, which hold non-Hermiticity induced
quartets of topological corner modes at the $\pi/2$, $\pi/3$ and $2\pi/3$
quasienergies.

\subsection{Square/Cubic-root non-Hermitian FTIs}
\label{subsec:TI}

\begin{figure}
	\begin{centering}
		\includegraphics[scale=0.49]{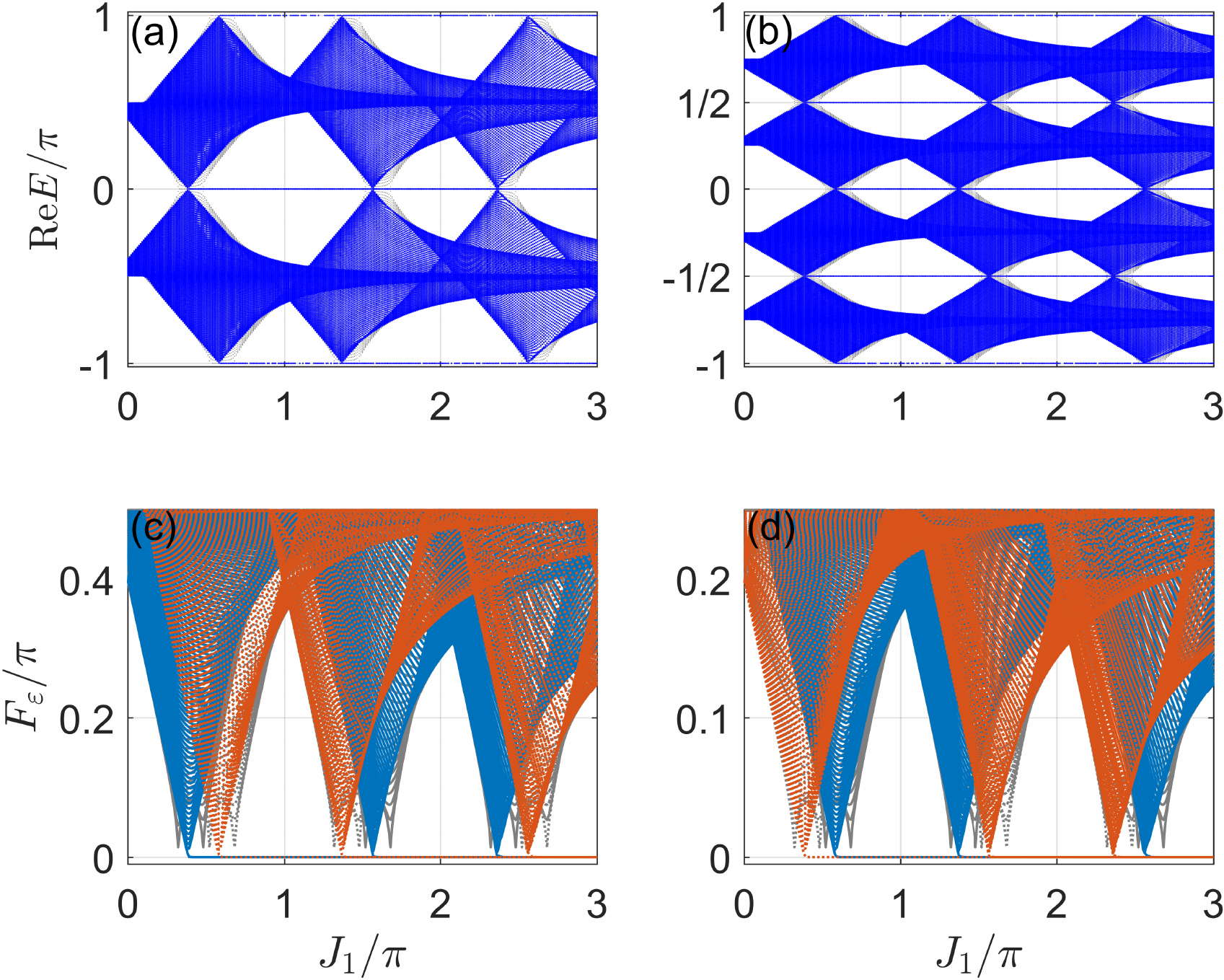}
		\par\end{centering}
	\caption{Floquet spectrum $E$ and gap function $F_{\varepsilon}$ of $U_{1/2}$
		{[}Eq.~(\ref{eq:U1ov2M1}){]} and $U$ versus $J_{1}$ under OBC and PBC.
		Other system parameters are set as $(J_{2},\mu,\lambda)=(0.5\pi,0.4\pi,0.25)$
		and the length of lattice is $L=400$. (a) and (b) show the 
		values of the real part of $E$ for the parent and square-root models
		described by $U$ and $U_{1/2}$ under the OBC (PBC) in blue (grey) dots, respectively. The blue solid and
		red dotted lines denote the gap functions $F_{0}$ and $F_{\pi}$
		of $U$ in (c), and the gap functions $F_{0}$ ($=F_{\pi}$) and $F_{\pi/2}$ of
		$U_{1/2}$ in (d) under the OBC. Grey solid and dotted lines denote the same gap
		functions under PBC.\label{fig:E0E1ov2M1}}
\end{figure}

We now apply the procedure in Sec.~\ref{sec:The} to find the square-
and cubic-roots of the first model in 
Sec.~\ref{sec:Mod}. In the lattice
representation, the square-root Floquet system is obtained
by identifying $U_1=e^{-iH_1/2}$ and $U_2=e^{-iH_2/2}$ in Eq.~(\ref{eq:Uhalf}), where the $H_1$
and $H_2$ are given by Eqs.~(\ref{eq:H1}) and (\ref{eq:H2}), respectively.
The Floquet operator $U_{1/2}$ is then derived following Eq.~(\ref{eq:Ut1}),
i.e.,
\begin{equation}
	U_{1/2}=\begin{pmatrix}0 & -e^{-iH_{1}/2}\\
		e^{-iH_{2}/2} & 0
	\end{pmatrix}.\label{eq:U1ov2M1}
\end{equation}
To obtain the cubic-root model, we may identify ${\tilde H}_{1}={\tilde H}_{3}=3H_{1}/4$ and ${\tilde H}_{2}=3H_{2}/2$ in Eq.~(\ref{eq:U1ov3}),
where $H_1$ and $H_2$ are defined by Eqs.~(\ref{eq:H1}) and (\ref{eq:H2}), respectively. This then leads to the Floquet operator
\begin{equation}
	U_{1/3}=\begin{pmatrix}0 & e^{-iH_{1}/4} & 0\\
		0 & 0 & e^{-iH_{2}/2}\\
		e^{-iH_{1}/4} & 0 & 0
	\end{pmatrix}.\label{eq:U1ov3M1}
\end{equation}
Solving the eigenvalue equations $U_{1/2(1/3)}|\psi\rangle=e^{-iE}|\psi\rangle$
under the OBC, with $E$ being the quasienergy, provides us with all
bulk and edge states of the square- (cubic-) root Floquet system.

\begin{figure*}
	\begin{centering}
		\includegraphics[scale=0.44]{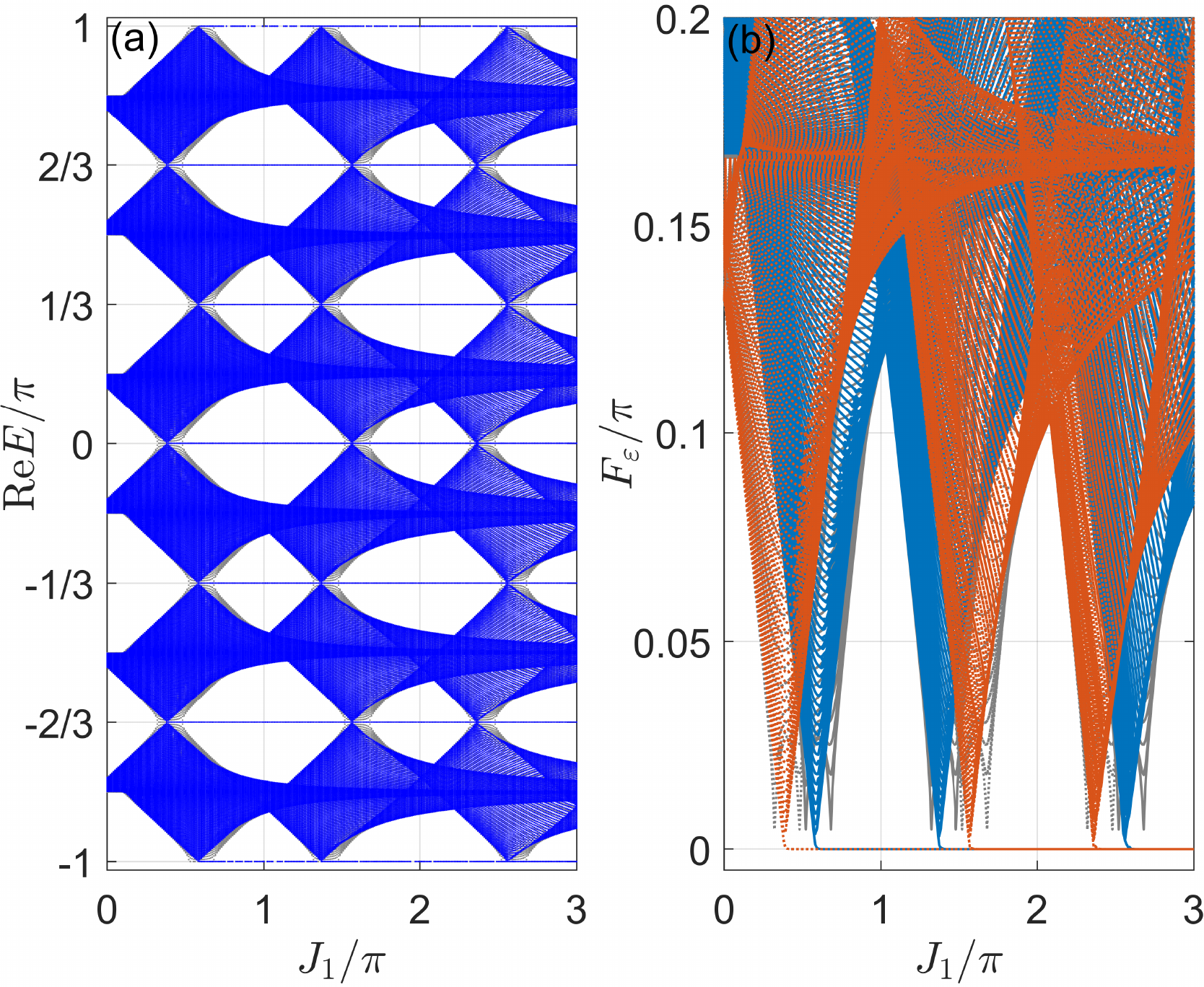}\includegraphics[scale=0.44]{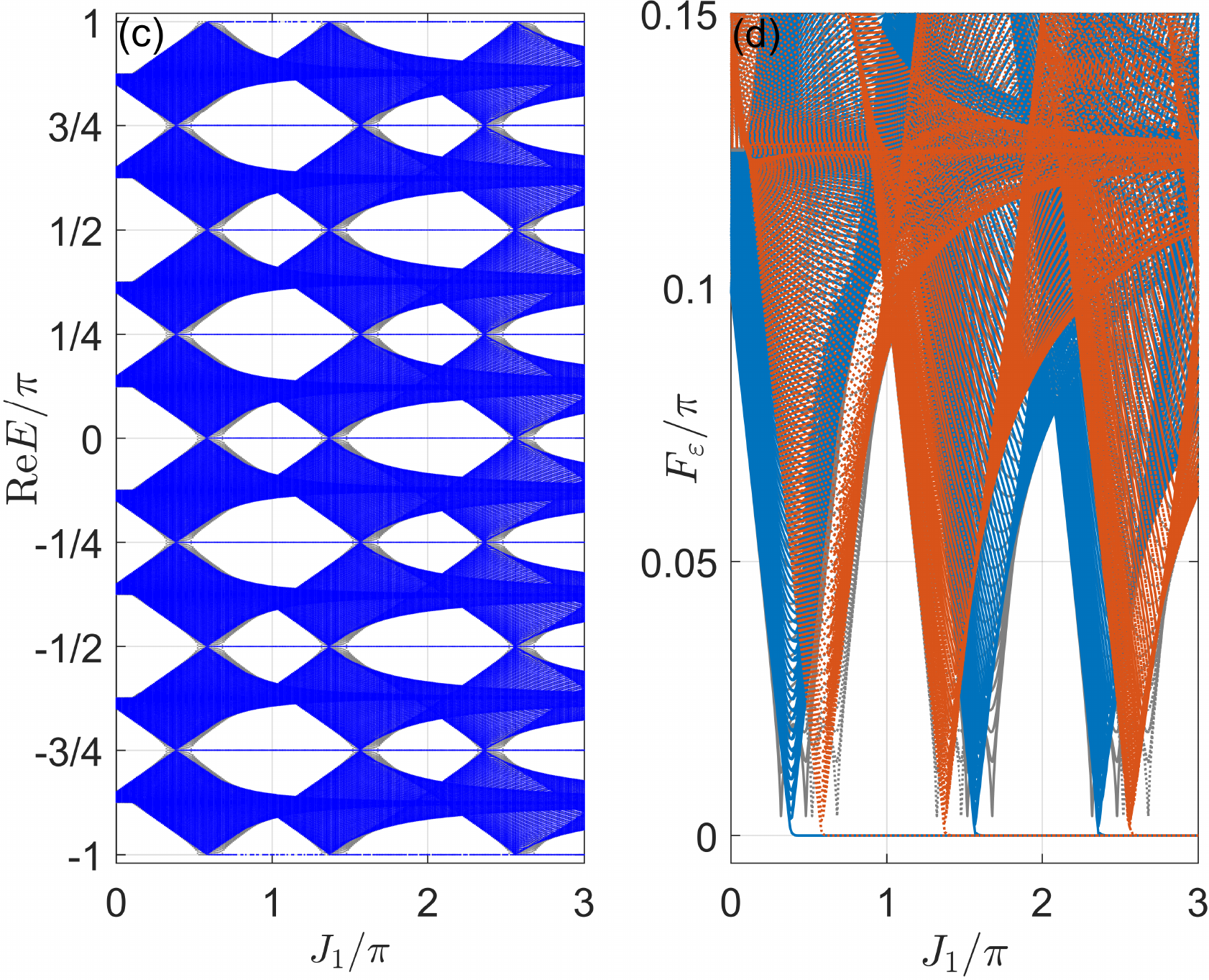}
		\par\end{centering}
	\caption{Floquet spectrum $E$ and gap function $F_{\varepsilon}$ of $U_{1/3}$
		{[}Eq.~(\ref{eq:U1ov3M1}){]} and $U_{1/4}$ {[}obtained following
		Eqs.~(\ref{eq:U1ov2M1}), (\ref{eq:H1ov4}) and (\ref{eq:U1ov4}){]}
		versus $J_{1}$ under both PBC and OBC. Other system parameters and the lattice size are
		the same as those used in Fig.~\ref{fig:E0E1ov2M1}. (a) and (c) show
		the real parts of $E$ for the cubic- and fourth-root
		models described by $U_{1/3}$ and $U_{1/4}$, respectively, under the OBC (blue dots) and PBC (grey dots in the background). The blue
		solid and red dotted lines denote the gap functions $F_{\pi/3}$ ($=F_{\pi}$)
		and $F_{2\pi/3}$ ($=F_{0}$) of $U_{1/3}$ in (b), and the gap functions $F_{\pi/4}$ ($=F_{3\pi/4}$)
		and $F_{\pi/2}$ ($=F_{0}=F_{\pi}$) of $U_{1/4}$ in (d) under the OBC. Corresponding gap functions under the PBC are given by the grey solid and dotted lines in (b) and (d).\label{fig:E1ov3E1ov4M1}}
\end{figure*}

As an important note, if there is a pair of degenerate edge modes
with zero-quasienergy in the parent model $U=e^{-iH_{1}/2}e^{-iH_{2}/2}$,
their quasienergies will be shifted to $\pi$ in $U_{1/2}^{2}$ according
to Eq.~(\ref{eq:U1ov2SQ}), yielding edge modes at quasienergies $E=\pm\pi/2$
in the system described by $U_{1/2}$. On the other hand, if a pair
of degenerate edge modes appears at $E=\pi$ in the parent
model, their quasienergies will become $0$ (mod $2\pi$) in $U_{1/2}^{2}$,
leading to edge states at $E=0,\pm\pi$ in the square-root model.
Following the same routine, we deduce that the edge modes at zero
($\pi$) quasienergy in $U=e^{-iH_{2}/4}e^{-iH_{1}/2}e^{-iH_{2}/4}$
generate edge states with $E=0,\pm2\pi/3$ ($E=\pm\pi/3,\pm\pi$) in the
system described by $U_{1/3}$ after taking the cubic root. Now if
we could relate the numbers of zero and $\pi$ edge modes in the parent
system $U$ to its topological invariants, these invariants should
also predict the numbers of zero, $\pi/2$, $\pi/3$, $2\pi/3$ and
$\pi$ modes in the square- and cubic-root systems if the symmetry
that protects their quantization is not broken during the process
of taking roots.

To showcase the fractional-quasienergy
edge modes in the spectrum in a more transparent manner, we introduce
the gap function $F_{\varepsilon}$ with respect to a quasienergy
$\varepsilon$, which is defined as
\begin{equation}
	F_{\varepsilon}=\sqrt{({\rm Re}E-\varepsilon)^{2}+({\rm Im}E)^{2}}.\label{eq:GF}
\end{equation}
Note that the $E$ in Eq.~(\ref{eq:GF}) is the collection of all
quasienergies obtained by diagonalizing the Floquet operator of the
system under consideration. It is clear that once there is an edge
state with real quasienergy $\varepsilon$ that resides in a gap
on the complex plane, we would have $F_{\varepsilon}=0$ for that
state and $F_{\varepsilon}>0$ for all other bulk states. To locate
the expected edge states of $U_{1/2}$ and $U_{1/3}$, we choose $\varepsilon=0,\pi/2,\pi$
and $\varepsilon=0,\pi/3,2\pi/3,\pi$ for them, respectively, in the
following numerical calculations.

In Fig.~\ref{fig:E0E1ov2M1}, we present the quasienergy (Floquet) spectrum and gap functions of the first model in Sec.~\ref{sec:Mod} and its square-root descendant
under both the PBC and OBC. The quasienergies and gap functions of the parent model
in Figs.~\ref{fig:E0E1ov2M1}(a) and \ref{fig:E0E1ov2M1}(c) are reproduced
from Ref.~\cite{NHFTI8}. A clear distinction between the spectrum under PBC (gray dots in the background) and OBC (blue dots) can be observed especially around the phase transition points,
implying the presence of NHSE in the system. To retrieve the bulk-edge
correspondence, a pair of open-boundary winding numbers $(\nu_{0},\nu_{\pi})$
is introduced in Ref.~\cite{NHFTI8} and reviewed in Sec.~\ref{sec:app1}, which correctly counts the number
of zero- and $\pi$-quasienergy edge modes $n_{0}$ and $n_{\pi}$
in the parent model through the relation $(n_{0},n_{\pi})=2(|\nu_{0}|,|\nu_{\pi}|)$.
Here $n_{E}$ denotes the number of edge states at the quasienergy
$E$. According to our square-root procedure, the edge modes at the quasienergies
$E=0,\pm\pi$ ($E=\pm\pi/2$) are generated by taking the square-root
of the $\pi$ (zero) Floquet edge modes. Therefore, we arrive at the
following bulk-edge correspondence for the square-root
FTIs described by $U_{1/2}$, i.e.,
\begin{equation}
	n_{\pi/2}=2|\nu_{0}|,\qquad n_{0}=n_{\pi}=2|\nu_{\pi}|,\label{eq:U1ov2BBCM1}
\end{equation}
where $n_{\pi/2}$ means the number of degenerate edge states at $E=\pm\pi/2$.
These relations are readily confirmed by comparing the spectrum and
gap functions presented in Figs.~\ref{fig:E0E1ov2M1}(b,d) and 
Figs.~\ref{fig:E0E1ov2M1}(a,c). Notably, with the increase of hopping amplitude
$J_{1}$, we observe a series of gap closing and topological phase
transitions in the square-root model. After each transition, the number
of edge modes $n_{0}$, $n_\pi$ or $n_{\pi/2}$ is found to be increased
by $2$ even in the presence of NHSE. Specifically, we find $(n_{\pi/2},n_{0},n_{\pi})=(0,0,0)$,
$(2,0,0)$, $(2,2,2)$, $(2,4,4)$, $(4,4,4)$, $(6,4,4)$, $(6,6,6)$ with the
increase of $J_{1}$ in Fig.~\ref{fig:E0E1ov2M1}(d), meanwhile the
winding numbers are $(\nu_{0},\nu_{\pi})=(0,0)$, $(1,0)$, $(1,-1)$,
$(1,-2)$, $(2,-2)$, $(3,-2)$, $(3,-3)$ according to the calculation
reported in Ref.~\cite{NHFTI8}. This process could continue with
the further increase of $J_{1}$. We can thus in principle obtain
arbitrarily many topological edge modes at fractional quasienergies
$E=\pm\pi/2$ in our square-root non-Hermitian Floquet system. This
highlights the universal advantage of Floquet engineering in generating
unique nonequilibrium states with strong topological signatures.

\begin{figure*}
\begin{centering}
\includegraphics[scale=0.29]{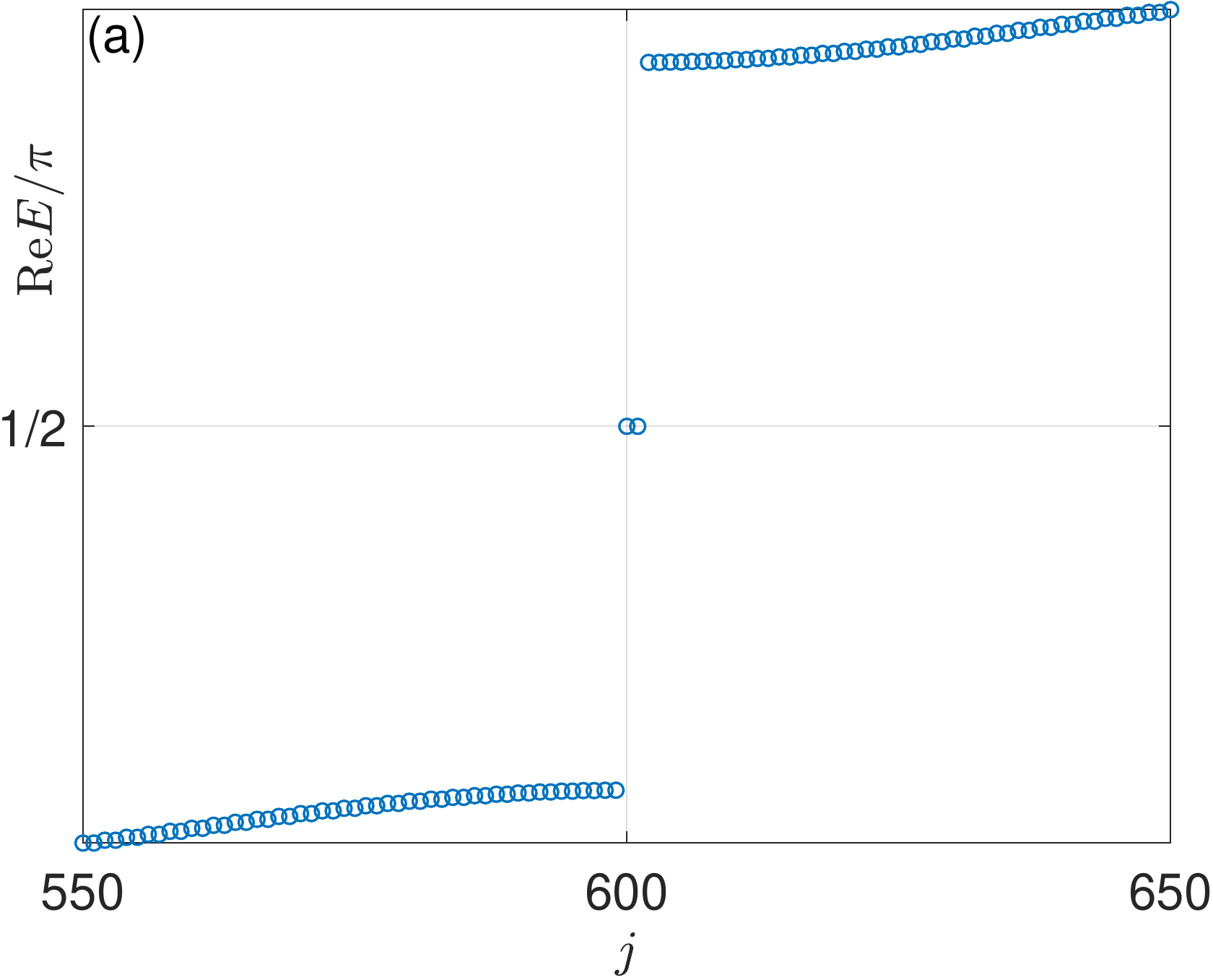}\includegraphics[scale=0.29]{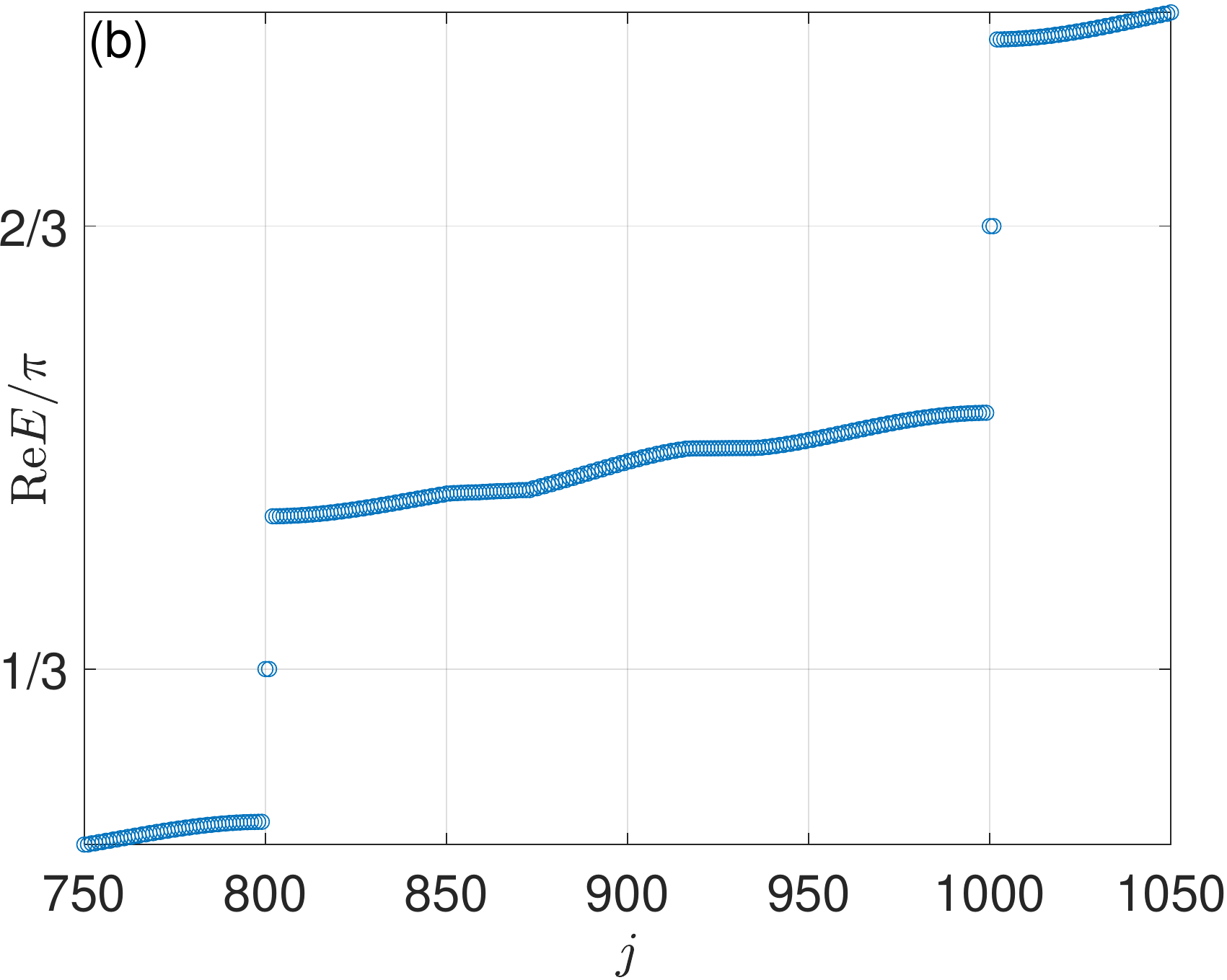}\includegraphics[scale=0.29]{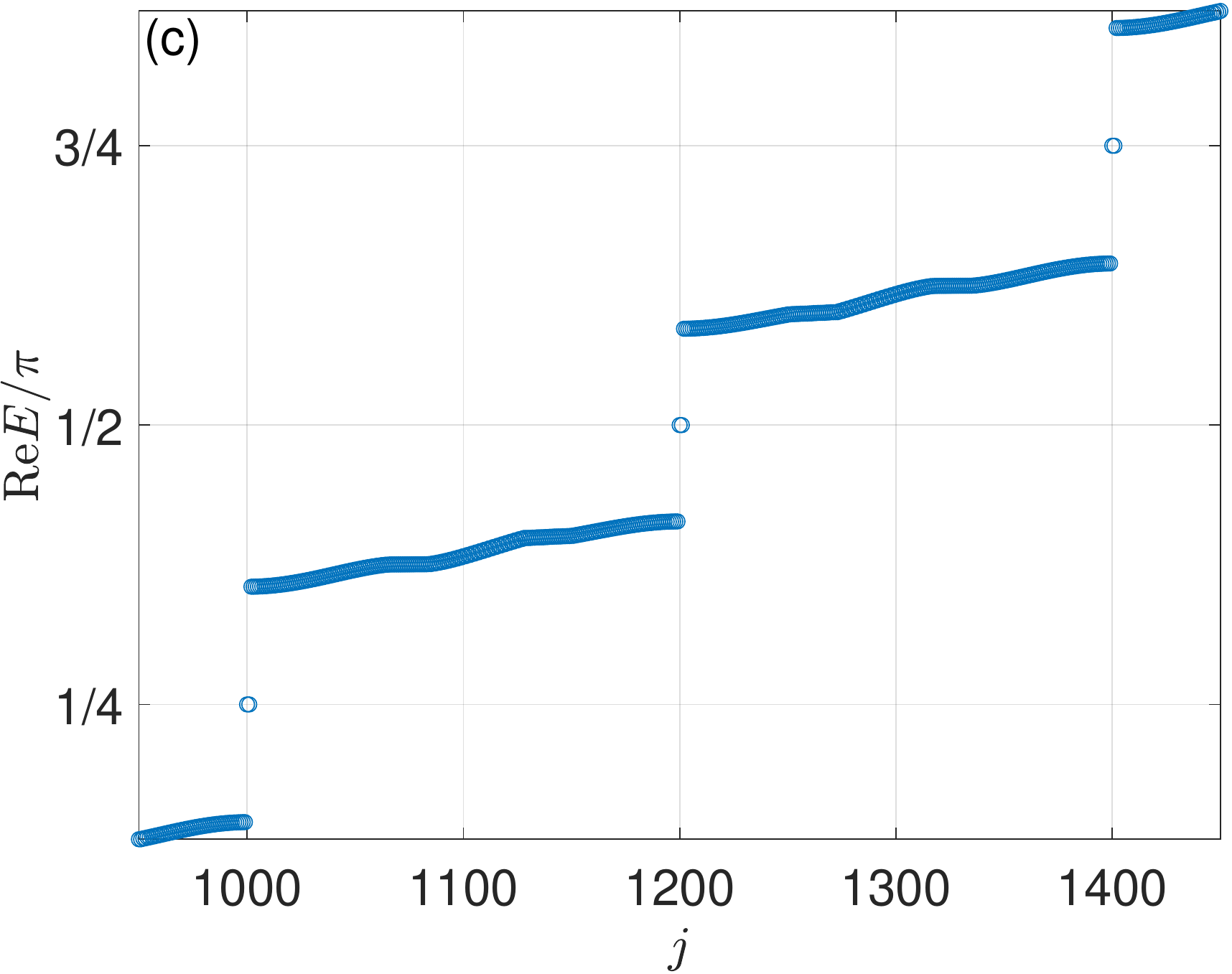}
\par\end{centering}
\begin{centering}
\includegraphics[scale=0.28]{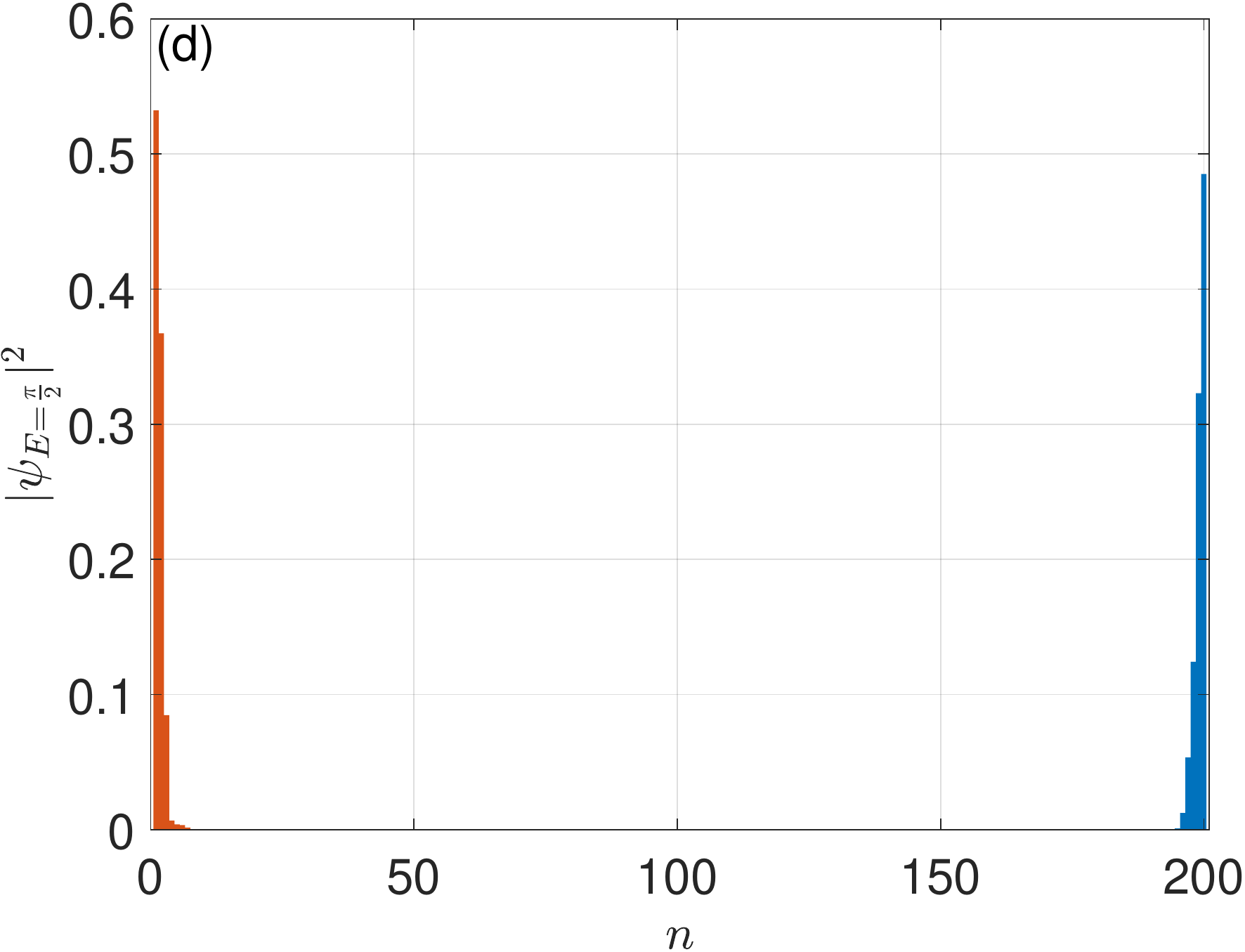}\includegraphics[scale=0.28]{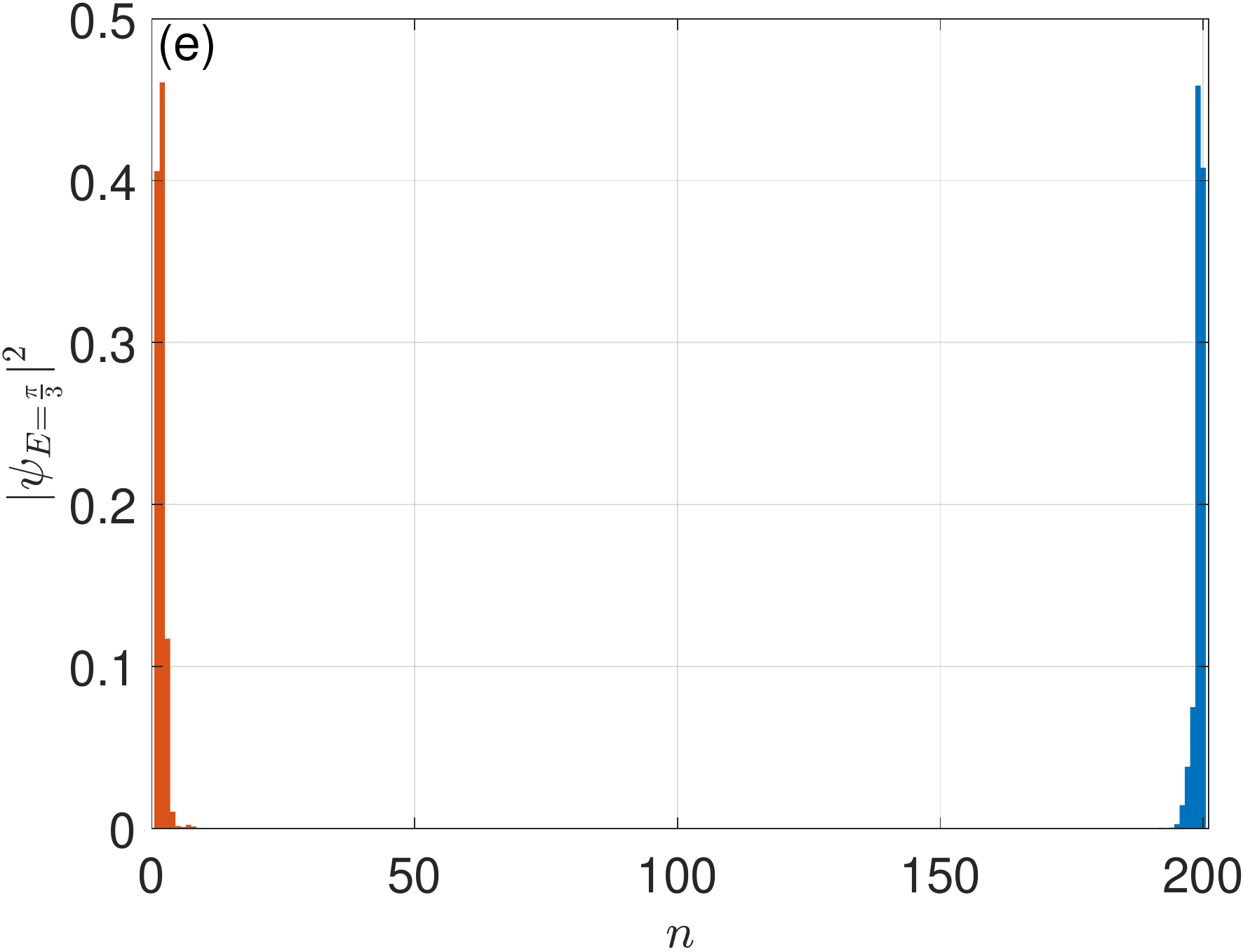}\includegraphics[scale=0.28]{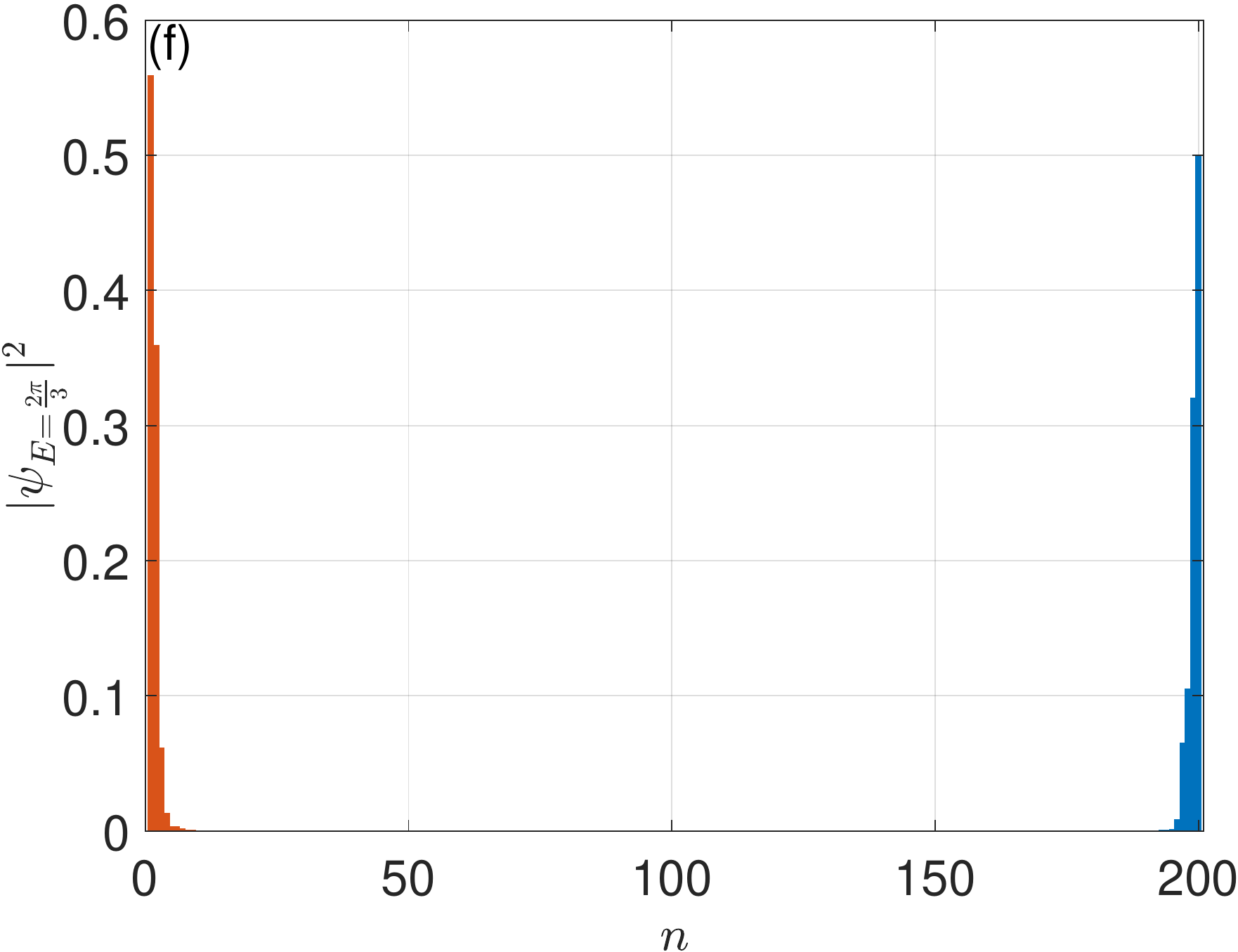}
\par\end{centering}
\begin{centering}
\includegraphics[scale=0.28]{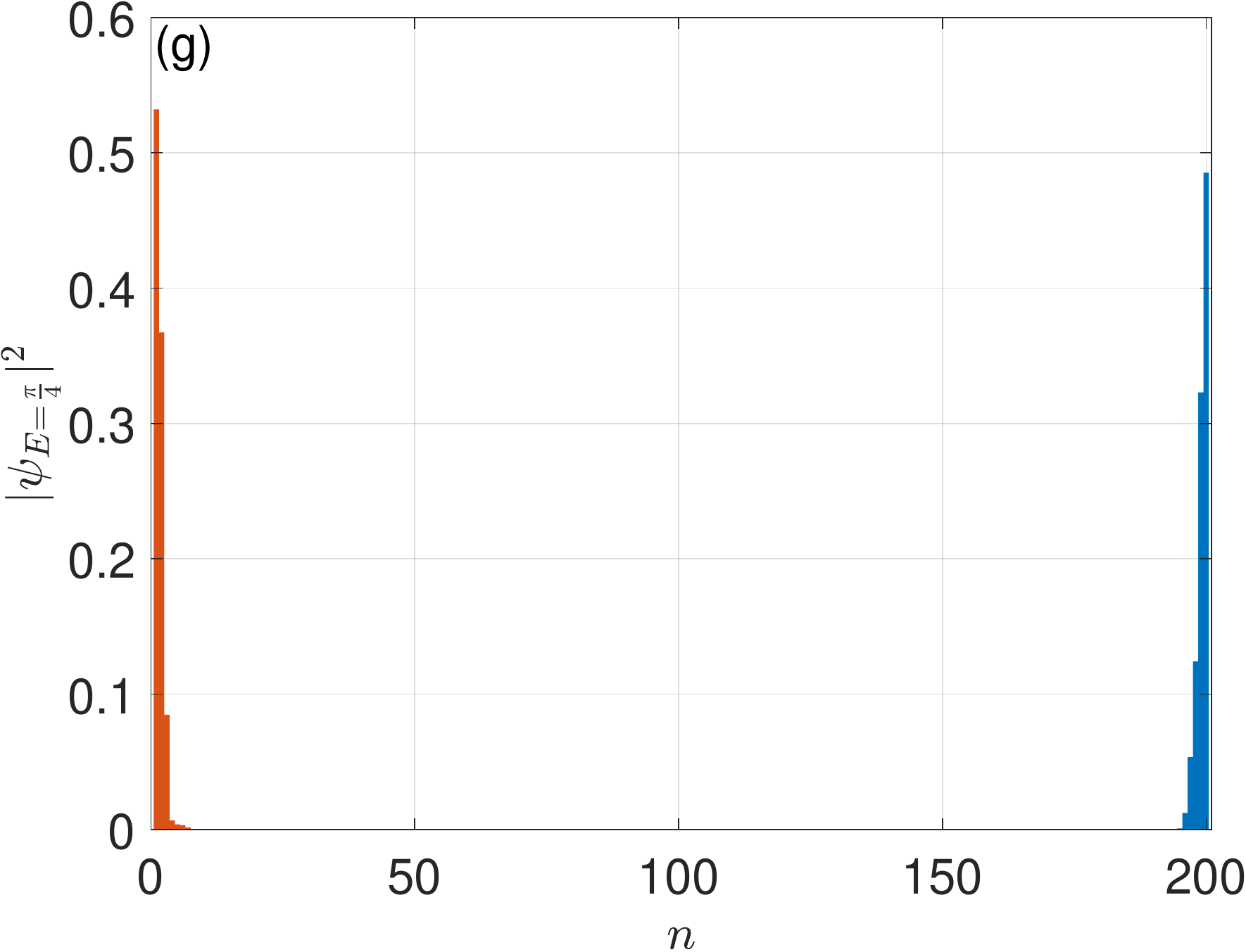}\includegraphics[scale=0.28]{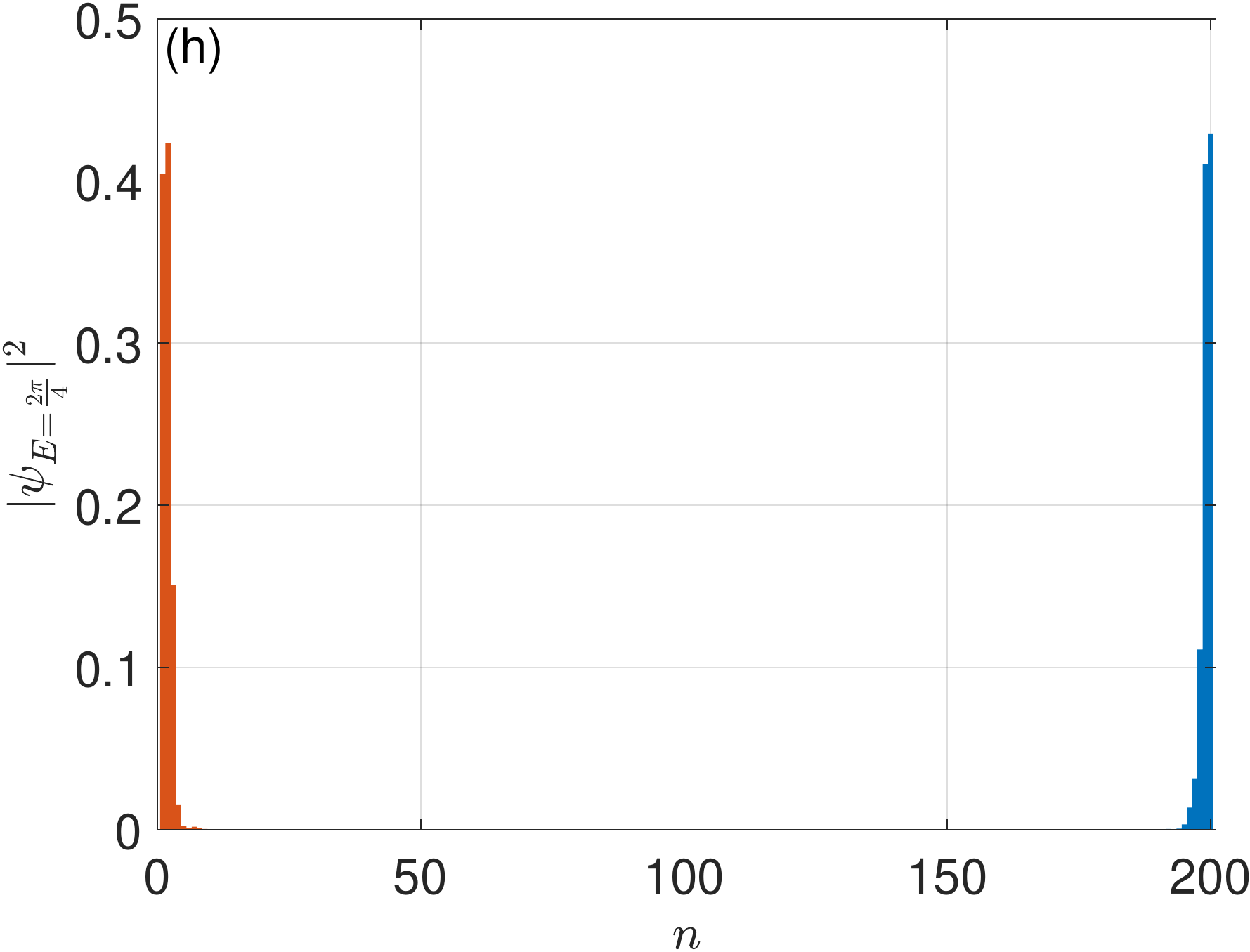}\includegraphics[scale=0.28]{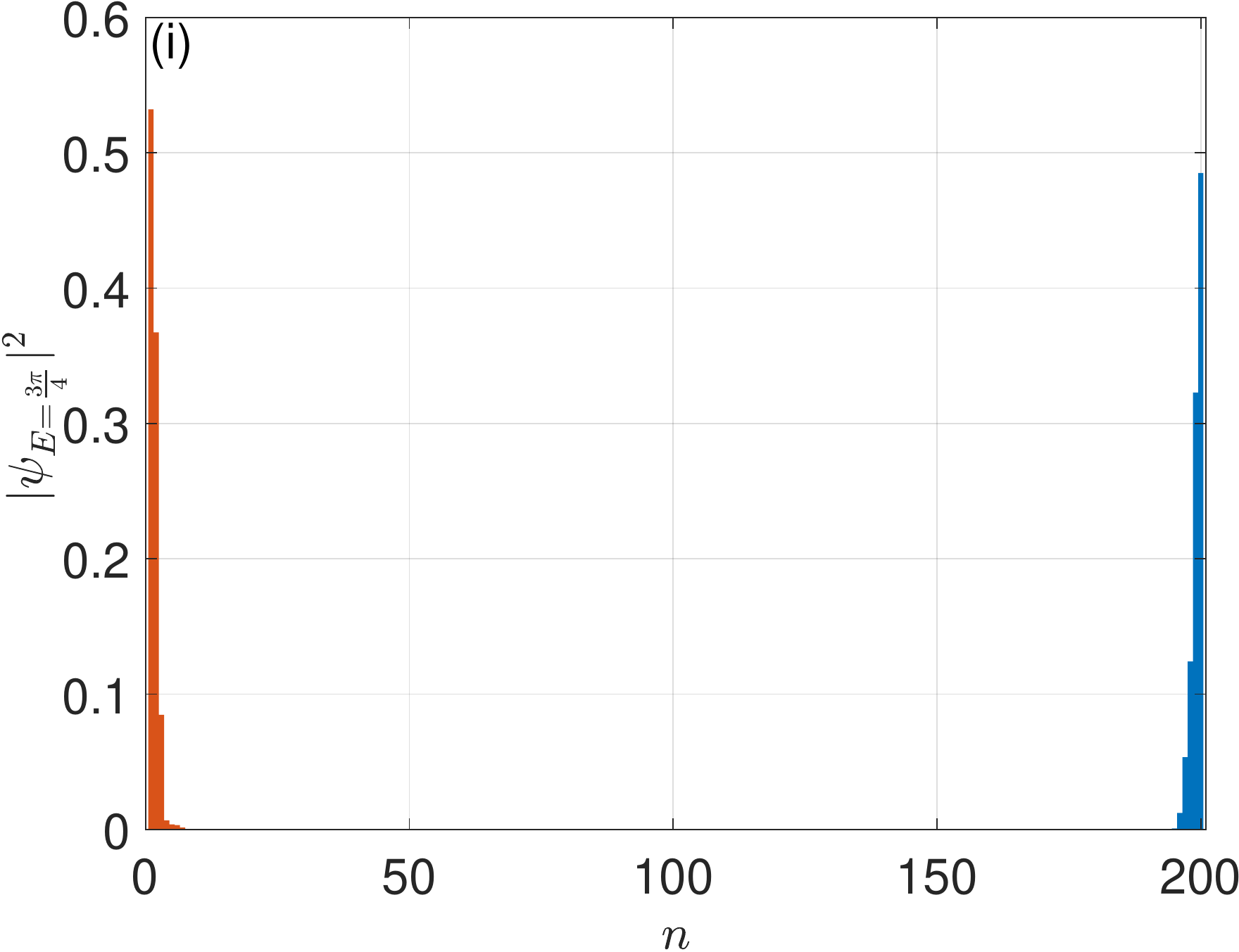}
\par\end{centering}
\begin{centering}
\includegraphics[scale=0.28]{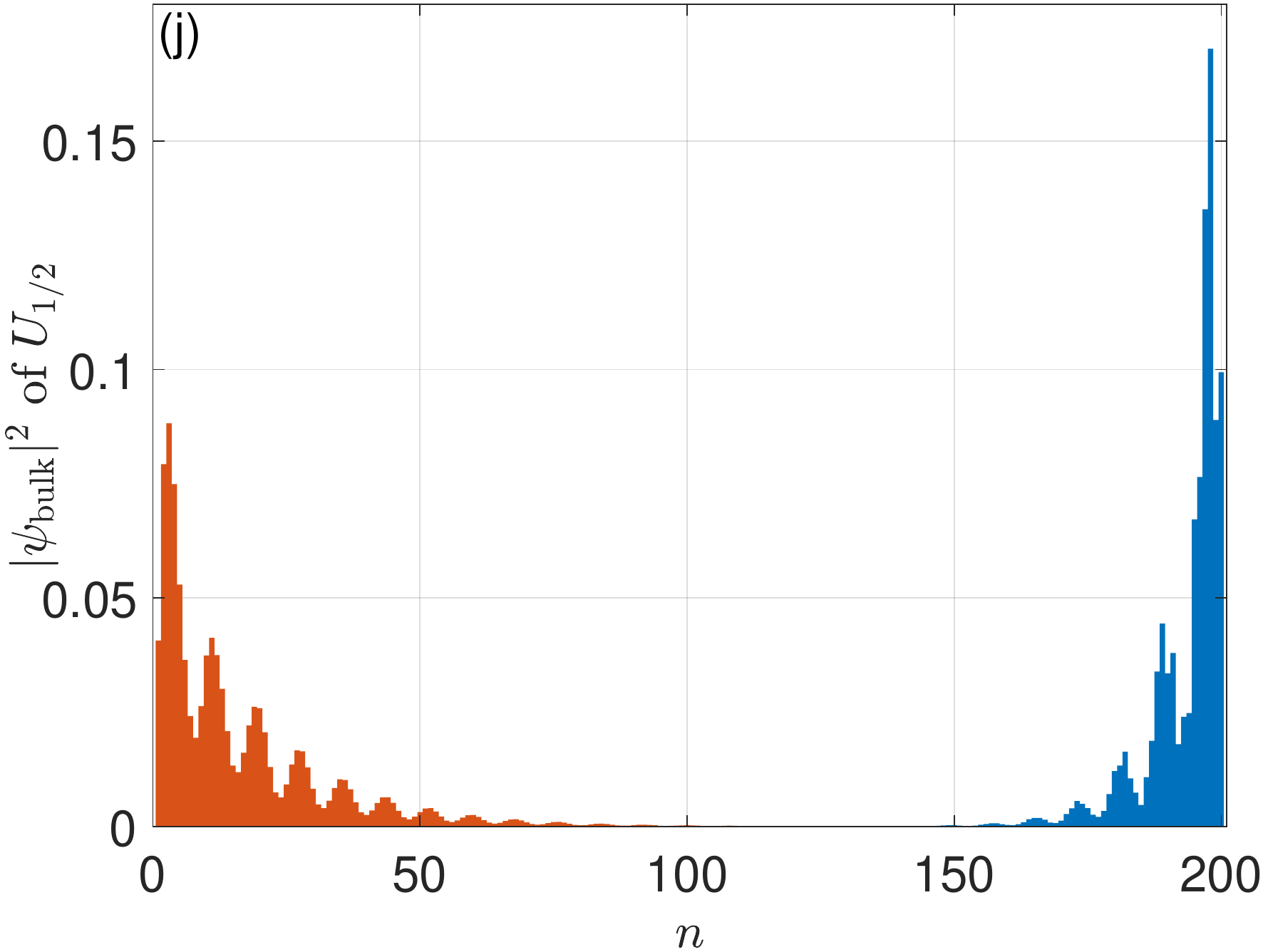}\includegraphics[scale=0.28]{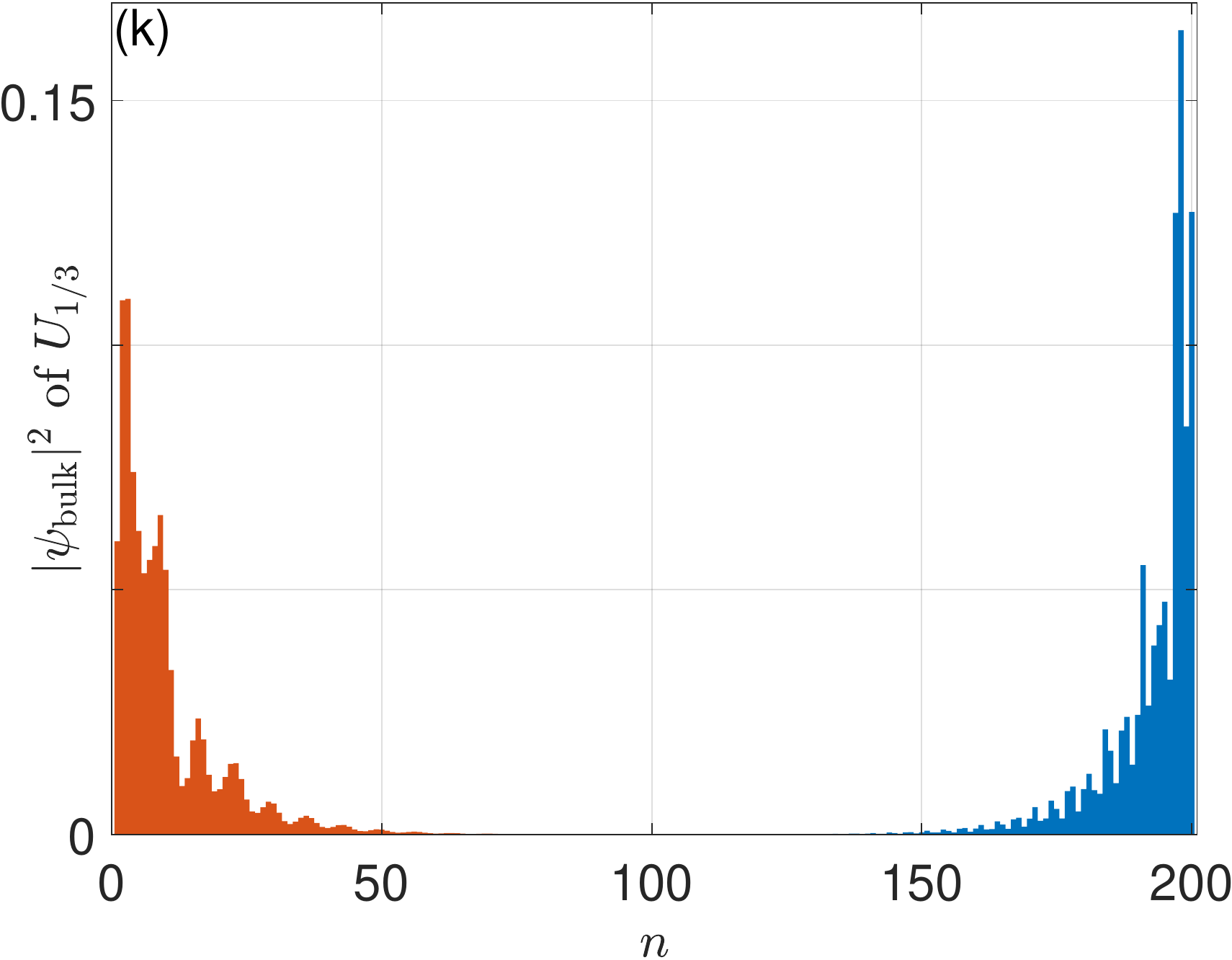}\includegraphics[scale=0.28]{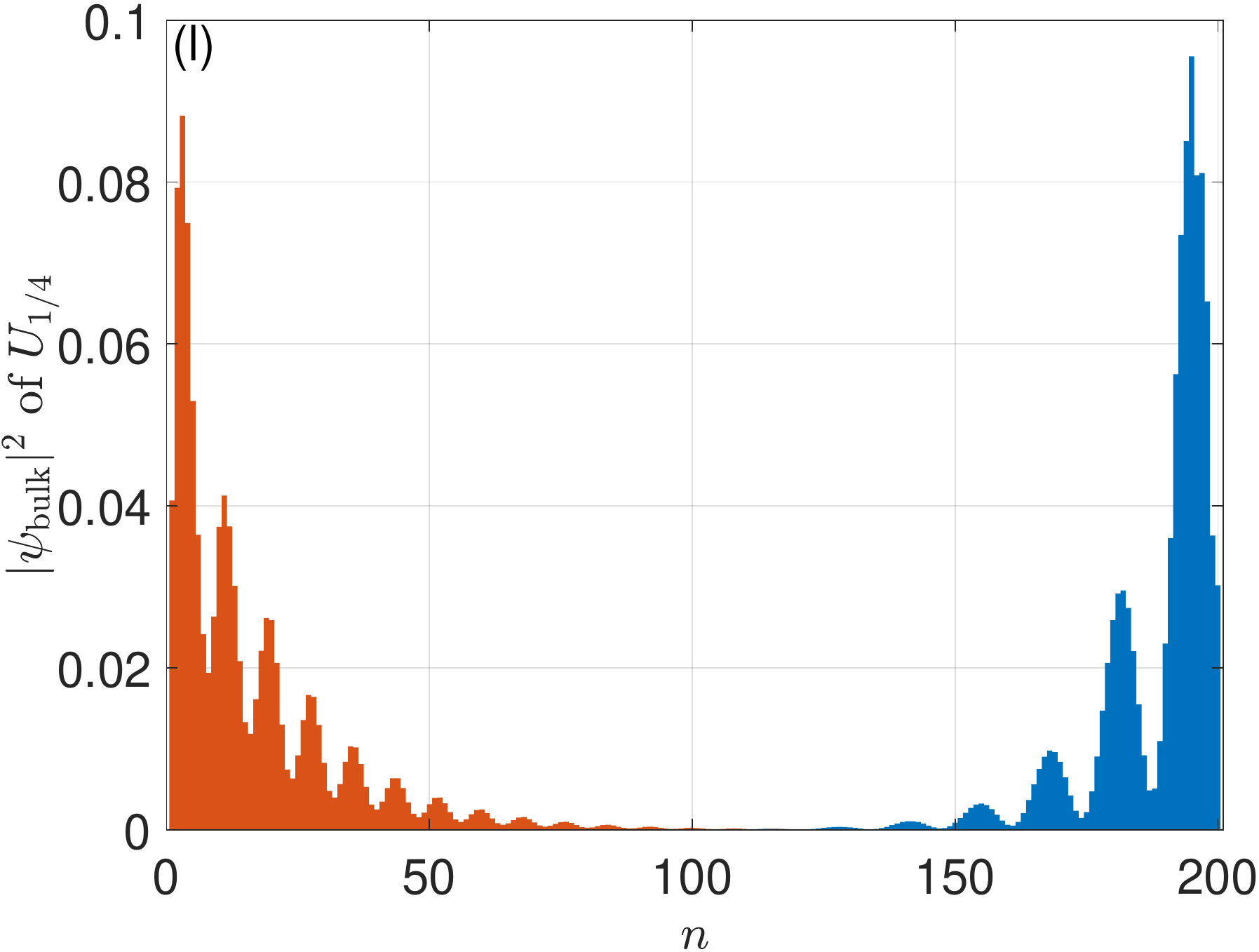}
\par\end{centering}
	\caption{The real part of Floquet spectrum, fractional-quasienergy edge modes and bulk skin modes of the $q$th-root non-Hermitian FTIs for $q=2,3,4$. $j$ and $n$ are state and unit cell indices. System parameters are $(J_1,J_2,\mu,\lambda)=(\pi,0.5\pi,0.4\pi,0.25)$. The length of lattice is $L=400$. (a), (b) and (c) show the Floquet spectrum of $U_{1/2}$, $U_{1/3}$ and $U_{1/4}$, zoomed in around $E=\pi/2$, $E=(\pi/3,2\pi/3)$ and $E=(\pi/4,2\pi/4,3\pi/4)$, respectively. (d) shows the degenerate edge modes of $U_{1/2}$ with $E=\pi/2$. (e) and (f) show the degenerate edge modes of $U_{1/3}$ with $E=\pi/3$ and $2\pi/3$. (g), (h) and (i) show the degenerate edge modes of $U_{1/4}$ with $E=\pi/4$, $2\pi/4$ and $3\pi/4$. (j), (k) and (l) show three pairs of typical bulk states for $U_{1/2}$, $U_{1/3}$ and $U_{1/4}$, respectively, which are piled up around the boundaries and thus represent non-Hermitian Floquet skin modes.\label{fig:EProb1}}
\end{figure*}

In Figs.~\ref{fig:E1ov3E1ov4M1}(a) and \ref{fig:E1ov3E1ov4M1}(b),
we further show the Floquet spectrum and gap function of the cubic-root non-Hermitian
FTI. In addition to edge states at the quasienergies $E=0,\pm\pi$, we also
observe degenerate edge modes at fractional quasienergies
$E=\pm\pi/3,\pm2\pi/3$. Recall that Eq.~(\ref{eq:U1ov3M1}) cubes to a block diagonal matrix consisting of multiple copies of Floquet operator of the parent model $U$. Therefore, the edge states at quasienergies $(0,\pm2\pi/3)$ {[}$(\pm\pi/3,\pm\pi)${]} are indeed descendants of the zero {[}$\pi${]} edge modes in the parent model, whose numbers are counted by $\nu_{0}$ {[}$\nu_{\pi}${]} \cite{NHFTI8}. We then obtain the bulk-edge correspondence
for cubic-root non-Hermitian FTIs as
\begin{equation}
	n_{0}=n_{2\pi/3}=2|\nu_{0}|,\qquad n_{\pi/3}=n_{\pi}=2|\nu_{\pi}|.\label{eq:U1ov3BBCM1}
\end{equation}
With the increase of $J_{1}$, the cubic-root system could also undergo
a series of topological phase transitions, with each of them being
accompanied by the increase of either $n_{0}$ and $n_{2\pi/3}$ or $n_{\pi/3}$ and $n_{\pi}$
by two. We can thus obtain arbitrarily many $\pi/3$ and $2\pi/3$
degenerate edge modes by tuning the single driving parameter $J_{1}$
even in the presence of NHSE. Since it has been demonstrated that
Floquet edge states could be utilized to construct boundary discrete time crystals
(DTCs)~\cite{FQC1,FTC1}, we expect the emergence of unique non-Hermitian
Floquet boundary DTCs through the superposition of $(\pm\pi/3,\pm2\pi/3)$ edge modes
and other edge states in the cubic-root FTIs. 

For completeness, we present the spectrum and gap function in
Figs.~\ref{fig:E1ov3E1ov4M1}(c) and \ref{fig:E1ov3E1ov4M1}(d) for
the fourth-root non-Hermitian FTI, which is constructed by applying
the procedure in Eqs.~(\ref{eq:H1ov4}) and (\ref{eq:U1ov4}) to
the first model in Sec.~\ref{sec:Mod}. The resulting system
holds Floquet edge states at $E=\pm\ell\pi/4$ with $\ell=0,...,4$.
Similar to our analysis of $U_{1/2}$ and $U_{1/3}$, the number of
these edge modes are related to the bulk topological invariants $(\nu_{0},\nu_{\pi})$
of the parent Floquet system via
\begin{equation}
	n_{\pi/4}=n_{3\pi/4}=2|\nu_{0}|,\qquad n_{0}=n_{\pi/2}=n_{\pi}=2|\nu_{\pi}|.\label{eq:U1ov4BBCM1}
\end{equation}
More precisely, we find the values of $(n_{E},n_{E'})$ for the rooted
model to change in the sequence $(0,0)$, $(2,0)$, $(2,2)$, $(2,4)$,
$(4,4)$, $(6,4)$, $(6,6)$ for $E=0,2\pi/3$ ($E=\pi/4,3\pi/4$)
and $E'=\pi/3,\pi$ ($E'=0,\pi/2,\pi$) in Fig.~\ref{fig:E1ov3E1ov4M1}(b)
(Fig.~\ref{fig:E1ov3E1ov4M1}(d)) with the increase of $J_{1}$, while
the winding numbers of the parent model change as $(\nu_{0},\nu_{\pi})=(0,0)$,
$(1,0)$, $(1,-1)$, $(1,-2)$, $(2,-2)$, $(3,-2)$, $(3,-3)$ during
the process \cite{NHFTI8}, confirming the relations in Eqs.~(\ref{eq:U1ov3BBCM1})
and (\ref{eq:U1ov4BBCM1}). 
In Fig.~\ref{fig:EProb1}, we present examples of degenerate edge modes at fractional quasienergies and bulk
skin modes that coexist with these topological edge states in the systems described by $U_{1/q}$ for $q=2,3,4$.
The numbers of edge modes found there coincide with our theoretical predictions.
All these results reveal the power of
our strategy in constructing $q$th-root FTIs
for any $q\in\mathbb{Z}^+$. 
In the following subsection, we will further demonstrate the applicability of the same routine in the construction of $q$th-root FSOTIs.

\subsection{Square/Cubic-root non-Hermitian Floquet second order topological insulators}
\label{subsec:SOTI}

\begin{figure*}
	\begin{centering}
		\includegraphics[scale=0.44]{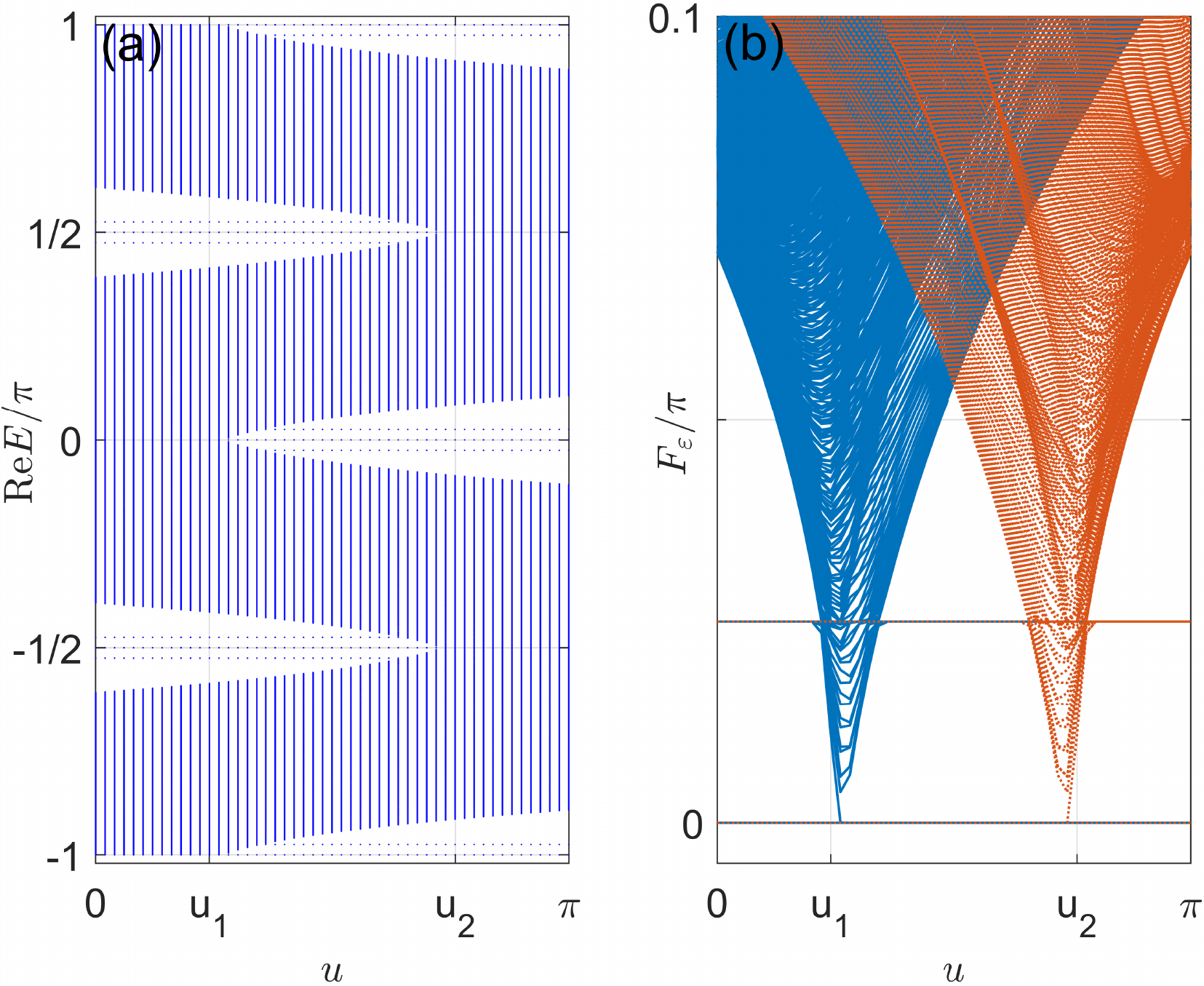}\includegraphics[scale=0.44]{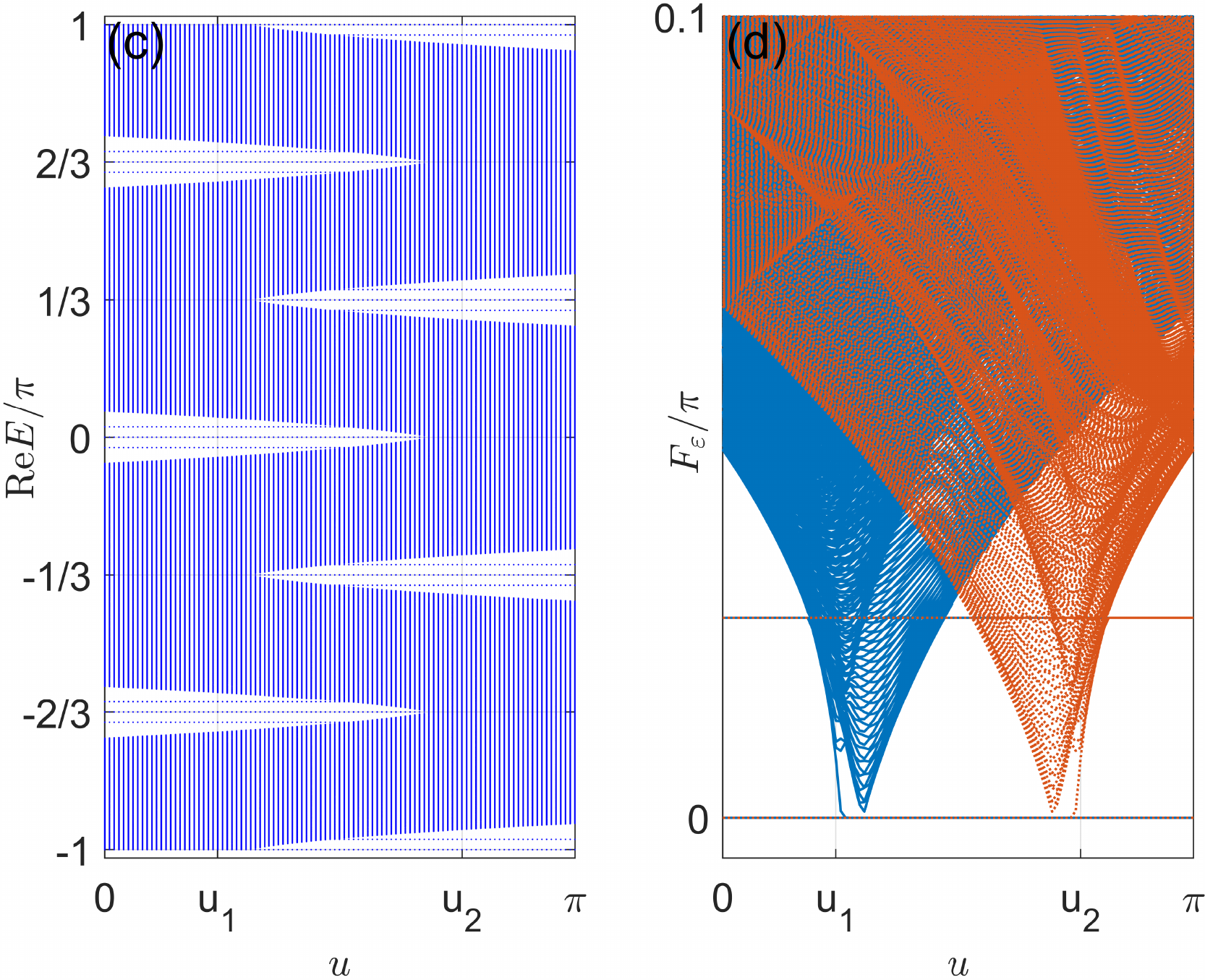}
		\par\end{centering}
	\caption{Floquet spectrum $E$ and gap function $F_{\varepsilon}$ of ${\cal U}_{1/2}$
		{[}Eq.~(\ref{eq:U1ov2M2}){]} and ${\cal U}_{1/3}$ {[}Eq.~(\ref{eq:U1ov3M2}){]}
		versus $u$. Other system parameters are set as $(J_{1},J_{2},\Delta,v)=(0.5\pi,5\pi,0.05\pi,0.5)$.
		The lattice size is $L=2000$ along each spatial direction. (a) and (c) show
		the values of the real part of $E$ for the square- and cubic-root
		models described by ${\cal U}_{1/2}$ and ${\cal U}_{1/3}$, respectively.
		The blue solid and red dotted lines denote the gap functions $F_{0}$ ($=F_{\pi}$)
		and $F_{\pi/2}$ of ${\cal U}_{1/2}$ in (b), and the gap functions $F_{\pi/3}$ ($=F_{\pi}$)
		and $F_{2\pi/3}$ ($=F_{0}$) of ${\cal U}_{1/3}$ in (d). $u_{1}$ and $u_{2}$
		denote critical values of $u$ where the spectral gap closes and the
		number of corner modes changes across the topological phase transition,
		which are obtained from Eq.~(\ref{eq:GaplessM2}).\label{fig:E1ov2E1ov3M2}}
\end{figure*}

\begin{figure}
	\begin{centering}
		\includegraphics[scale=0.5]{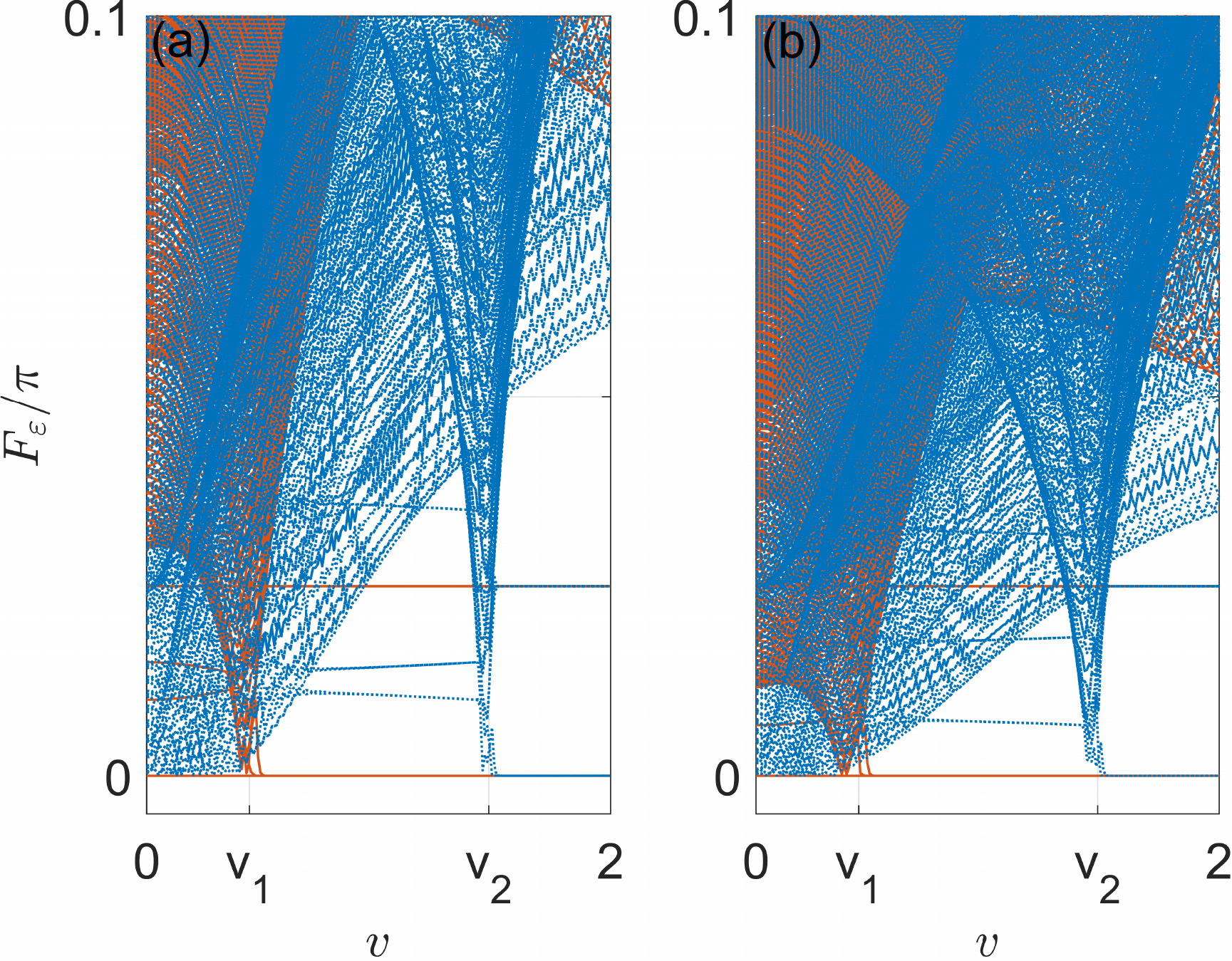}
		\par\end{centering}
	\caption{Gap function $F_{\varepsilon}$ of ${\cal U}_{1/2}$ {[}Eq.~(\ref{eq:U1ov2M2}){]}
		and ${\cal U}_{1/3}$ {[}Eq.~(\ref{eq:U1ov3M2}){]} versus $v$. Other
		system parameters are $(J_{1},J_{2},\Delta,u)=(2.5\pi,3\pi,0.05\pi,0)$
		and the length of lattice is $L=2000$ in each spatial dimension. The blue
		dotted and red solid lines denote the gap functions $F_{0}$ ($=F_{\pi}$) and
		$F_{\pi/2}$ of ${\cal U}_{1/2}$ in (a), and the gap functions $F_{\pi/3}$ ($=F_{\pi}$)
		and $F_{2\pi/3}$ ($=F_{0}$) of ${\cal U}_{1/3}$ in (b). $v_{1}$ and $v_{2}$
		are imaginary parts of $\mu=u+iv$ where the spectral gap closes and the
		number of corner modes change through topological phase transitions,
		obtained by solving Eq.~(\ref{eq:GaplessM2}).\label{fig:F1ov2F1ov3M2}}
\end{figure}

We next generate square- and cubic-root non-Hermitian Floquet second-order topological insulators  (FSOTIs)
by applying our theory in Sec.~\ref{sec:The} 
to the second model in Sec.~\ref{sec:Mod}.
To find the square-root model, we identify $U_1=e^{-i\mathcal{H}_1/2}$ and $U_2=e^{-i\mathcal{H}_2/2}$ in Eq.~(\ref{eq:Uhalf}), where the $\mathcal{H}_1$ and $\mathcal{H}_2$ are given by Eq.~(\ref{eq:H12}) in Sec.~\ref{sec:Mod}.
The resulting square-root Floquet operator reads
\begin{equation}
	{\cal U}_{1/2}=\begin{pmatrix}0 & -e^{-i{\cal H}_{1}/2}\\
		e^{-i{\cal H}_{2}/2} & 0
	\end{pmatrix}.\label{eq:U1ov2M2}
\end{equation}
Similarly, to obtain the cubic-root model, we identify ${\tilde H}_1={\tilde H}_3=3{\mathcal H}_1/4$ and ${\tilde H}_2=3{\mathcal H}_2/2$, where the $\mathcal{H}_1$ and $\mathcal{H}_2$ are given by Eq.~(\ref{eq:H12}).
This leads to the cubic-root Floquet operator
\begin{equation}
	{\cal U}_{1/3}=\begin{pmatrix}0 & e^{-i{\cal H}_{1}/4} & 0\\
		0 & 0 & e^{-i{\cal H}_{2}/2}\\
		e^{-i{\cal H}_{1}/4} & 0 & 0
	\end{pmatrix}.\label{eq:U1ov3M2}
\end{equation}
Recall that the parent Floquet system ${\cal U}=e^{-i\frac{1}{2}{\cal H}_{1}}e^{-i\frac{1}{2}{\cal H}_{2}}$ possesses multiple and non-Hermiticity induced fourfold degenerate corner modes at zero and $\pi$ quasienergies, which are obtained by solving the eigenvalue equation ${\cal U}|\psi\rangle=e^{-iE}|\psi\rangle$. The numbers of these corner modes are related to a pair of topological
invariants introduced in Ref.~\cite{NHFTI5} (see also Sec.~\ref{sec:app2}). Since the process of taking
the square (cubic) root of ${\cal U}$ does not break the protecting
CS of these corner modes, we expect the topological invariants
of ${\cal U}$ to be able to predict the numbers of corner modes at
the $(0,\pi/2,\pi)$ {[}$(0,\pi/3,2\pi/3,\pi)${]} quasienergies of
${\cal U}_{1/2}$ {[}${\cal U}_{1/3}${]}.

\begin{figure*}
\begin{centering}
\includegraphics[scale=0.3]{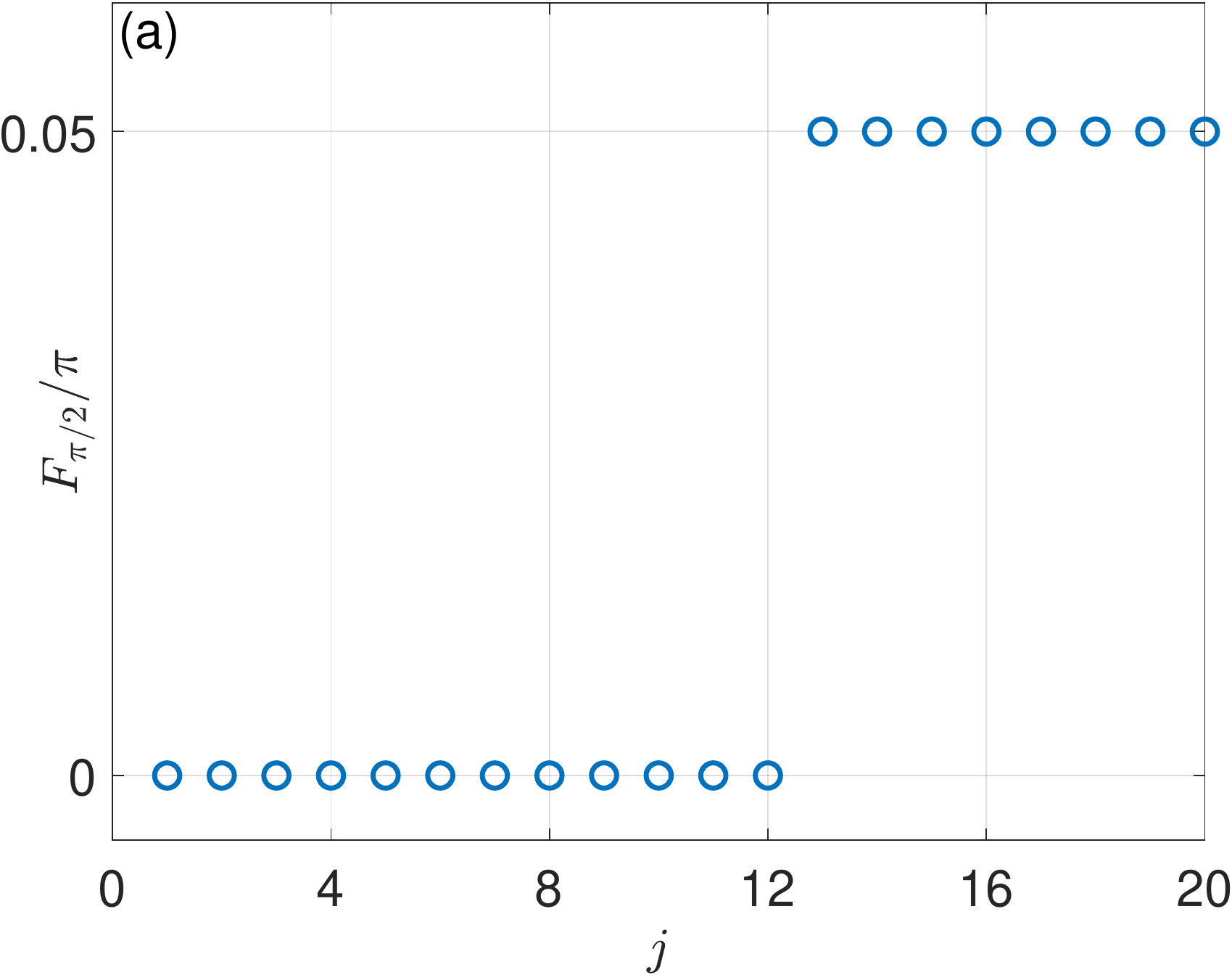}\includegraphics[scale=0.3]{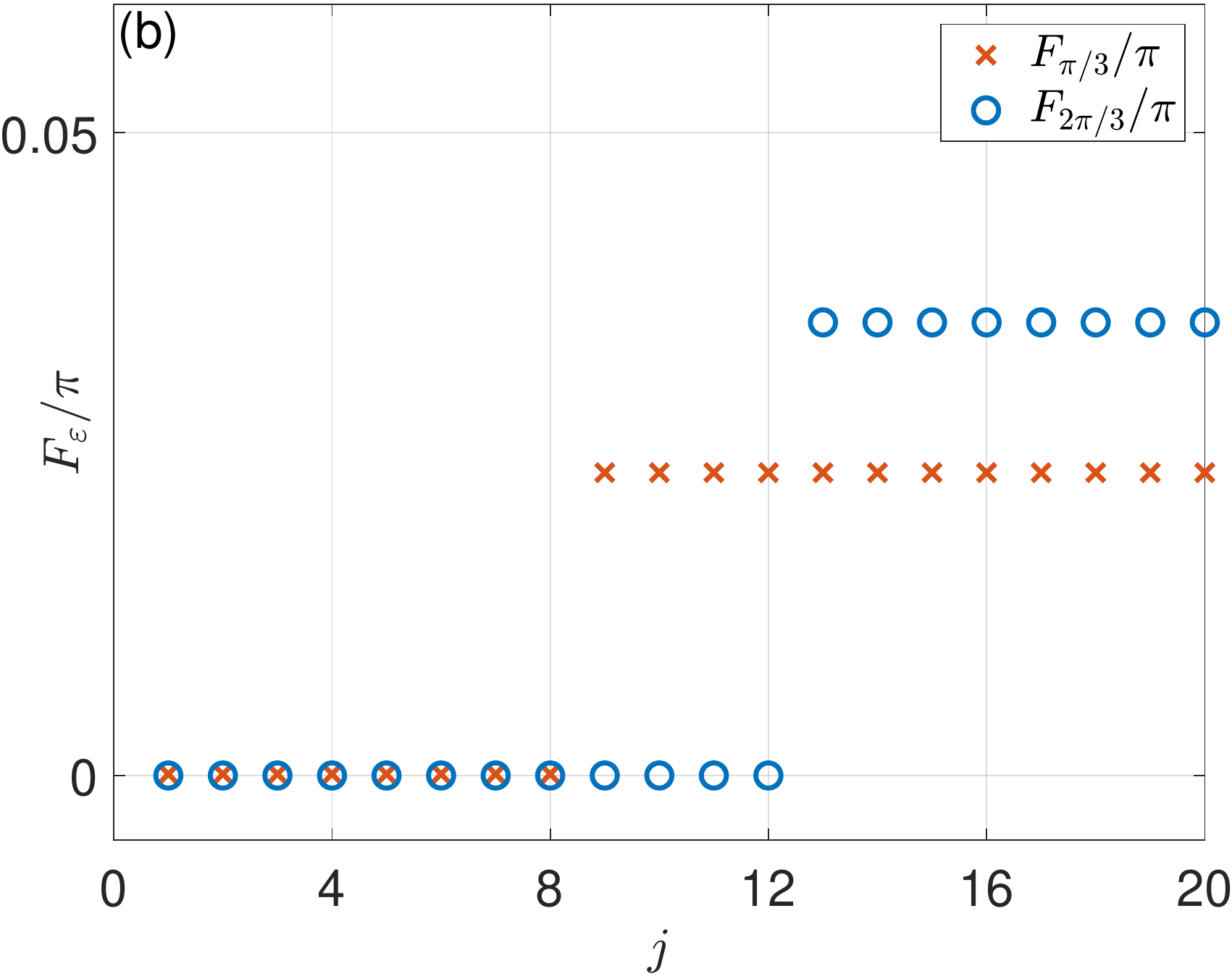}
\par\end{centering}
\begin{centering}
\includegraphics[scale=0.3]{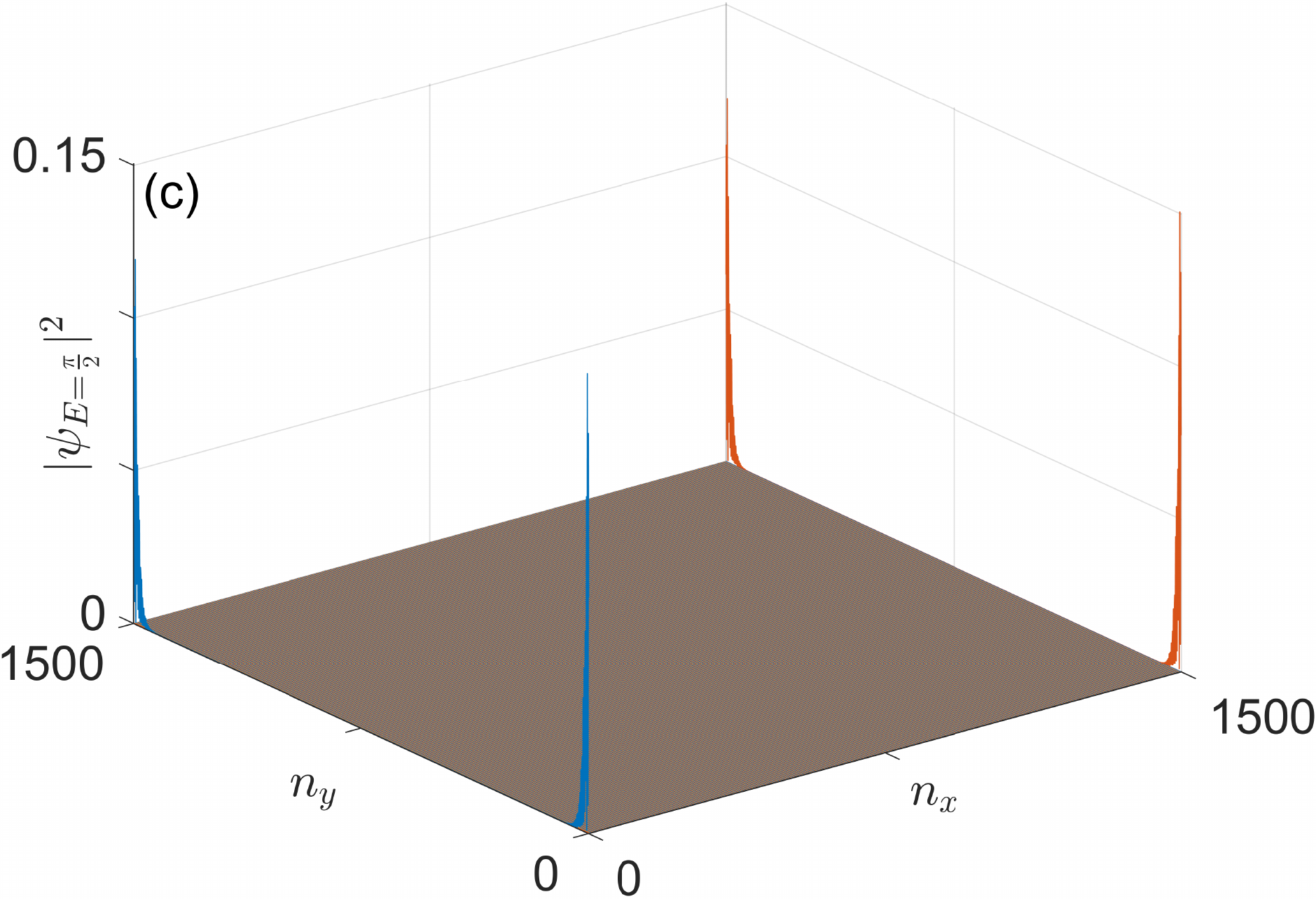}\includegraphics[scale=0.3]{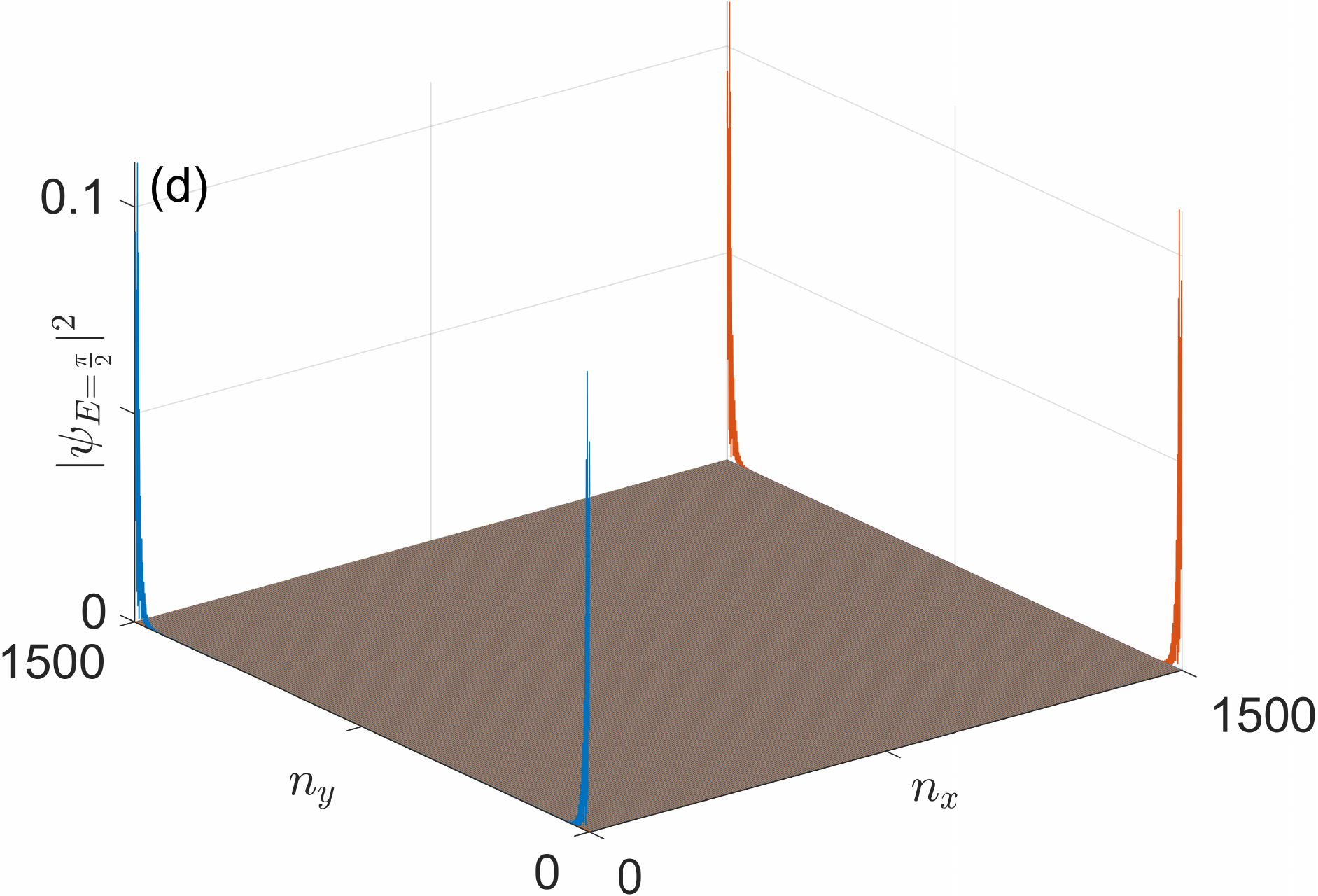}
\par\end{centering}
\begin{centering}
\includegraphics[scale=0.3]{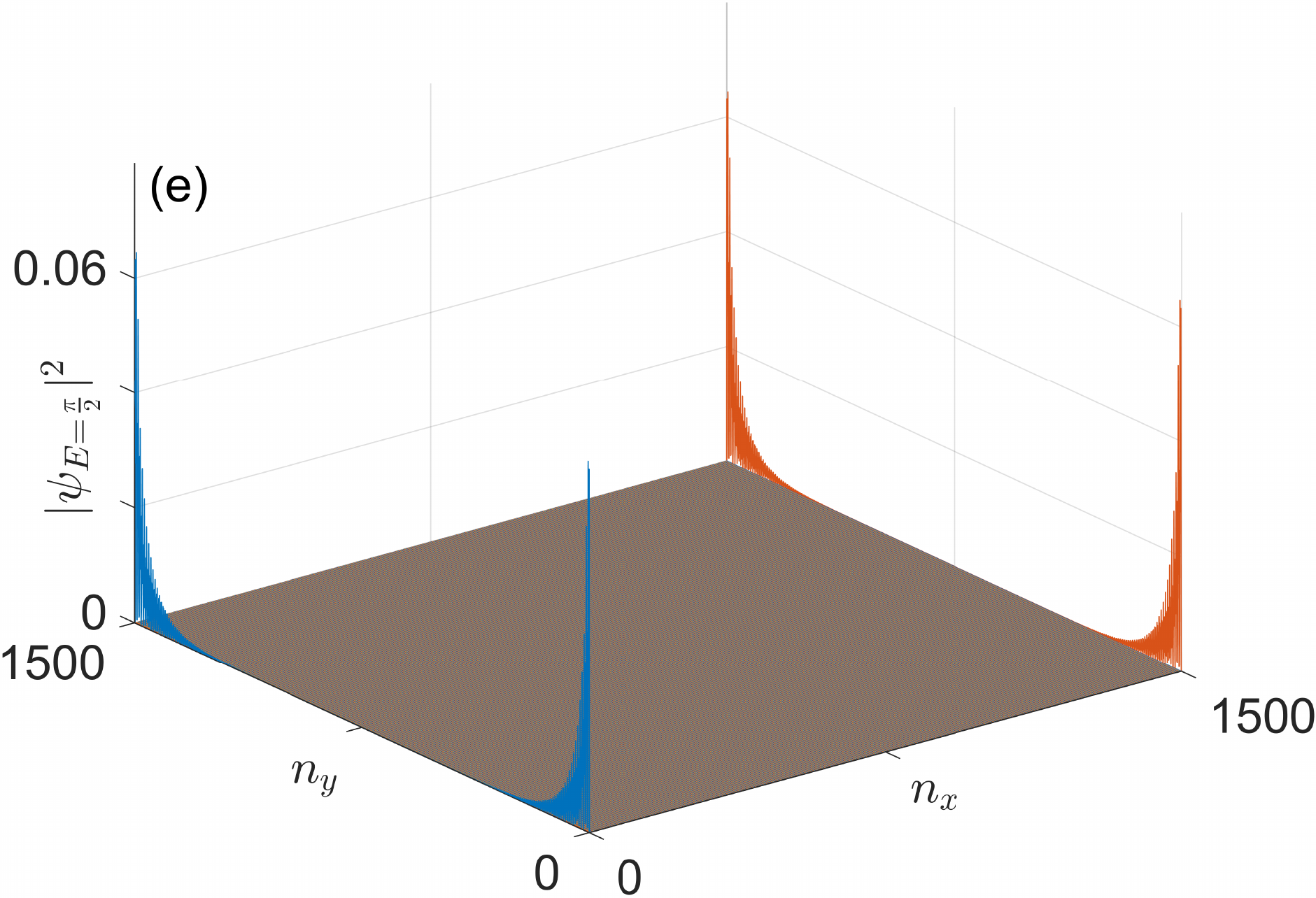}\includegraphics[scale=0.3]{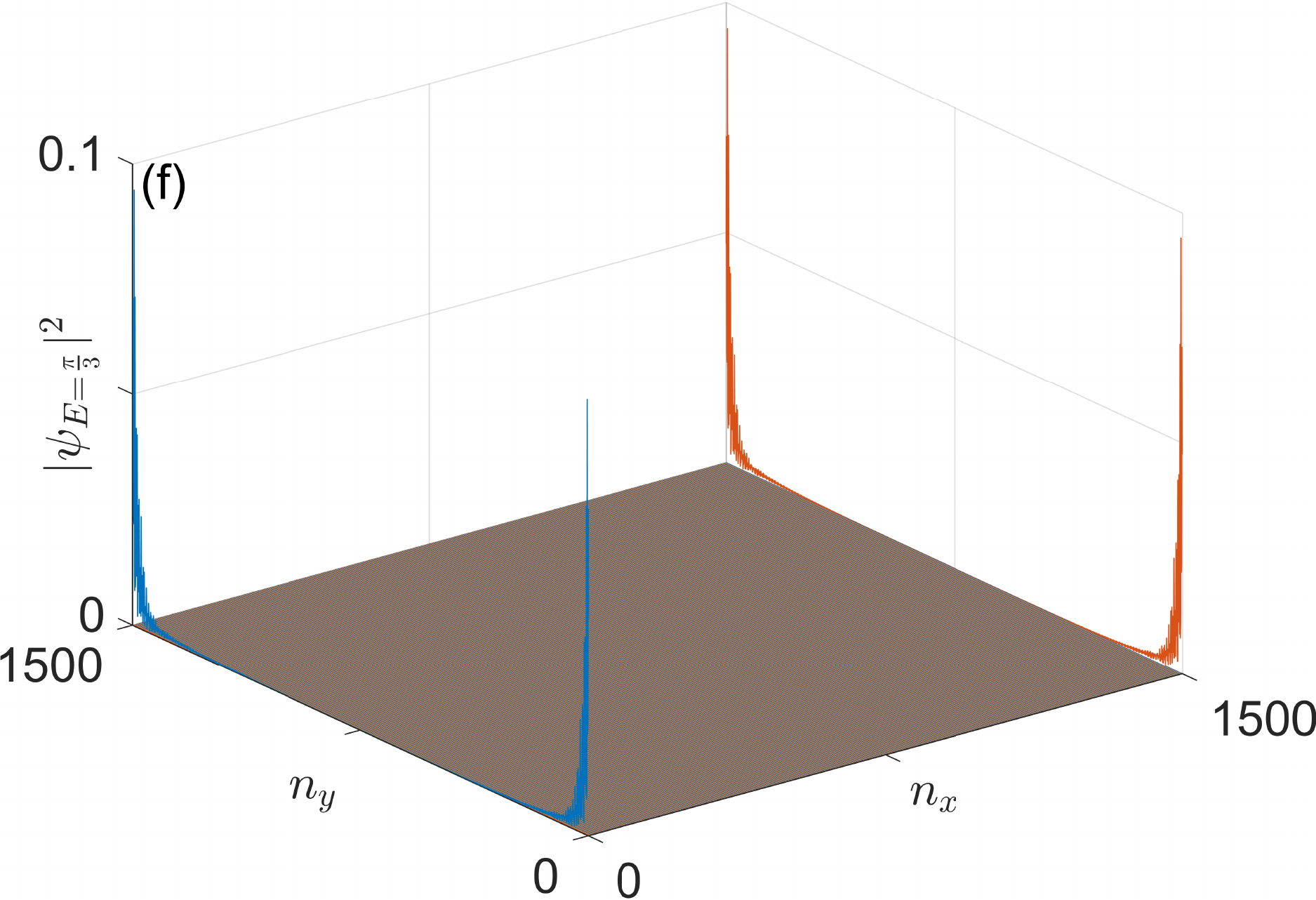}
\par\end{centering}
\begin{centering}
\includegraphics[scale=0.3]{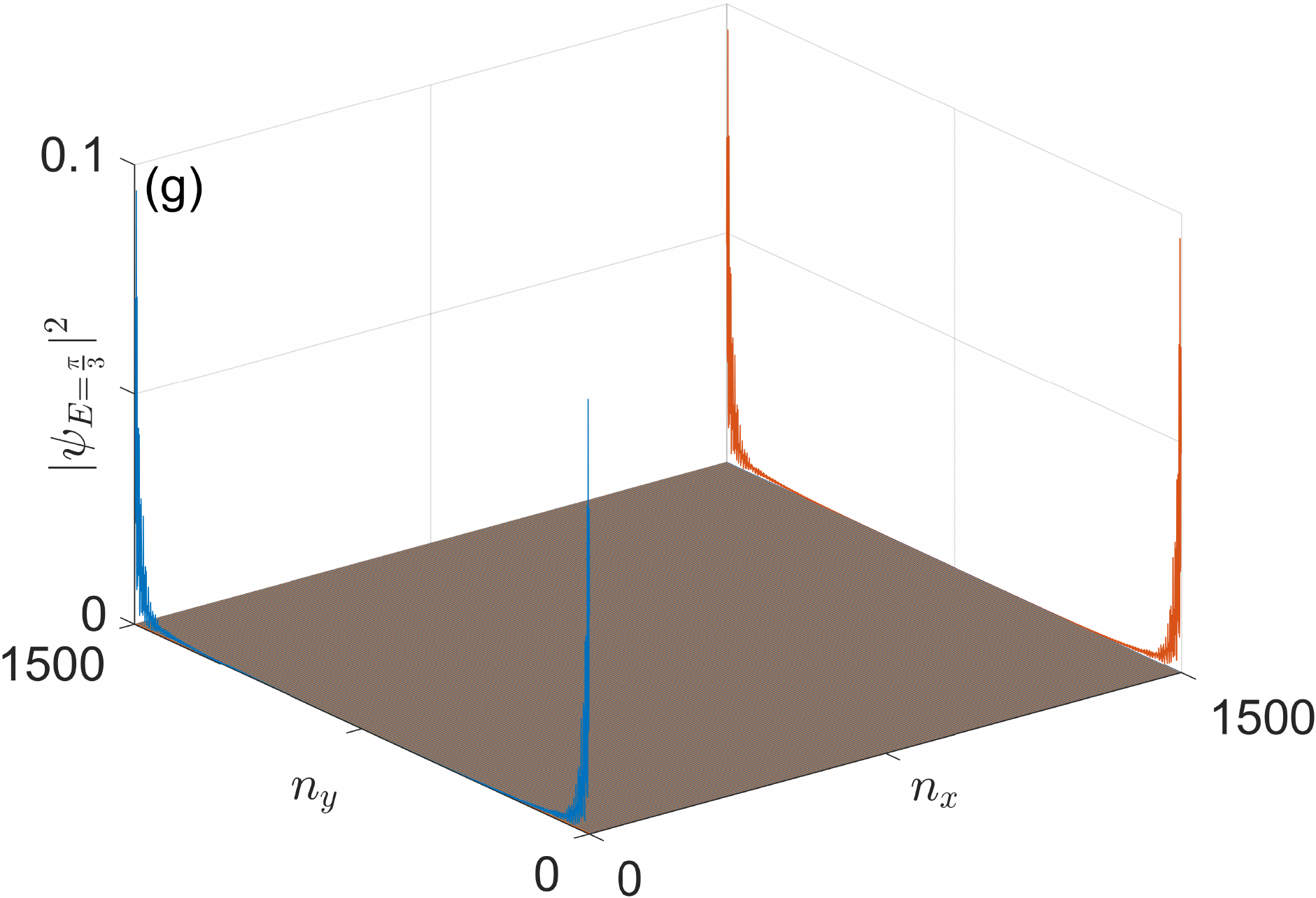}\includegraphics[scale=0.3]{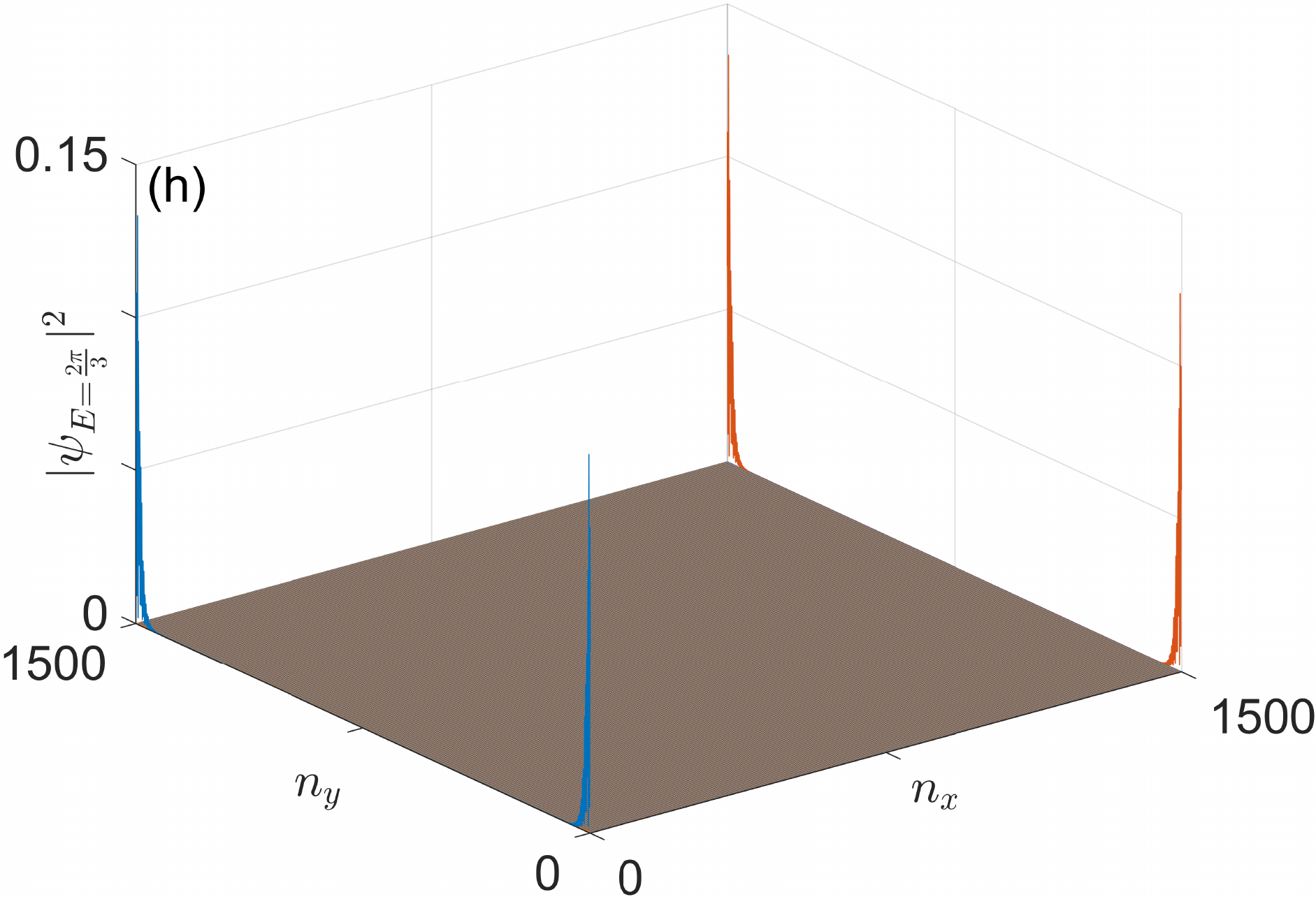}
\par\end{centering}
\begin{centering}
\includegraphics[scale=0.3]{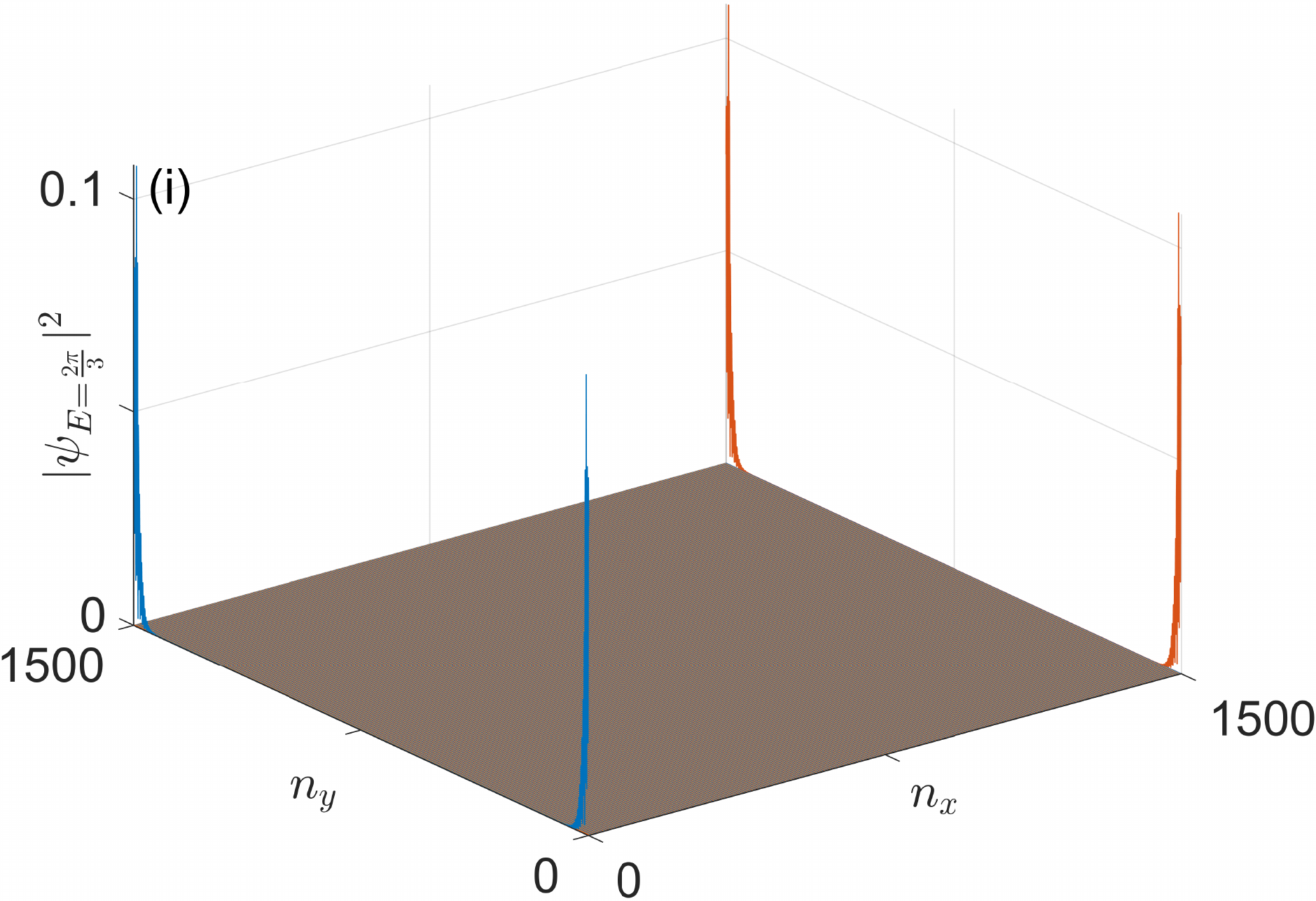}\includegraphics[scale=0.3]{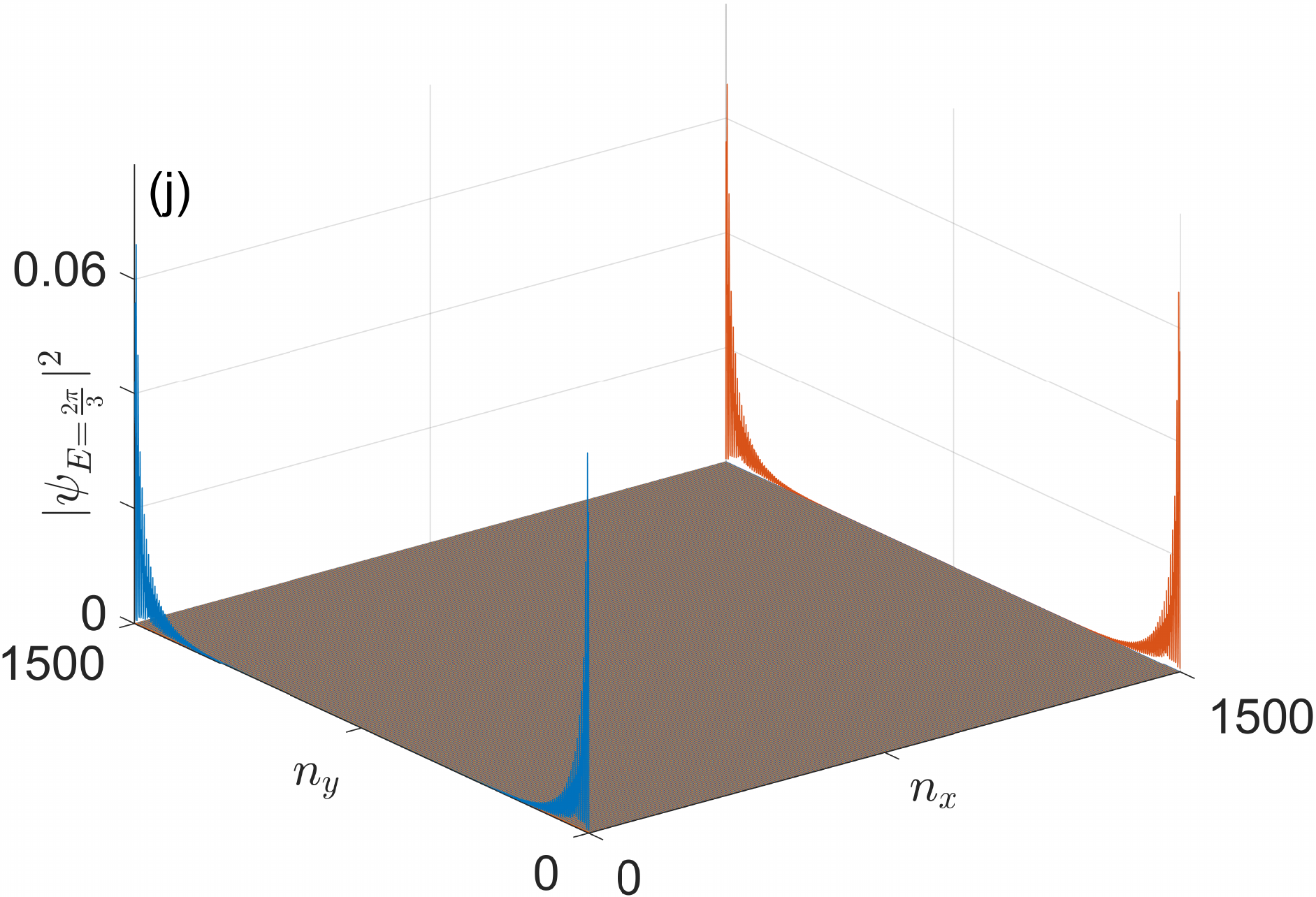}
\par\end{centering}
\caption{Gap functions and probability distributions of corner modes for ${\cal U}_{1/2}$ in (a), (c)--(e) and for ${\cal U}_{1/3}$ in (b), (f)--(j). $j$ and $n_{x,y}$ are state and unit cell indices. System parameters are $(J_1,J_2,\Delta,\mu)=(2.5\pi,3\pi,0.05\pi,2i)$. The lattice size is $L_x=L_y=3000$. (a) and (b) show the gap function $F_{\varepsilon}$, which is zoomed in around $F_{\varepsilon}=0$ for $\varepsilon=\pi/2$, $\pi/3$ and $2\pi/3$. (c)--(e) show the twelve corner modes of ${\cal U}_{1/2}$ at the quasienergy $\pi/2$ (with $F_{\pi/2}=0$). (f), (g) show the eight corner modes of ${\cal U}_{1/3}$ at the quasienergy $\pi/3$ (with $F_{\pi/3}=0$). (h)--(j) show the twelve corner modes of ${\cal U}_{1/3}$ at the quasienergy $2\pi/3$ (with $F_{2\pi/3}=0$).\label{fig:EProb2}}
\end{figure*}

The spectra of ${\cal U}_{1/2}$ and ${\cal U}_{1/3}$
are presented in Fig.~\ref{fig:E1ov2E1ov3M2}. In Figs.~\ref{fig:E1ov2E1ov3M2}(a)
and \ref{fig:E1ov2E1ov3M2}(c), we observe states
at the fractional-quasienergies $E=\pm\pi/2$ and $E=\pm\pi/3,\pm2\pi/3$ for the square-root and cubic-root systems respectively, whose
spatial profiles are found to be localized around the four corners
of the lattice. The numbers of these corner modes can further be controlled
by changing the real part of onsite potential $u$, as can be seen
clearly in Figs.~\ref{fig:E1ov2E1ov3M2}(b) and \ref{fig:E1ov2E1ov3M2}(d).
The critical values $(u_{1},u_{2})$ where the number of corner modes
change correspond to topological phase transition points, which are
determined by the gapless condition of the parent model \cite{NHFTI5},
i.e.,
\begin{equation}
	\cos\left[J_{1}\sqrt{1-(n\pi-u)^{2}/J_{2}^{2}}\right]\cosh v=\pm1.\label{eq:GaplessM2}
\end{equation}
Moreover, since the $\pi/2$ (zero and $\pi$) modes of ${\cal U}_{1/2}$
are inherited from the zero ($\pi$) modes of the parent model ${\cal U}$,
their numbers are determined by the winding numbers $(\nu_{0},\nu_{\pi})$
of the second model in Sec.~\ref{sec:Mod} (see also Sec.~\ref{sec:app2}) through
the bulk-corner correspondence relations
\begin{equation}
	n_{\pi/2}=4|\nu_{0}|,\qquad n_{0}=n_{\pi}=4|\nu_{\pi}|.\label{eq:U1ov2BBCM2}
\end{equation}
Similarly, as the zero and $2\pi/3$ ($\pi/3$ and $\pi$) corner
modes of ${\cal U}_{1/3}$ are inherited from the zero ($\pi$) corner
modes of ${\cal U}$, we have the following relations to determine
their numbers from the bulk invariants of the parent model, i.e.,
\begin{equation}
	n_{0}=n_{2\pi/3}=4|\nu_{0}|,\qquad n_{\pi/3}=n_{\pi}=4|\nu_{\pi}|.\label{eq:U1ov3BBCM2}
\end{equation}
In the regions $u\in(0,u_{1})$, $(u_{1},u_{2})$, $(u_{2},\pi)$, 
find winding numbers $(\nu_{0},\nu_{\pi})=(5,4)$, $(5,5)$, $(4,5)$,
and the number of corner modes $(n_{\pi/2},n_{0},n_\pi)=(20,16,16)$, $(20,20,20)$,
$(16,20,20)$ {[}$(n_{0},n_{2\pi/3},n_{\pi/3},n_{\pi})=(20,20,16,16)$, $(20,20,20,20)$,
$(16,16,20,20)${]} for ${\cal U}_{1/2}$ {[}${\cal U}_{1/3}${]}, which
confirm the Eqs.~(\ref{eq:U1ov2BBCM2}) and (\ref{eq:U1ov3BBCM2}).

A very intriguing feature of the square- and cubic-root FSOTIs
studied here is that the non-Hermitian effect could induce more topological
corner modes than in the Hermitian limit. To demonstrate this, we
investigate the gap functions of ${\cal U}_{1/2}$ and ${\cal U}_{1/3}$
versus the gain and loss strength $v$ in Figs.~\ref{fig:F1ov2F1ov3M2}(a)
and \ref{fig:F1ov2F1ov3M2}(b). In both figures, $v_{1}$ and $v_{2}$
denote critical values of $v$ where the system undergoes topological
phase transitions. Their specific values are determined by the gapless
condition of the parent model in Eq.~(\ref{eq:GaplessM2}).
With the increase of $v$ from $0$ to $2$, we find that in the three
regions $v\in(0,v_{1})$, $(v_{1},v_{2})$, $(v_{2},2)$, the winding
numbers of the parent model are $(\nu_{0},\nu_{\pi})=(1,0)$, $(3,0)$,
$(3,-2)$ \cite{NHFTI5}, whereas the number of corner modes are $(n_{\pi/2},n_{0},n_{\pi})=(4,0,0)$,
$(12,0,0)$, $(12,8,8)$ in Fig.~\ref{fig:F1ov2F1ov3M2}(a) for the square-root
model ${\cal U}_{1/2}$, and $(n_{0},n_{2\pi/3},n_{\pi/3},n_{\pi})=(4,4,0,0)$,
$(12,12,0,0)$, $(12,12,8,8)$ in Fig.~\ref{fig:F1ov2F1ov3M2}(b) for the cubic-root system ${\cal U}_{1/3}$. These results are clearly consistent with
the bulk-corner correspondence relations for square- and cubic-root
FSOTIs as stated in Eqs.~(\ref{eq:U1ov2BBCM2}) and (\ref{eq:U1ov3BBCM2}).

Note in passing that the horizontal lines appearing at $F_\varepsilon\neq0$ in Figs.~\ref{fig:E1ov2E1ov3M2}(b), \ref{fig:E1ov2E1ov3M2}(d) and Fig.~\ref{fig:F1ov2F1ov3M2} are related to eigenmodes formed by coupling edge states along the $y$-direction and bulk states along the $x$-direction of the lattice. As the 1D chains along $y$ possess a chiral symmetry, the coupling of their degenerate edge modes with the bulk states along $x$ can yield degenerate states in 2D that are protected by the chiral symmetry of a 1D subsystem. We may thus regard these edge states as weak edge states caused by weak topology. In the 2D system, their number is sensitive to the system size along $x$ and their quasienergies are sensitive to the choice of system parameters. Comparatively, the numbers and quasienergies of corner modes are solely protected by the chiral symmetry and topological invariants of the 2D system, making them robust to the changes of system size and parameters before encountering a phase transition.

Finally, in Fig.~\ref{fig:EProb2}, we present the gap functions and spatial profiles of Floquet corner modes at the quasienergies $\pi/2$, $\pi/3$ and $2\pi/3$. Their numbers are found to precisely coincide with the bulk-corner correspondence relations in  Eqs.~(\ref{eq:U1ov2BBCM2}) and (\ref{eq:U1ov3BBCM2}). The non-Hermiticity enriched higher-order topology in rooted Floquet systems may also assist us to engineer unique DTCs and quantum computing schemes with the multiple quartets of corner modes at different fractional-quasienergies that are robust to the perturbation of environment.

\section{Conclusion}
\label{sec:Sum}
In summary, we proposed a systematic approach to construct the $q$th-root of any periodically driven system and presented $2^{n}$th-
and $3^{n}$th-root Floquet topological insulators as explicit examples. The latter are shown to exhibit degenerate edge/corner modes at fractional
quasienergies $\frac{\pi}{2^{n}}(0,1,...,2^{n})$ and $\frac{\pi}{3^{n}}(0,1,...,3^{n})$,
whose topological nature is inherited from their $2^{n}$th-
and $3^{n}$th-power parent systems. Square- and cubic-root non-Hermitian Floquet topological insulators with
multiple and tunable topological edge/corner states at quasienergies
$\pi/2$, $\pi/3$ and $2\pi/3$ were investigated in details. Further connections were made between
the number of these edge/corner modes and the bulk topological invariants
of parent systems, yielding the bulk-edge/corner correspondence in
two classes of rooted Floquet topological insulators. Intriguingly, non-Hermitian
effects are found to induce more corner modes with fractional-quasienergies
and generate multiple edge states coexisting with the non-Hermitian skin effect in rooted
systems. Our discoveries thus uncover a unique class of topological
phases that originates from the cooperation among driving, non-Hermiticity
and the process of taking the nontrivial roots of Floquet systems. 

From the experimental perspective, the obtained systems from our $q$th-rooting procedure are expected to be implementable with the same setups employed for realizing their parent models. To this end, the additional degrees of freedom required in our $q$th-rooting procedure can be principally implemented by coupling multiple copies of the parent system. In the context of the non-Hermitian Floquet first and second order topological insulators explored in this paper, their corresponding square- and cubic-root systems can thus be realized in principle via setups like photonic quantum walks \cite{NHFQC2,NHQW1,NHQW2,NHQW3,NHQW4}. For example, the anisotropic hopping amplitude and non-Hermitian lattice potential can be implemented by introducing controlled optical losses with acousto-optical modulators in coupled optical fibre loops \cite{NHFQC2}. Moreover, the winding numbers used in characterizing their topology can in principle be experimentally probed via measuring mean chiral displacements \cite{NHFTI2,NHFTI4,MCD} or time-averaged spin textures \cite{NHFTI3,spint}, which can also be conducted in similar photonic setups \cite{NHQW1,NHQW2,NHQW3,NHQW4}.

In future work, it would be interesting to apply our scheme to realize
$q$th-root chiral Floquet topological insulators and gapless topological
phases in higher spatial dimensions. The application of our approach to systems
with many-body interactions is also expected to be fruitful. In particular, it was recently shown that the interplay between interaction and periodic driving may promote $2\pi/{2^n}$ modes into $Z_{2^n}$ parafermions \cite{Fparafermion}. Other families of $2\pi/q$ modes obtained in this work thus open avenues for exploring different types of Floquet parafermions not covered in Ref.~\cite{Fparafermion}. In particular, $Z_3$ parafermions, which are expected to arise in systems with $2\pi/3$ modes when subjected to appropriate interactions, form a main ingredient for constructing the powerful Fibonacci anyons \cite{z3para} that enable topologically protected universal quantum computation.

\section*{Acknowledgements}

\paragraph{Author contributions}
L.Z. and R.W.B. contributed equally to this work.

\paragraph{Funding information}
L.Z. is supported by the National Natural Science Foundation of China (Grant No. 11905211), the Young Talents Project at Ocean University of China (Grant No. 861801013196), and the Applied Research Project of Postdoctoral Fellows in Qingdao (Grant No. 861905040009). 
R.W.B. is supported by the Australian Research Council Centre of Excellence for Engineered Quantum Systems (EQUS, CE170100009).

\begin{appendix}

\section{Topological invariants of the non-Hermitian FTI}
\label{sec:app1}
Here we briefly recap the open-boundary winding numbers (OBWNs) introduced in Ref.~\cite{NHFTI8}.
They will be used to establish the bulk-edge correspondence for the first parent model in Sec.~\ref{sec:Mod}
and its $q$th-root descendants in Sec.~\ref{subsec:TI}. We first consider the dynamics of the model in two symmetric time frames,
where the Floquet operator $U=e^{-iH_1/2}e^{-iH_2/2}$ is transformed to $U_a=e^{-iH_2/4}e^{-iH_1/2}e^{-iH_2/4}$ and 
$U_b=e^{-iH_1/4}e^{-iH_2/2}e^{-iH_1/4}$. Next we define the $Q$-matrix in the time frame $\alpha$ ($=a,b$) as 
$Q_\alpha=\sum_j(|\psi^+_{\alpha j}\rangle\langle{\tilde \psi}^+_{\alpha j}|-|\psi^-_{\alpha j}\rangle\langle{\tilde \psi}^-_{\alpha j}|)$.
The right and left Floquet eigenvectors $|\psi^\pm_{\alpha j}\rangle$ and $\langle{\tilde \psi}^\pm_{\alpha j}|$ satisfy the eigenvalue equations
$U_\alpha|\psi^\pm_{\alpha j}\rangle=e^{-i(\pm E_j)}|\psi^\pm_{\alpha j}\rangle$ 
and $\langle{\tilde \psi}^\pm_{\alpha j}|U_\alpha=\langle{\tilde \psi}^\pm_{\alpha j}|e^{-i(\pm E_j)}$.
An OBWN for $U_\alpha$ is then defined as $\nu_\alpha={\rm Tr_B}(\Gamma Q_\alpha[Q_\alpha,X])/L_{\rm B}$.
Here $\Gamma$ is the chiral symmetry (CS) operator. $X$ is the unit cell position operator.
For a system with $L$ lattice sites, we decompose it into a bulk region and two edge regions
at the left and right. The trace ${\rm Tr_B}$ is taken over the bulk region, which contains $L_{\rm B}$ lattice sites.
The length of each edge region is $L_{E}=(L-L_{\rm B})/2$, which should be chosen properly in order to avoid the
obstruction of non-Hermitian skin effect.
Finally, we define two OBWNs for a 1D non-Hermitian FTI with CS as $\nu_0=(\nu_a+\nu_b)/2$ and $\nu_\pi=(\nu_a-\nu_b)/2$.
These winding numbers can only take integer values. They are further related to the numbers of Floquet edge modes at zero and $\pi$
quasienergies $n_0$ and $n_\pi$ through the relations $(n_0,n_\pi)=2(|\nu_0|,|\nu_\pi|)$. Following our analysis in the
main text, $(\nu_0,\nu_\pi)$ could also count the numbers of fractional-quasienergy edge modes in the $q$th-root descendants of
the parent model $U$.

\section{Topological invariants of the non-Hermitian FSOTI}
\label{sec:app2}
Here we summarize the definition of bulk winding numbers for the second parent model in Sec.~\ref{sec:Mod} of main text.
Following Ref.~\cite{NHFTI5}, we first transform the Floquet operator ${\cal U}$ into two symmetric
time frames $a$ and $b$ by shifting the initial time of evolution from $t=0$ to $t=1/4$ and $t=3/4$ respectively.
The Floquet operators in these time frames take the forms ${\cal U}_a=e^{-i{\cal H}_2/4}e^{-i{\cal H}_1/2}e^{-i{\cal H}_2/4}$
and ${\cal U}_b=e^{-i{\cal H}_1/4}e^{-i{\cal H}_2/2}e^{-i{\cal H}_1/4}$.
Performing Fourier transforms from position to momentum representations, we obtain
${\cal U}_\alpha=\sum_{k_x,k_y}|k_x,k_y\rangle{\mathfrak U}_{\alpha}(k_x,k_y)\langle k_x,k_y|$ with $\alpha=a,b$.
In the tensor product form, we have ${\mathfrak U}_\alpha(k_x,k_y)={\cal U}_0(k_x)\otimes{\cal U}_\alpha(k_y)$, where
${\cal U}_0(k_x)=e^{-i{\cal H}_x(k_x)}$, 
\begin{equation}
{\cal U}_a(k_y)=e^{-i{\cal H}_{y2}(k_y)/4}e^{-i{\cal H}_{y1}(k_y)/2}e^{-i{\cal H}_{y2}(k_y)/4},
\end{equation}
\begin{equation}
{\cal U}_b(k_y)=e^{-i{\cal H}_{y1}(k_y)/4}e^{-i{\cal H}_{y2}(k_y)/2}e^{-i{\cal H}_{y1}(k_y)/4}.
\end{equation}
The ${\cal H}_x(k_x)$, ${\cal H}_{y1}(k_y)$ and ${\cal H}_{y2}(k_y)$ are Fourier transforms of the
Eqs.~(\ref{eq:Hx})--(\ref{eq:Hy2}) in the main text. ${\mathfrak U}_{\alpha}(k_x,k_y)$ has the CS
in the sense that $\Gamma{\mathfrak U}_{\alpha}(k_x,k_y)\Gamma={\mathfrak U}_{\alpha}^{-1}(k_x,k_y)$ for $\alpha=a,b$,
where $\Gamma=\sigma_z\otimes\sigma_y$. In our model, ${\cal U}_0$ simply describes the evolution operator of
a Su-Schrieffer-Heeger model in its topological flat band limit, which possesses a winding number $w=1$.
Taking the Taylor expansion of ${\cal U}_\alpha(k_y)$ yields
${\cal U}_\alpha(k_y)=\cos({\cal E})-i(d_{\alpha x}\sigma_x+d_{\alpha z}\sigma_z)$,
for which another winding number can be defined as 
$w_\alpha=\int^{\pi}_{-\pi}\frac{dk_y}{2\pi}\frac{d_{\alpha x}\partial_{k_y}d_{\alpha z}-d_{\alpha z}\partial_{k_y}d_{\alpha x}}{d_{\alpha x}^2+d_{\alpha z}^2}$ for $\alpha=a,b$.
Put together, we obtain a pair of winding numbers $(\nu_a,\nu_b)=w\times (w_a,w_b)$ for the Floquet operators $({\cal U}_a,{\cal U}_b)$.
Their combination results in the integer topological invariants $\nu_0=(\nu_a+\nu_b)/2$ and $\nu_\pi=(\nu_a-\nu_b)/2$ of the
two-dimensional parent system ${\cal U}$, which are related to the numbers of Floquet corner modes at zero and $\pi$
quasienergies $n_0$ and $n_\pi$ through the relations $(n_0,n_\pi)=4(|\nu_0|,|\nu_\pi|)$ \cite{NHFTI5}.
Following the analysis in the main text, $(\nu_0,\nu_\pi)$ could also determine the numbers of 
fractional-quasienergy corner modes in the $q$th-root descendants of the parent model ${\cal U}$
so long as the chiral symmetry $\Gamma$ is preserved.

\section{Stability to disorder}\label{sec:app3}

\begin{figure*}
\begin{centering}
\includegraphics[scale=0.7]{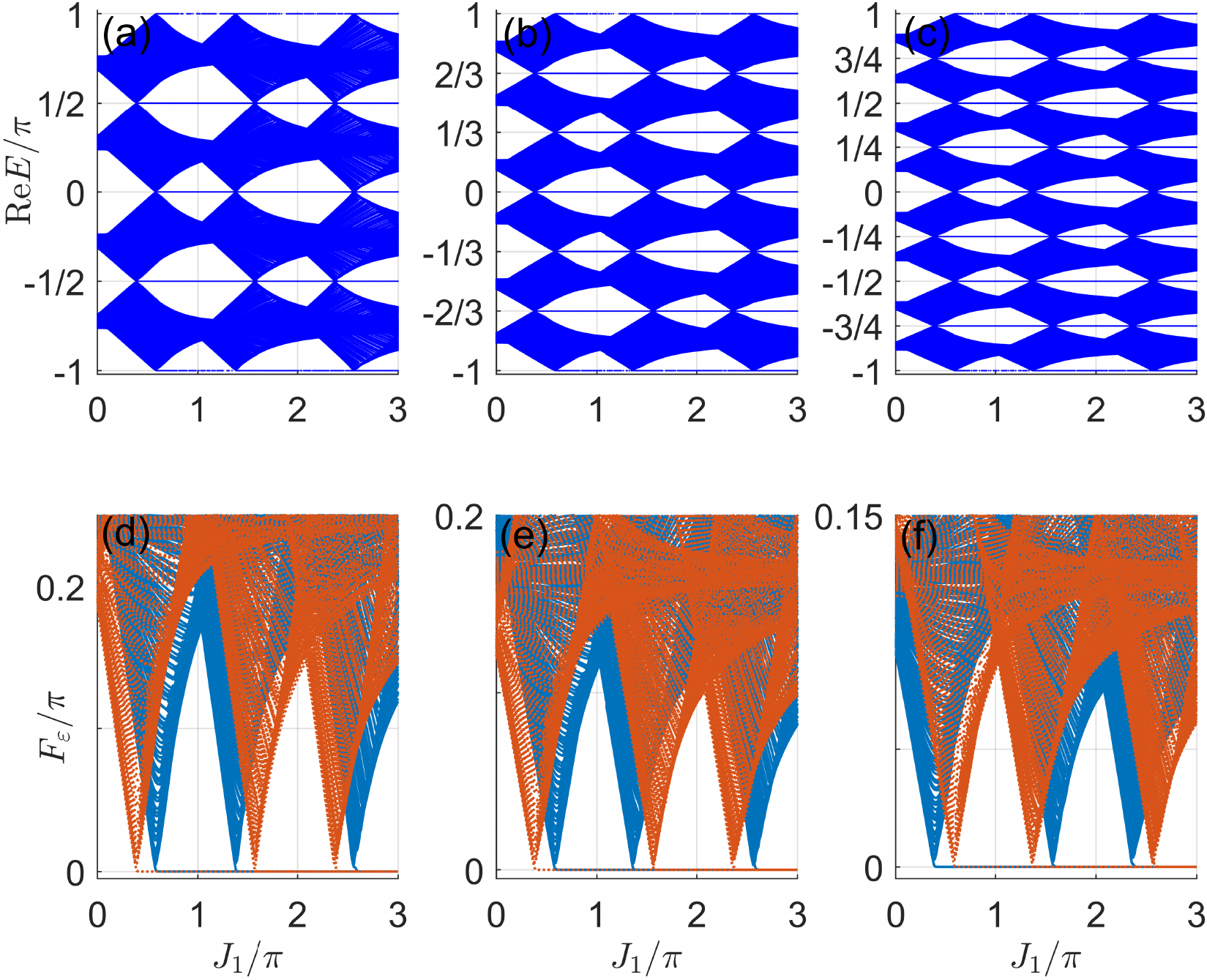}
\par\end{centering}
\caption{Floquet spectrum $E$ [(a)--(c)] and gap function $F_{\varepsilon}$ [(d)--(f)] of $U_{1/2}$
		{[}Eq.~(\ref{eq:U1ov2M1}){]}, $U_{1/3}$ {[}Eq.~(\ref{eq:U1ov3M1}){]} and $U_{1/4}$ {[}obtained from
		Eqs.~(\ref{eq:U1ov2M1}), (\ref{eq:H1ov4}) and (\ref{eq:U1ov4}){]} versus $J_{1}$ with disorder and under the OBC. The size of lattice, other system parameters and color schemes used for all panels are the same as those used for Figs.~\ref{fig:E0E1ov2M1} and \ref{fig:E1ov3E1ov4M1}.\label{fig:App1}}
\end{figure*}

\begin{figure*}
\begin{centering}
\includegraphics[scale=0.7]{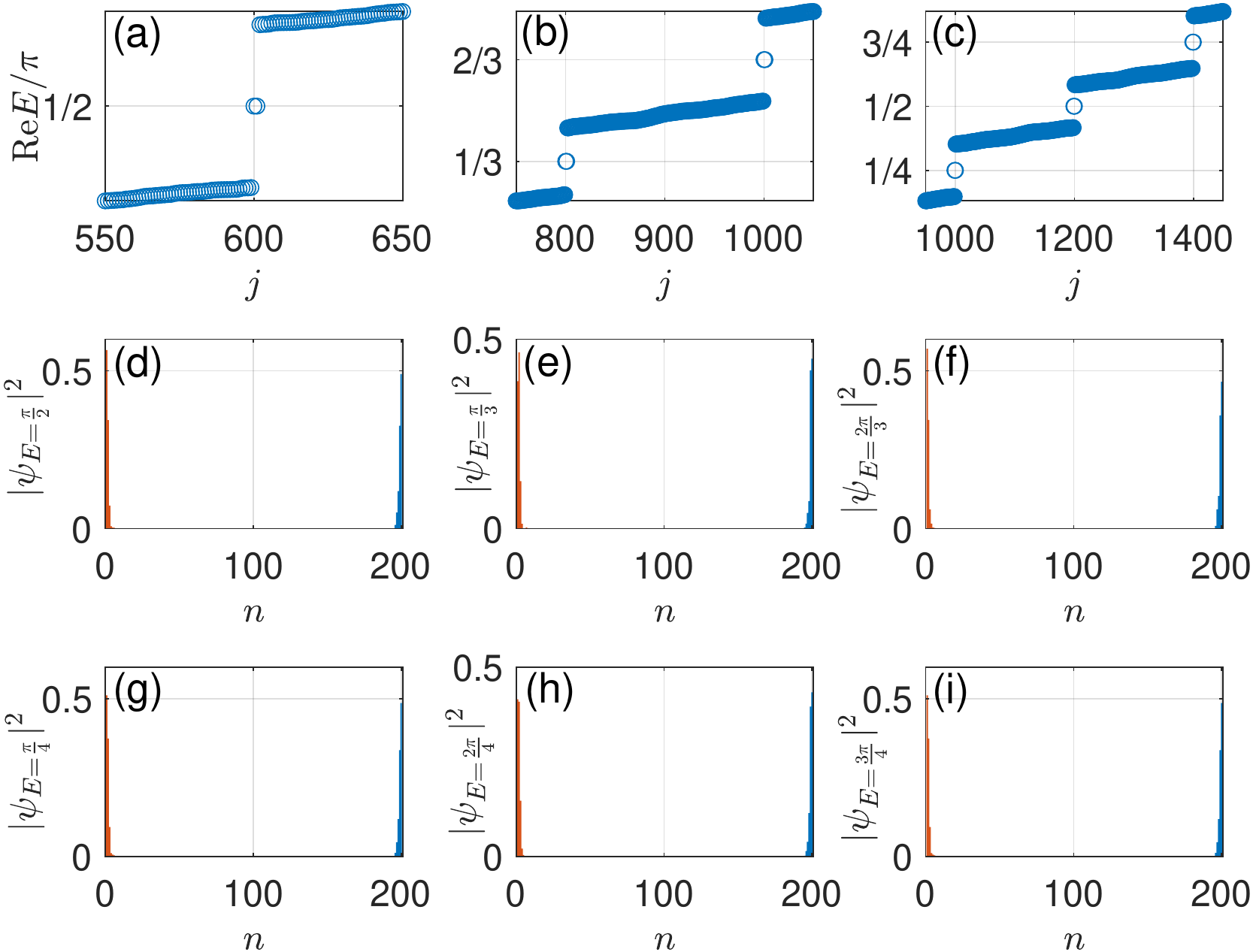}
\par\end{centering}
\caption{The real part of Floquet spectrum and fractional-quasienergy edge modes of the $q$th-root non-Hermitian FTIs for $q=2,3,4$ with disorder. The notations, length of lattice and system parameters are the same as those used in Fig.~\ref{fig:EProb1}.\label{fig:App2}}
\end{figure*}

\begin{figure*}
\begin{centering}
\includegraphics[scale=0.7]{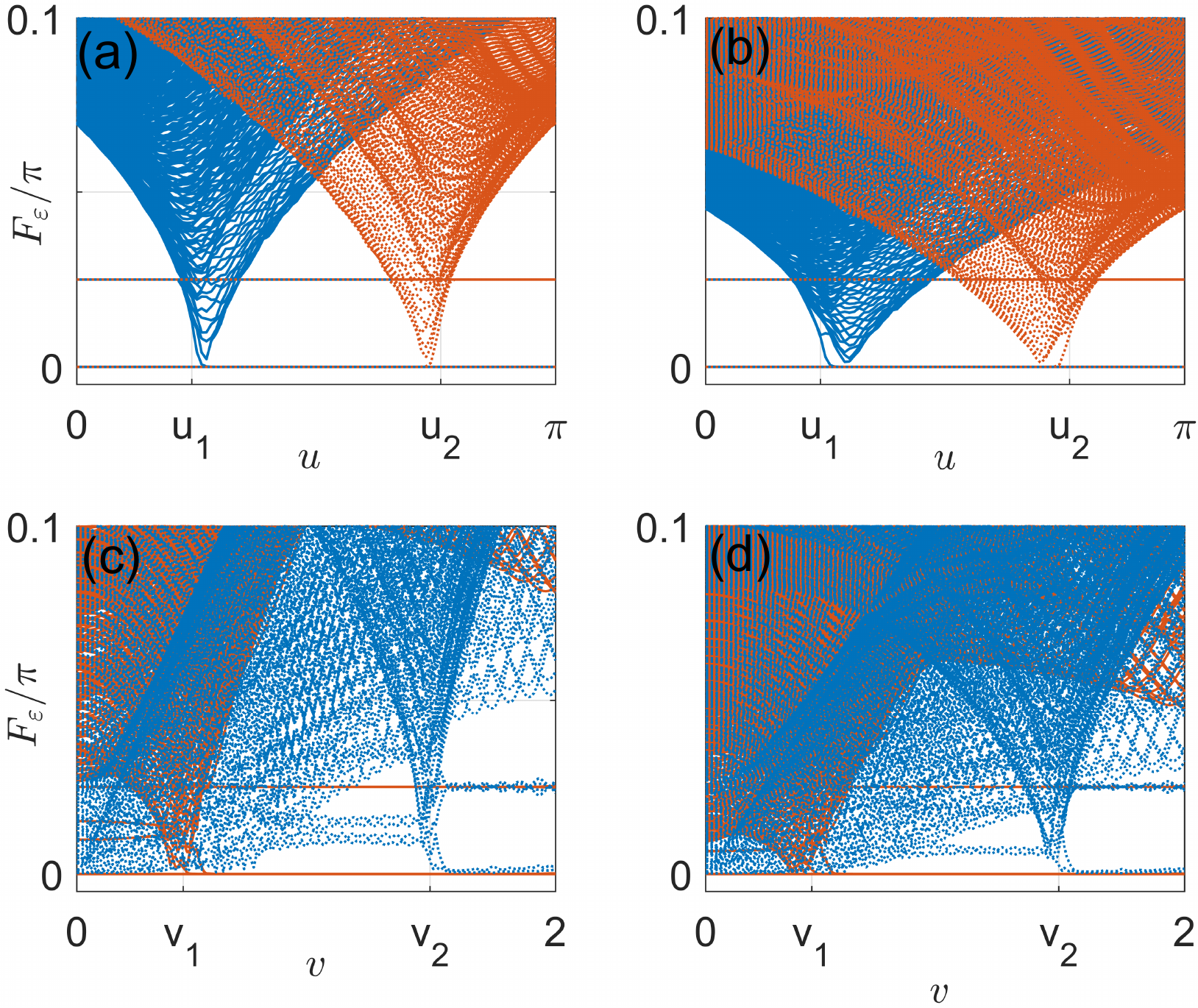}
\par\end{centering}
\caption{Gap function $F_{\varepsilon}$ of ${\cal U}_{1/2}$ {[}Eq.~(\ref{eq:U1ov2M2}){]} in (a), (c)
		and ${\cal U}_{1/3}$ {[}Eq.~(\ref{eq:U1ov3M2}){]} in (b), (d) versus $u$ and $v$ with disorder. The size of lattice and other system parameters are the same as those used in Figs.~\ref{fig:E1ov2E1ov3M2} and \ref{fig:F1ov2F1ov3M2}.\label{fig:App3}}
\end{figure*}

\begin{figure*}
\begin{centering}
\includegraphics[scale=0.7]{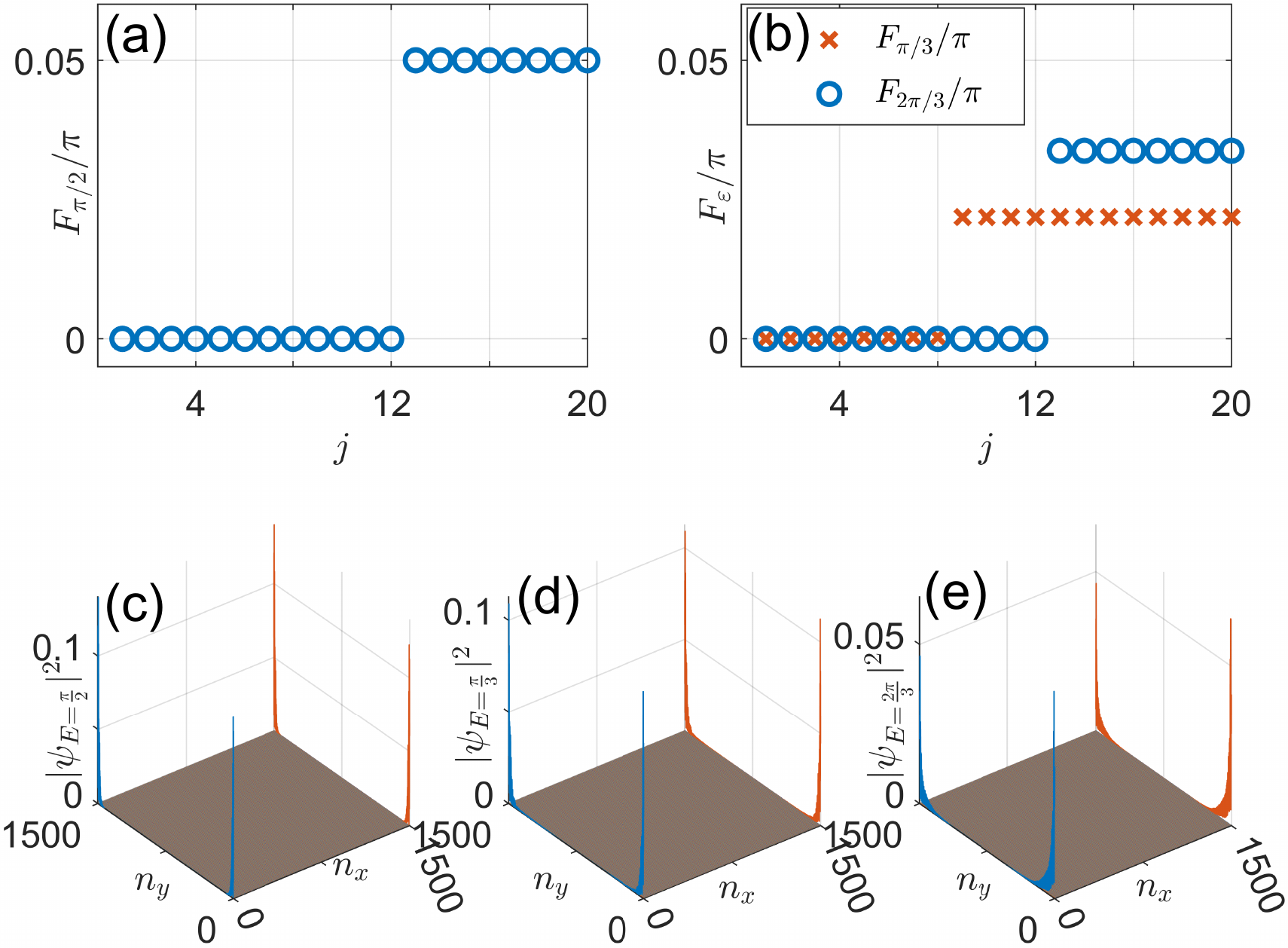}
\par\end{centering}
\caption{Gap functions and probability distributions of corner modes for ${\cal U}_{1/2}$ in (a), (c) and for ${\cal U}_{1/3}$ in (b), (d), (e). The notations, system parameters and size of lattice are the same as those used in Fig.~\ref{fig:EProb2}. (c) shows four out of the twelve $\pi/2$ corner modes. (d) shows four out of the eight $\pi/3$ corner modes. (e) shows four out of the twelve $2\pi/3$ corner modes.\label{fig:App4}}
\end{figure*}

In this section, we demonstrate the stability of $q$th-root Floquet topological phases to disorder through numerical calculations. For the non-Hermitian FTI, we add random intracell coupling terms $H_{1d}=\sum_n W_{n}|n\rangle\langle n|\otimes\sigma_y$ to $H_1$ and $H_{2d}=\sum_n W'_{n}|n\rangle\langle n|\otimes\sigma_x$ to $H_2$ in Eqs.~(\ref{eq:H1}) and (\ref{eq:H2}), respectively. Here $W_n,W'_n$ take different values for different unit cells $n$ and vary randomly in the range of $[-W,W]$. The form of disorder terms $H_{1d}$ and $H_{2d}$ are chosen to be general enough and also to ensure that the chiral symmetries of the parent and the $q$th-root systems are preserved. In Fig.~\ref{fig:App1}, we show the quasienergy spectrum and gap functions of the $q$th-root non-Hermitian FTI for $q=2,3,4$ with the disorder amplitude $W=0.2$ (comparable with the minimal energy scale of the clean system). The results show that the degenerate edge modes at different fractional quasienergies in the bulk spectrum gaps are well preserved under the impact of disorder. In Fig.~\ref{fig:App2}, we further show the $E=\pi/2,\pi/3,2\pi/3,\pi/4,2\pi/4,3\pi/4$ edge modes and their spatial profiles for $q=2,3,4$ with the same disorder amplitude $W=0.2$. It is clear that these fractional quasienergy edge modes indeed survive in the disordered system and are well localized around the sample boundary. The results presented in Figs.~\ref{fig:App1} and \ref{fig:App2} are obtained for one disorder realization. We also checked a number of other disorder realizations in the calculation and found no observable difference.

For the 2D model, we introduce disorder by adding ${\cal H}_{1d}={\mathbb I}_x\otimes\sum_{n}{\cal W}_n|n\rangle\langle n|\otimes\sigma_z$ and ${\cal H}_{2d}={\mathbb I}_x\otimes\sum_{n}{\cal W}'_n|n\rangle\langle n|\otimes\sigma_x$ to ${\cal H}_{y1}$ and ${\cal H}_{y2}$ in Eqs.~(\ref{eq:Hy1}) and (\ref{eq:Hy2}), respectively. The ${\cal W}_n$ and ${\cal W}'_n$ take different values for different cell indices $n$ and vary randomly in the range of $[-{\cal W},{\cal W}]$. We also choose the disorder terms in ${\cal H}_{1d}$ and ${\cal H}_{2d}$ to be general enough and to make sure that the chiral symmetries of the parent and the $q$th-root models are retained. In Figs.~\ref{fig:App3} and \ref{fig:App4}, we present the gap functions and the spatial profiles of fractional-quasienergy corner modes of the $q$th-root non-Hermitian FSOTIs for $q=2,3$ and the disorder amplitude ${\cal W}=0.2$ (comparable with the minimal energy scale of the clean system). The numerical results clearly suggest that the $q$th-root second order topological phases and their accompanying corner states in our system are robust to perturbations induced by symmetry-preserving disorder. The results presented in Figs.~\ref{fig:App3} and \ref{fig:App4} are obtained for one disorder realization. We also checked a number of other disorder realizations in the calculation and found no observable difference.


\begin{thebibliography}{99}
	
	\bibitem{FTPRev1} T. Kitagawa, \textit{Topological phenomena in quantum walks},
	Quantum Inf. Process. \textbf{11}, 1107 (2012), \doi{10.1007/s11128-012-0425-4}.
	
	\bibitem{FTPRev2} J. Cayssol, B. D\'ora, F. Simon, and R. Moessner,
	\textit{Floquet topological insulators}, Phys. Status Solidi RRL \textbf{7},
	101 (2013), \doi{10.1002/pssr.201206451}.
	
	\bibitem{FTPRev3} A. Eckardt, \textit{Atomic quantum gases in periodically
	driven optical lattices}, Rev. Mod. Phys. \textbf{89}, 011004 (2017), \doi{10.1103/RevModPhys.89.011004}.
	
	\bibitem{FTPRev4} T. Oka and S. Kitamura, \textit{Floquet Engineering of Quantum
	Materials}, Annu. Rev. Condens. Matter Phys. \textbf{10}, 387 (2019), \doi{10.1146/annurev-conmatphys-031218-013423}.
	
	\bibitem{FTPRev5} M. S. Rudner and N. H. Lindner, \textit{Band structure engineering
	and non-equilibrium dynamics in Floquet topological insulators}, Nat.
	Rev. Phys. \textbf{2}, 229 (2020), \doi{10.1038/s42254-020-0170-z}.
	
	\bibitem{FTPRev6} F. Harper, R. Roy, M. S. Rudner, and S. L. Sondhi,
	\textit{Topology and Broken Symmetry in Floquet Systems}, Annu. Rev. Condens.
	Matter Phys. \textbf{11}, 345 (2020), \doi{10.1146/annurev-conmatphys-031218-013721}.
	
	\bibitem{FTPClass1} F. Nathan and M. S. Rudner, \textit{Topological singularities
	and the general classification of Floquet-Bloch systems}, New J. Phys.
	\textbf{17}, 125014 (2015), \doi{10.1088/1367-2630/17/12/125014}.
	
	\bibitem{FTPClass2} R. Roy and F. Harper, \textit{Periodic table for Floquet
	topological insulators}, Phys. Rev. B \textbf{96}, 155118 (2017), \doi{10.1103/PhysRevB.96.155118}.
	
	\bibitem{FTPClass3} S. Yao, Z. Yan, and Z. Wang, \textit{Topological invariants
	of Floquet systems: General formulation, special properties, and Floquet
	topological defects}, Phys. Rev. B \textbf{96}, 195303 (2017), \doi{10.1103/PhysRevB.96.195303}.
	
	\bibitem{FTPClass4} D. V. Else and C. Nayak, \textit{Classification of topological
	phases in periodically driven interacting systems}, Phys. Rev. B \textbf{93},
	201103(R) (2016), \doi{10.1103/PhysRevB.93.201103}.
	
	\bibitem{FTPClass5} A. C. Potter, T. Morimoto, and A. Vishwanath,
	\textit{Classification of Interacting Topological Floquet Phases in One Dimension},
	Phys. Rev. X \textbf{6}, 041001 (2016), \doi{10.1103/PhysRevX.6.041001}.
	
	\bibitem{FTPClass6} C. Zhang and M. Levin, \textit{Classification of interacting
	Floquet phases with $U(1)$ symmetry in two dimensions}, Phys. Rev.
	B \textbf{103}, 064302 (2021), \doi{10.1103/PhysRevB.103.064302}.
	
	\bibitem{FCAExp1} G. Jotzu, M. Messer, R. Desbuquois, M. Lebrat, T.
	Uehlinger, D. Greif, and T. Esslinger, \textit{Experimental realization of
	the topological Haldane model with ultracold fermions}, Nature \textbf{515},
	237-240 (2014), \doi{10.1038/nature13915}.
	
	\bibitem{FCAExp2} L. Asteria, D. T. Tran, T. Ozawa, M. Tarnowski,
	B. S. Rem, N. Fl\"aschner, K. Sengstock, N. Goldman, and C. Weitenberg,
	\textit{Measuring quantized circular dichroism in ultracold topological matter},
	Nat. Phys. \textbf{15}, 449-454 (2019), \doi{10.1038/s41567-019-0417-8}.
	
	\bibitem{FCAExp3} K. Wintersperger, C. Braun, F. N. \"Unal, A. Eckardt,
	M. D. Liberto, N. Goldman, I. Bloch, and M. Aidelsburger, \textit{Realization
	of an anomalous Floquet topological system with ultracold atoms}, Nat.
	Phys. \textbf{16}, 1058-1063 (2020), \doi{10.1038/s41567-020-0949-y}.
	
	\bibitem{FPHExp1} T. Kitagawa, M. A. Broome, A. Fedrizzi, M. S. Rudner,
	E. Berg, I. Kassal, A. Aspuru-Guzik, E. Demler, and A. G. White, \textit{Observation
	of topologically protected bound states in photonic quantum walks},
	Nat. Commun. \textbf{3}, 882 (2012), \doi{10.1038/ncomms1872}.
	
	\bibitem{FPHExp2} M. C. Rechtsman, J. M. Zeuner, Y. Plotnik, Y. Lumer,
	D. Podolsky, F. Dreisow, S. Nolte, M. Segev, and A. Szameit, Photonic
	\textit{Floquet topological insulators}, Nature\textbf{496}, 196-200 (2013), \doi{10.1038/nature12066}.
	
	\bibitem{FPHExp3} W. Hu, J. C. Pillay, K. Wu, M. Pasek, P. P. Shum,
	and Y. D. Chong, \textit{Measurement of a Topological Edge Invariant in a
	Microwave Network}, Phys. Rev. X \textbf{5}, 011012 (2015), \doi{10.1103/PhysRevX.5.011012}.
	
	\bibitem{FSSExp1} Y. Wang, H. Steinberg, P. Jarillo-Herrero, and N.
	Gedik, \textit{Observation of Floquet-Bloch States on the Surface of a Topological
	Insulator}, Science \textbf{342}, 453-457 (2013), \doi{10.1126/science.1239834}.
	
	\bibitem{FSSExp2} J. W. McIver, B. Schulte, F.-U. Stein, T. Matsuyama,
	G. Jotzu, G. Meier, and A. Cavalleri,
	\textit{Light-induced anomalous Hall effect in graphene}, Nat. Phys. \textbf{16},
	38-41 (2020), \doi{10.1038/s41567-019-0698-y}.
	
	\bibitem{FSSExp3} B. Chen, S. Li, X. Hou, F. Ge, F. Zhou, P. Qian,
	F. Mei, S. Jia, N. Xu, and H. Shen, \textit{Digital quantum simulation of
	Floquet topological phases with a solid-state quantum simulator}, Photon.
	Res. \textbf{9}, 81-87 (2021), \doi{10.1364/PRJ.404163}.
	
	\bibitem{FQC1} R. W. Bomantara and J. Gong, \textit{Simulation of Non-Abelian
	Braiding in Majorana Time Crystals}, Phys. Rev. Lett. \textbf{120},
	230405 (2018), \doi{10.1103/PhysRevLett.120.230405}.
	
	\bibitem{FQC2} R. W. Bomantara and J. Gong, \textit{Quantum computation via
	Floquet topological edge modes}, Phys. Rev. B \textbf{98}, 165421 (2018), \doi{10.1103/PhysRevB.98.165421}.
	
	\bibitem{FQC3} R. W. Bomantara and J. Gong, \textit{Measurement-only quantum
	computation with Floquet Majorana corner modes}, Phys. Rev. B \textbf{101},
	085401 (2020), \doi{10.1103/PhysRevB.101.085401}.
	
	\bibitem{SRTP1} J. Arkinstall, M. H. Teimourpour, L. Feng, R. El-Ganainy,
	and H. Schomerus, \textit{Topological tight-binding models from nontrivial
	square roots}, Phys. Rev. B \textbf{95}, 165109 (2017), \doi{10.1103/PhysRevB.95.165109}.
	
	\bibitem{Gordon} W. Gordon, \textit{Der Comptoneffekt nach der Schr\"odingerschen
	Theorie}, Zeitschrift F\"ur Physik \textbf{40}, 117-133 (1926), \doi{10.1007/BF01390840}.
	
	\bibitem{Klein} O. Klein, \textit{Elektrodynamik und Wellenmechanik vom Standpunkt
	des Korrespondenzprinzips}, Zeitschrift f\"ur Physik \textbf{41}, 407-422
	(1927), \doi{10.1007/BF01400205}.
	
	\bibitem{Dirac} P. A. M. Dirac, \textit{The quantum theory of the electron},
	Proc. R. Soc. London A \textbf{117}, 610 (1928), \doi{10.1098/rspa.1928.0023}.
	
	\bibitem{SRTP2} G. Pelegr\'i, A. M. Marques, R. G. Dias, A. J. Daley,
	V. Ahufinger, and J. Mompart, \textit{Topological edge states with ultracold
	atoms carrying orbital angular momentum in a diamond chain}, Phys.
	Rev. A \textbf{99}, 023612 (2019), \doi{10.1103/PhysRevA.99.023612}.
	
	\bibitem{SRTP3} M. Kremer, I. Petrides, E. Meyer, M. Heinrich, O.
	Zilberberg, and A. Szameit, \textit{A square-root topological insulator with
	non-quantized indices realized with photonic Aharonov-Bohm cages},
	Nat. Commun. \textbf{11}, 907 (2020), \doi{10.1038/s41467-020-14692-4}.
	
	\bibitem{SRTP4} L. Song, H. Yang, Y. Cao, and P. Yan, \textit{Realization
	of the square-root higher-order topological insulator in electric
	circuits}, Nano Lett. \textbf{20}, 7566 (2020), \doi{10.1021/acs.nanolett.0c03049}.
	
	\bibitem{SRTP5} M. Yan, X. Huang, L. Luo, J. Lu, W. Deng, and Z. Liu,
	\textit{Acoustic square-root topological states}, Phys. Rev. B \textbf{102},
	180102(R) (2020), \doi{10.1103/PhysRevB.102.180102}.
	
	\bibitem{SRTP6} T. Mizoguchi, Y. Kuno, and Y. Hatsugai, \textit{Square-root
	higher-order topological insulator on a decorated honeycomb lattice},
	Phys. Rev. A \textbf{102}, 033527 (2020), \doi{10.1103/PhysRevA.102.033527}.
	
	\bibitem{SRTP8} S. Ke, D. Zhao, J. Fu, Q. Liao, B. Wang, and P. Lu,
	\textit{Topological Edge Modes in Non-Hermitian Photonic Aharonov-Bohm Cages},
	IEEE J. Sel. Top. Quantum Electron. \textbf{26}, 4401008 (2020), \doi{10.1109/JSTQE.2020.3010586}.
	
	\bibitem{SRTP7} M. Ezawa, \textit{Systematic construction of square-root topological
	insulators and superconductors}, Phys. Rev. Res. \textbf{2}, 033397
	(2020), \doi{10.1103/PhysRevResearch.2.033397}.
	
	\bibitem{SRTP13} A. M. Marques, L. Madail, and R. G. Dias, \textit{One-dimensional
	$2^{n}$-root topological insulators and superconductors}, Phys. Rev.
	B \textbf{103}, 235425 (2021), \doi{10.1103/PhysRevB.103.235425}.
	
	\bibitem{SRTP14} A. M. Marques and R. G. Dias, \textit{$2^{n}$-root weak,
	Chern, and higher-order topological insulators, and $2^{n}$-root
	topological semimetals}, Phys. Rev. B \textbf{104}, 165410 (2021), \doi{10.1103/PhysRevB.104.165410}.
	
	\bibitem{SRTP11} T. Yoshida, T. Mizoguchi, Y. Kuno, and Y. Hatsugai,
	\textit{Square-root topological phase with time-reversal and particle-hole
	symmetry}, Phys. Rev. B \textbf{103}, 235130 (2021), \doi{10.1103/PhysRevB.103.235130}.
	
	\bibitem{SRTP9} T. Mizoguchi, T. Yoshida, and Y. Hatsugai, \textit{Square-root
	topological semimetals}, Phys. Rev. B \textbf{103}, 045136 (2021), \doi{10.1103/PhysRevB.103.045136}.
	
	\bibitem{SRTP10} Z. Lin, S. Ke, X. Zhu, and X. Li, \textit{Square-root non-Bloch
	topological insulators in non-Hermitian ring resonators}, Opt. Express
	\textbf{29}, 8462 (2021), \doi{10.1364/OE.419852}.
	
	\bibitem{SRTP12} R. G. Dias and A. M. Marques, \textit{Matryoshka approach
	to sinecosine topological models}, Phys. Rev. B \textbf{103}, 245112
	(2021), \doi{10.1103/PhysRevB.103.245112}.
	
	\bibitem{AFTI1} M. S. Rudner, N. H. Lindner, E. Berg, and M. Levin,
	\textit{Anomalous Edge States and the Bulk-Edge Correspondence for Periodically
	Driven Two-Dimensional Systems}, Phys. Rev. X \textbf{3}, 031005 (2013), \doi{10.1103/PhysRevX.3.031005}.
	
	\bibitem{AFTI2} P. Titum, E. Berg, M. S. Rudner, G. Refael, and N. H. Lindner, 
	\textit{Anomalous Floquet-Anderson Insulator as a Nonadiabatic
	Quantized Charge Pump}, Phys. Rev. X \textbf{6}, 021013 (2016), \doi{10.1103/PhysRevX.6.021013}.
	
	\bibitem{AFTI3} L. Zhou and J. Gong, \textit{Recipe for creating an arbitrary
	number of Floquet chiral edge states}, Phys. Rev. B \textbf{97}, 245430
	(2018), \doi{10.1103/PhysRevB.97.245430}.
	
	\bibitem{FSRTP1} R. W. Bomantara, \textit{Square-root Floquet topological phases and time crystals}, arXiv:2111.14327, \doi{10.48550/arXiv.2111.14327}.
	
	\bibitem{NHRev1} V. M. M. Alvarez, J. E. B. Vargas, M. Berdakin, 
	and L. E. F. F. Torres, \textit{Topological states of non-Hermitian systems}, Eur. Phys. J. Spec. Top. {\bf 227}, 1295 (2018), \doi{10.1140/epjst/e2018-800091-5}.
	
	\bibitem{NHRev2} A. Ghatak and T. Das, \textit{New topological invariants in non-Hermitian systems},
	J. Phys.: Condens. Matter {\bf 31}, 263001 (2019), \doi{10.1088/1361-648X/ab11b3}.
	
	\bibitem{NHRev3} Y. Ashida, Z. Gong, and M. Ueda, \textit{Non-Hermitian physics}, Adv. Phys. {\bf 69}, 249-435 (2020), \doi{10.1080/00018732.2021.1876991}.
	
	\bibitem{NHRev4} C. Coulais, R. Fleury, and J. van Wezel, \textit{Topology and broken Hermiticity},
	Nat. Phys. {\bf 17}, 9-13 (2021), \doi{10.1038/s41567-020-01093-z}.
	
	\bibitem{NHRev5} E. J. Bergholtz, J. C. Budich, and F. K. Kunst, \textit{Exceptional topology of non-Hermitian systems},
	Rev. Mod. Phys. {\bf 93}, 015005 (2021), \doi{10.1103/RevModPhys.93.015005}.
	
	\bibitem{NHSE1} V. M. M. Alvarez, J. E. B. Vargas, and L. E. F. F. Torres,
	\textit{Non-Hermitian robust edge states in one dimension: Anomalous localization and eigenspace condensation at exceptional points},
	Phys. Rev. B {\bf 97}, 121401(R) (2018), \doi{10.1103/PhysRevB.97.121401}.
	
	\bibitem{NHSE2} S. Yao and Z. Wang, \textit{Edge States and Topological Invariants
	of Non-Hermitian Systems}, Phys. Rev. Lett. \textbf{121}, 086803 (2018), \doi{10.1103/PhysRevLett.121.086803}.
	
	\bibitem{NHSE3} F. Song, S. Yao, and Z. Wang, \textit{Non-Hermitian Topological
	Invariants in Real Space}, Phys. Rev. Lett. \textbf{123}, 246801 (2019), \doi{10.1103/PhysRevLett.123.246801}.
	
	\bibitem{NHSE4} K. Yokomizo and S. Murakami, \textit{Non-Bloch Band Theory
	of Non-Hermitian Systems}, Phys. Rev. Lett. \textbf{123}, 066404 (2019), \doi{10.1103/PhysRevLett.123.066404}.
	
	\bibitem{NHSE5} C. H. Lee and R. Thomale, \textit{Anatomy of skin modes and
	topology in non-Hermitian systems}, Phys. Rev. B \textbf{99}, 201103(R)
	(2019), \doi{10.1103/PhysRevB.99.201103}.
	
	\bibitem{NHSE6} C. H. Lee, L. Li, and J. Gong, \textit{Hybrid Higher-Order Skin-Topological Modes in Nonreciprocal Systems},
	Phys. Rev. Lett. {\bf 123}, 016805 (2019), \doi{10.1103/PhysRevLett.123.016805}.
	
	\bibitem{NHSE7} L. Li, S. Mu, C. H. Lee, and J. Gong, \textit{Quantized classical response from spectral winding topology},
	Nat. Commun. {\bf 12}, 5294 (2021), \doi{10.1038/s41467-021-25626-z}.
	
	\bibitem{NHFTI1} L. Zhou and J. Gong, \textit{Non-Hermitian Floquet topological
	phases with arbitrarily many real-quasienergy edge states}, Phys. Rev.
	B \textbf{98}, 205417 (2018), \doi{10.1103/PhysRevB.98.205417}.
	
	\bibitem{NHFTI2} Z. Turker, S. Tombuloglu, and C. Yuce, \textit{PT symmetric
	Floquet topological phase in SSH model}, Phys. Lett. A, \textbf{382},
	2013 (2018), \doi{10.1016/j.physleta.2018.05.015}. 
	
	\bibitem{NHFTI3} L. Zhou and J. Pan, \textit{Non-Hermitian Floquet topological
	phases in the double-kicked rotor}, Phys. Rev. A \textbf{100}, 053608
	(2019), \doi{10.1103/PhysRevA.100.053608}.
	
	\bibitem{NHFTI4} L. Zhou, \textit{Dynamical characterization of non-Hermitian
	Floquet topological phases in one dimension}, Phys. Rev. B \textbf{100},
	184314 (2019), \doi{10.1103/PhysRevB.100.184314}.
	
	\bibitem{NHFTI5} J. Pan and L. Zhou \textit{Non-Hermitian Floquet second order
	topological insulators in periodically quenched lattices}, Phys. Rev.
	B \textbf{102}, 094305 (2020), \doi{10.1103/PhysRevB.102.094305}.
	
	\bibitem{NHFTI6} L. Zhou, \textit{Non-Hermitian Floquet phases with even-integer
	topological invariants in a periodically quenched two-leg ladder},
	Entropy \textbf{22}, 746 (2020), \doi{10.3390/e22070746}.
	
	\bibitem{NHFTI7} X. Zhang and J. Gong, \textit{Non-Hermitian Floquet topological
	phases: Exceptional points, coalescent edge modes, and the skin effect},
	Phys. Rev. B \textbf{101}, 045415 (2020), \doi{10.1103/PhysRevB.101.045415}.
	
	\bibitem{NHFTI9} H. Wu and J. An, \textit{Floquet topological phases of non-Hermitian
	systems}, Phys. Rev. B \textbf{102}, 041119(R) (2020), \doi{10.1103/PhysRevB.102.041119}.
	
	\bibitem{NHFTI8} L. Zhou, Y. Gu, and J. Gong, \textit{Dual topological characterization
	of non-Hermitian Floquet phases}, Phys. Rev. B \textbf{103}, L041404
	(2021), \doi{10.1103/PhysRevB.103.L041404}.
	
	\bibitem{NHFTI10} Y. Cao, Y. Li, and X. Yang, \textit{Non-Hermitian bulk-boundary
	correspondence in a periodically driven system}, Phys. Rev. B \textbf{103},
	075126 (2021), \doi{10.1103/PhysRevB.103.075126}.
	
	\bibitem{NHFTI11} S. Wu, W. Song, S. Gao, Y. Chen, S. Zhu, and T.
	Li, \textit{Floquet $\pi$ mode engineering in non-Hermitian waveguide lattices},
	Phys. Rev. Res. \textbf{3}, 023211 (2021), \doi{10.1103/PhysRevResearch.3.023211}.
	
	\bibitem{NHFTI12} V. M. Vyas and D. Roy, \textit{Topological aspects of periodically
	driven non-Hermitian Su-Schrieffer-Heeger model}, Phys. Rev. B \textbf{103},
	075441 (2021), \doi{10.1103/PhysRevB.103.075441}.
	
	\bibitem{NHFTSC1} M. van Caspel, S. E. T. Arze, and I. P. Castillo,
	\textit{Dynamical signatures of topological order in the driven-dissipative
	Kitaev chain}, SciPost Phys. \textbf{6}, 026 (2019), \doi{10.21468/SciPostPhys.6.2.026}.
	
	\bibitem{NHFTSC2} L. Zhou, \textit{Non-Hermitian Floquet topological superconductors
	with multiple Majorana edge modes}, Phys. Rev. B \textbf{101}, 014306
	(2020), \doi{10.1103/PhysRevB.101.014306}.
	
	\bibitem{NHFSM1} P. He and Z. Huang, \textit{Floquet engineering and simulating
	exceptional rings with a quantum spin system}, Phys. Rev. A \textbf{102},
	062201 (2020), \doi{10.1103/PhysRevA.102.062201}.
	
	\bibitem{NHFSM2} A. Banerjee and A. Narayan, \textit{Controlling exceptional
	points with light}, Phys. Rev. B \textbf{102}, 205423 (2020), \doi{10.1103/PhysRevB.102.205423}.
	
	\bibitem{NHFSM3} D. Chowdhury, A. Banerjee, and A. Narayan, \textit{Light-driven
	Lifshitz transitions in non-Hermitian multi-Weyl semimetals}, Phys.
	Rev. A \textbf{103}, L051101 (2021), \doi{10.1103/PhysRevA.103.L051101}.
	
	\bibitem{NHFSM4} D. Chowdhury, A. Banerjee, and A. Narayan, \textit{Exceptional
	hexagonal warping effect in multi-Weyl semimetals}, Phys. Rev. B \textbf{105},
	075133 (2022), \doi{10.1103/PhysRevB.105.075133}.
	
	\bibitem{NHFQC1} L. Zhou, \textit{Floquet engineering of topological localization
	transitions and mobility edges in one-dimensional non-Hermitian quasicrystals},
	Phys. Rev. Res. \textbf{3}, 033184 (2021), \doi{10.1103/PhysRevResearch.3.033184}.
	
	\bibitem{NHFQC3} L. Zhou and W. Han, \textit{Driving-induced multiple PT-symmetry breaking 
	and reentrant localization transitions in non-Hermitian Floquet quasicrystals}, arXiv:2203.03995, \doi{10.48550/arXiv.2203.03995}.
	
	\bibitem{NHFQC2} S. Weidemann, M. Kremer, S. Longhi, and A. Szameit,
	\textit{Topological triple phase transition in non-Hermitian Floquet quasicrystals},
	Nature \textbf{601}, 354-359 (2022), \doi{10.1038/s41586-021-04253-0}.
	
	\bibitem{subchiral} A. M. Marques and R. G. Dias, \textit{One-dimensional topological insulators with noncentered inversion symmetry axis}, Phys.~Rev.~B {\bf 100}, 041104(R) (2019), \doi{10.1103/PhysRevB.100.041104} 
	
	\bibitem{FTC1} B. Wang, J. Quan, J. Han, X. Shen, H. Wu, and Y. Pan,
	\textit{Observation of Photonic Topological Floquet Time Crystals, Laser Photonics}
	Rev. \textbf{2022}, 2100469 (2022), \doi{10.1002/lpor.202100469}.
	
	\bibitem{NHQW1} X. Zhan, L. Xiao, Z. Bian, K. Wang, X. Qiu, B. C. Sanders, W. Yi, and P. Xue, \textit{Detecting Topological Invariants in Nonunitary Discrete-Time Quantum Walks}, Phys. Rev. Lett. {\bf 119}, 130501 (2017), \doi{10.1103/PhysRevLett.119.130501}.
	
	\bibitem{NHQW2} L. Xiao, X. Zhan, Z. H. Bian, K. K. Wang, X. Zhang, X. P. Wang, J. Li, K. Mochizuki, D. Kim, N. Kawakami, W. Yi, H. Obuse, B. C. Sanders, and P. Xue, \textit{Observation of topological edge states in parity-time-symmetric quantum walks}, Nature Phys. {\bf 13}, 1117-1123 (2017), \doi{10.1038/nphys4204}.
	
	\bibitem{NHQW3} K. Wang, X. Qiu, L. Xiao, X. Zhan, Z. Bian, B. C. Sanders, W. Yi, and P. Xue, \textit{Observation of emergent momentum-time skyrmions in parity-time-symmetric non-unitary quench dynamics}, Nat. Commun. {\bf 10}, 2293 (2019), \doi{10.1038/s41467-019-10252-7}.
	
	\bibitem{NHQW4} L. Xiao, T. Deng, K. Wang, Z. Wang, W. Yi, and P. Xue, \textit{Observation of Non-Bloch Parity-Time Symmetry and Exceptional Points}, Phys. Rev. Lett. {\bf 126}, 230402 (2021), \doi{10.1103/PhysRevLett.126.230402}.
	
	\bibitem{MCD}  F.~Cardano, A.~D.~Errico, A.~Dauphin, M.~Maffei, B.~Piccirillo, C.~de~Lisio, G.~D.~Filippis, V.~Cataudella, E.~Santamato, L.~Marrucci, M.~Lewenstein, and P.~Massignan, \textit{Detection of Zak phases and topological invariants in a chiral quantum walk of twisted photons}, Nat. Commun. {\bf 8}, 15516 (2017), \doi{10.1038/ncomms15516}.
	
	\bibitem{spint} B.~Zhu, Y.~Ke, H.~Zhong, and C.~Lee, \textit{Dynamic winding number for exploring band topology}, Phys. Rev. Res. {\bf 2}, 023043 (2020), \doi{10.1103/PhysRevResearch.2.023043}.
	
	\bibitem{Fparafermion} R.~W.~Bomantara, \textit{$Z_4$ parafermion $\pm \pi/2$ modes in an interacting periodically driven superconducting chain}, Phys. Rev. B {\bf 104}, L121410 (2021), \doi{10.1103/PhysRevB.104.L121410}. 
	
	\bibitem{z3para} E.~M.~Stoudenmire, D.~J.~Clarke, R.~S.~K.~Mong, and J.~Alicea, \textit{Assembling Fibonacci anyons from a $Z_3$ parafermion lattice model}, Phys. Rev. B {\bf 91}, 235112 (2015), \doi{10.1103/PhysRevB.91.235112}. 

\end{thebibliography}
\end{appendix}



\end{document}